\documentclass[11pt]{article}
\usepackage[utf8]{inputenc}
\usepackage{amssymb,amsmath}
\usepackage{bm} 
\usepackage{booktabs} 
\usepackage{array}
\usepackage{latexsym}
\usepackage{graphicx}
\usepackage{color}
\usepackage{datetime}
\usepackage[nosort]{cite}
\usepackage{verbatim}
\usepackage{enumerate}
\usepackage{chngpage} 
\usepackage{mathrsfs}
\usepackage{pgfplots}
\usepackage{euscript}
\usepackage{psfrag}
\usepackage{wrapfig}
\usepackage{subcaption}

\usepackage{subcaption}
\usepackage{datetime}

\usepackage[nosort]{cite}
\usepackage{chngpage} 
\usepackage{setspace}
\usepackage{tensor}
\usepackage{physics}

\usepackage{multirow}
\usepackage{stmaryrd}

\usepackage{mciteplus}

\usepackage[colorlinks=true,      linkcolor=blue,      urlcolor=blue,      
            filecolor=blue,      citecolor=blue,       pdfstartview=FitH,     
						pdfpagemode=UseNone,      bookmarksopen=true]{hyperref}  
\usepackage[all]{hypcap}     

\definecolor{cardinal}{rgb}{0.6,0,0}
\definecolor{darkgreen}{rgb}{0,0.4,0}
\definecolor{golden}{rgb}{0.92, 0.7, 0}
\definecolor{midnight}{rgb}{0, 0, 0.5}
\definecolor{darkblue}{rgb}{0, 0, 0.7}
\definecolor{purple}{rgb}{0.5, 0, 0.5}
\newcommand{\Red}{\color{red}}

\def\oneone{\rlap 1\mkern4mu{\rm l}}

\def\IR{\mathbb{R}}

\def\cC{{\cal C}}

\def\cL{{\cal L}}

\def\d{{\rm d}}
\def\AdSSS{$AdS_4 \times S^2 \times S^2 \times \Sigma~$}

\numberwithin{equation}{section}

\usepackage{tikz} 
\usepackage{tikz-cd} 
\usetikzlibrary{decorations.pathreplacing} 
\usetikzlibrary{decorations.pathmorphing} 
\usetikzlibrary{decorations.markings} 
\usetikzlibrary{arrows} 
\usetikzlibrary{shapes} 
\usetikzlibrary{positioning} 

\tikzset{flavor/.style={regular polygon,regular polygon sides=4,inner sep=2.5pt, draw}}
\tikzset{gauge/.style={circle, draw,inner sep=2.5pt}}
\tikzstyle{NS5}=[draw, thick, red] 
\tikzstyle{D5h}=[draw, thick, blue] 
\tikzstyle{D3}=[draw, black] 
\tikzset{D5/.style={circle, draw=blue, inner sep=0pt, fill=blue, minimum size=2mm}}

\newcommand{\curvebellD}[2]{%
  \begin{tikzpicture}[baseline]%
    \begin{scope}[rotate=45]\begin{scope}[cm={1,0,-.7,1,(0,0)}]%
      \pgfmathtruncatemacro{\N}{#1-1}%
      \foreach \k in {0,...,\N}{%
        \draw[thick,blue]
          plot[domain=-3:3, samples=200]
            ({\x},{#2*exp(-\x*\x)+0.1*\k});
      }%
    \end{scope}\end{scope}%
  \end{tikzpicture}%
}\newcommand{\curvebellNS}[2]{%
  \begin{tikzpicture}[baseline]%
    \begin{scope}[rotate=90]\begin{scope}[cm={1,0,0,1,(0,0)}]%
      \pgfmathtruncatemacro{\N}{#1-1}%
      \foreach \k in {0,...,\N}{%
        \draw[thick,red]
          plot[domain=-3:3, samples=200]
            ({\x},{#2*exp(-\x*\x)+0.1*\k});
      }%
    \end{scope}\end{scope}%
  \end{tikzpicture}%
}
\newcommand{\cylNS}[1]{
\begin{tikzpicture}

\def\R{#1}      
\def\H{6}      
\def\k{0.35}   

\fill[red!15]
  (0,\R) -- (\H,\R)
  arc[start angle=90,end angle=-90,x radius=\k*\R,y radius=\R]
  -- (0,-\R)
  arc[start angle=-90,end angle=90,x radius=\k*\R,y radius=\R]
  -- cycle;

\fill[red!20] (0,0) ellipse [x radius=\k*\R, y radius=\R];
\fill[red!20] (\H,0) ellipse [x radius=\k*\R, y radius=\R];

\draw[dashed] (0,0) ++(90:\R) arc[start angle=90,end angle=270,x radius=\k*\R,y radius=\R];
\draw         (0,0) ++(270:\R) arc[start angle=270,end angle=90,x radius=\k*\R,y radius=\R];

\draw (\H,0) ellipse [x radius=\k*\R, y radius=\R];

\draw (0,\R) -- (\H,\R);
\draw (0,-\R) -- (\H,-\R);
\end{tikzpicture}}

\begin{document}

\phantom{AAA}
\vspace{-10mm}

\begin{flushright}
%
%
\end{flushright}

\vspace{-.9cm}

\begin{center}
{\huge {\bf  Zooming out of AdS$_4 \times$S$^2 \times$S$^2$:  }}\\
{\huge {\bf \vspace*{.25cm}  }}
{\huge {\bf   The Branes behind the CFT's}}

\vspace{1cm}

{\large{\bf { Iosif Bena$^{1}$, Antoine Bourget$^{1}$, Rapha\"el Dulac$^{1}$, Dimitrios Toulikas$^{2}$   \\ and  Nicholas P. Warner$^{1,3,4}$}}}

\vspace{1cm}

$^1$Institut de Physique Th\'eorique, \\
Universit\'e Paris-Saclay, CEA, CNRS,\\
Orme des Merisiers, Gif sur Yvette, 91191 CEDEX, France \\[12pt]

$^2$Department of Physics, \\ 
Ben-Gurion University of the Negev, \\
Beer Sheva 84105, Israel \\ [12pt]

\centerline{$^3$Department of Physics and Astronomy}
\centerline{and $^4$Department of Mathematics,}
\centerline{University of Southern California,} 
\centerline{Los Angeles, CA 90089, USA}

\vspace{10mm} 
{\footnotesize\upshape\ttfamily iosif.bena, antoine.bourget, raphael.dulac @ipht.fr, toulikas @post.bgu.ac.il,   warner @usc.edu} \\

\vspace{1cm}
 
\textsc{Abstract}

\end{center}

\begin{adjustwidth}{3mm}{3mm} 
 
We reveal the supersymmetric brane configurations that give rise to  AdS$_4\times$S$^2\times$S$^2$ supergravity solutions, which are holographic duals to three-dimensional $\mathcal{N}=4$ CFTs  or to conformal boundaries and domain walls of four-dimensional $\mathcal{N}=4$ SYM. We show that these solutions preserve the same Killing spinors as orthogonal D3, D5 and NS5 branes in flat space, and that the singular sources of these solutions correspond to semi-infinite D3-D5 and D3-NS5 spikes. 

We track these solutions all the way from the weak-coupling regime of parameters, where the branes do not backreact, to the supergravity regime. We explain how the AdS$_4$ factor arises from certain universal self-similar bending regions of the five-branes, whose steepness is the same as the weak-coupling linking numbers. We also propose a brane configuration that gives rise to the Janus interface solutions.

Our construction gives a clear geometric explanation of the Gaiotto-Witten ``good-bad-ugly'' classification of eight-supercharge theories: only good theories have five-branes that do not cross when back-reacting, and end up sourcing an AdS$_4\times$S$^2\times$S$^2$ solution.

 \vspace{-1.2mm}
\noindent
\end{adjustwidth}

\thispagestyle{empty}
\newpage


\tableofcontents

\section{Introduction}
\label{sec:Intro}

Supersymmetric supergravity solutions based on  warped $\rm AdS \times S\times S \times \Sigma$ geometries, where $\Sigma$ is a Riemann surface, have wide-ranging applications in string theory and holography  \cite{DHoker:2007zhm,DHoker:2007hhe,deBoer:1999gea,DHoker:2008lup,Bachas:2013vza,Apruzzi:2025znw,Heydeman:2025fde,Gaiotto:2009gz,Lozano:2019jza,Lozano:2018pcp}, especially in the study of holographic interfaces, domain walls, boundary CFT's \cite{Jensen:2016pah,Harvey:2025ttz}, double holography and Randall-Sundrum constructions \cite{Bachas:2011xa,Uhlemann:2021nhu}, AdS/BCFT correspondence \cite{Sugimoto:2023oul,He:2025due} and even entanglement islands \cite{Karch:2022rvr}.

However, the exact relation between these solutions and the putative underlying branes is often implicit or unclear. The existence of an AdS factor suggests that these solutions may come from zooming-in on a scaling limit of certain branes, but it can be very difficult to ``step back'' and see the whole picture of the branes whose back-reaction creates the $\rm AdS \times S\times S \times \Sigma$ geometry. The problem is akin to the ancient Buddhist story of the {\it Blind Men and an Elephant}\footnote{See, for example, \href{https://en.wikipedia.org/wiki/Blind_men_and_an_elephant}{Wikipedia}.}: we cannot see the whole if we are limited to the zoomed-in details. In particular, the scaling limits do not allow one to ascertain whether the $\rm AdS \times S\times S \times \Sigma$ solutions come from flat branes, from branes pulled by other branes, by branes with a curved world-volume, or by branes in transverse fields. As we will discuss, for $(2+1)$-dimensional CFTs, ambiguities in the brane configurations are intimately related to the problem of reconstructing UV limits from IR data.

In a series of papers \cite{Bena:2023rzm, Bena:2024dre, Bena:2025hxt}, four of the authors, with Houppe and Chakraborty, have reconstructed the branes underlying the AdS$_3 \times$S$^3 \times$S$^3 \times $$\Sigma$  solutions of 11-dimensional supergravity \cite{Bachas:2013vza,DHoker:2008lup,DHoker:2008rje}. They have shown that {\em all} solutions based on a simple Riemann surface come from particular scaling limits of a system of intersecting  M2, M5 and M5' branes preserving the eight supercharges of the ``straight'' branes \cite{Lunin:2007mj, Bena:2023rzm,Bena:2024dre}. More precisely, the $AdS_3$ radial direction is to be identified with the self-similar region of the spikes formed by back-reaction of semi-infinite M2 branes that end, and hence pull on, the orthogonal M5 and M5' branes. Furthermore, the $\gamma$ parameter that parametrizes the isometry superalgebra of the AdS$_3 \times$S$^3 \times$S$^3 \times $$\Sigma$ solutions was shown to govern the scaling of the M2-M5-M5' coordinates that gives rise to the $AdS_3$ solutions \cite{Bena:2025hxt}. Hence, solutions with different $\gamma$ truly represent different {\em Blind Men} observing different scaling regimes of the same M2-M5-M5' {\em Elephant}. 

The purpose of this paper is to reveal the supersymmetric branes that give rise to $AdS_4 \times S^2 \times S^2 \times \Sigma$ in Type IIB String Theory \cite{DHoker:2007zhm,DHoker:2007hhe,Assel:2011xz}. We will establish that all these solutions come from a scaling limit of a configuration of orthogonal D3, D5 and NS5 branes, preserving 8 supercharges. 

There are five classes of such solutions:
\begin{enumerate}[{\bf A.}]  
    \item When no asymptotic $AdS_5$ region is present, the solutions are dual to the various three-dimensional $\mathcal{N}=4$ CFTs that live in the infrared of the gauge theories on the D3 branes stretched between D5 branes and NS5 branes \cite{Bachas:2011xa,Assel:2011xz}.
    
    \item When one asymptotic $AdS_5$ region is present, the solution is dual to the 4d $\mathcal{N}=4$ SYM theory with various boundary conditions. These boundary conditions correspond to  the different ways in which D3 branes can end on D5 and NS5 branes, and were classified by Gaiotto and Witten \cite{Gaiotto:2008sa,Gaiotto:2008ak}.

    \item When two asymptotic $AdS_5$ regions of different central charges are present, the solution must have D5 and/or NS5 sources that account for the difference of the central charges, and describes a {\em Janus domain wall}. 
    
    \item When the solution has two asymptotic $AdS_5$ regions with the same central charge, and has no D5 or NS5 sources, it describes a {\em Janus interface}\cite{DHoker:2006qeo}.

    \item There are also multi-Janus solutions with multiple $AdS_5$ asymptotic regions \cite{DHoker:2007hhe}.

\end{enumerate} 

To reveal the configurations of D3, D5 and NS5 branes whose back-reaction gives rise to these $AdS_4 \times S^2 \times S^2 \times \Sigma$ solutions in a scaling regime, we will work in three regimes of parameters.

\begin{enumerate} [{\bf 1.}]

\item The first is the no-back-reaction regime, where one can read off the field theory on the D3 branes by simply describing which D3 branes end on which D5 and NS5 branes\cite{Hanany:1996ie}. 

\item The second is the partial-back-reaction regime, in which one treats some branes as probes in the background sourced by the others, and can identify various regions with scaling symmetry on the brane world-volume\cite{Callan:1998iq, Pelc:2000kb,Marolf:2000ci}.

\item The third is the fully back-reacted regime, where the supergravity solutions sourced by the branes admit one of the five scaling $AdS_4 \times S^2 \times S^2 \times \Sigma$ regions described above\cite{DHoker:2007zhm,DHoker:2007hhe}.

\end{enumerate}

We will begin with the fully back-reacted solutions, in regime {\bf 3}, which are determined by two harmonic functions on a Riemann surface, $\Sigma$, with the topology of an infinite strip. On the top and bottom sides of the strip there can be singularities corresponding to objects that have D5 or NS5 charges, as well as a non-trivial D3 Page charge.  Furthermore, at the $- \infty$ and $+ \infty$ ends of the strip, the divergence of the harmonic functions determines the asymptotics. 

Type {\bf A} solutions, which have no asymptotic $AdS_5$ region, are expected to come from the back-reaction of D3 branes stretched between D5 and NS branes. Type {\bf B} solutions have one asymptotic $AdS_5$ region, and hence are expected to come from the back-reaction of semi-infinite D3 branes ending on five-branes. Type {\bf C} and {\bf D} solutions describe a collection of infinite D3 branes (hence two asymptotic $AdS_5$ regions) together with (for Type-{\bf C} solutions only) some semi-infinite ones ending on five branes. 

Type {\bf E} solutions have multiple asymptotic regions and have a higher level of technical complexity. Therefore, we will largely focus on type-{\bf A}, {\bf B}, {\bf C} and {\bf D} solutions, in which the Riemann surface is an infinite strip. However, our results also apply to type {\bf E} solutions, and we will briefly comment on their brane realization.

The first step in our analysis is to relate the Riemann-surface coordinates and the $AdS_4$ radial variable of $AdS_4 \times S^2 \times S^2 \times \Sigma$ solutions to the coordinates used for writing 8-supercharge solutions describing orthogonal D3, D5 and NS5 branes. This will allow us to show that all 
$AdS_4 \times S^2 \times S^2 \times \Sigma$ solutions preserve the 8 Killing spinors of the D3-D5-NS5 solutions. Furthermore, it will allow us to show that these solutions come from a certain scaling limit of the full asymptotically-flat back-reacted D3-D5-NS5 solution. Because these solutions have a dilaton that is independent of the $AdS_4$ radius, it is not hard to argue that this scaling limit zooms in on a region where the physics is dominated by the D3 branes. This distinguishes these solutions from the $AdS_3 \times S^3 \times S^3 \times \Sigma$ solutions, which can come both from M2-dominated and from M5-dominated regions of the full solution \cite{Bena:2025hxt}.

The second step of our analysis is to place brane probes in the $AdS_4 \times S^2 \times S^2 \times \Sigma$ solutions to reveal the geometry of the D5 and NS5 sources of these solutions. We will show\footnote{This will be done by comparing the location of a probe brane in a solution with $n_5$ D5 and NS5 sources to a solution with $n_5+1$ sources.} that these D5 and NS5 sources correspond to infinite D3-D5 and D3-NS5 spikes when written in the coordinates adapted to flat orthogonal D3, D5 and NS5 branes. Furthermore, the $AdS_4$ radial direction is the direction  along the spike.

This poses an immediate puzzle: {\em How can  a solution with no $AdS_5$ asymptotic region, corresponding in regime {\bf 1} to D3 branes sandwiched between D5 and NS5 branes, have sources that are infinite D3-D5 and infinite D3-NS5 spikes?}

There are three insights in back-reacted branes that are needed to resolve this puzzle. First, it is not only branes ending on branes that can shape brane profiles. The B-field sourced by NS5 branes can also bend the D5 branes and the C-field sourced by D5 branes can bend the NS5 branes. For example, in regime {\bf 2} one can treat the D5 and D3 branes as probes of a fully-back-reacted multi-NS5 solution. The pullback of the B-field of the NS5 branes on the D5 world-volume gives rise to a position-dependent D3-brane charge density. This, ``B-field blast,'' in turn, can give rise to a very non-trivial shape for the D5-D3 brane probes \cite{Pelc:2000kb}\footnote{This happens in all partially-back-reacted Hanany-Witten systems \cite{Marolf:2000ci}.}. 

Second, as one changes the asymptotic positions of the branes, one finds that their shapes can change drastically, except for certain world-volume regions that exhibit a universal scaling behavior. The shapes of each brane in these regions is governed by a single quantity, which reflects the steepness of the brane profile. These ``steepness numbers''  count the number of D3 branes ending on the D5 (or NS5) brane from the right and from the left, as well as the B(or C)-field blasts the D5 (or NS5) brane receives from the NS5 (or D5) branes on the right and on the left. The scaling region is characterized by the steepness numbers, which remain invariant as one varies the asymptotic relative positions of the branes.  This indicates that these steepness numbers are universal (energy and coupling-independent) properties of this system.

The third is to realize that the region of the D3-D5-NS5 solutions that gives rise to the $AdS_4 \times S^2 \times S^2 \times \Sigma$ solutions comes from zooming in on these scaling regions of the D5 and NS5 branes. In particular,  what appears as an infinite spike in the zoomed-in region can, in fact, be a {\em finite} region of the  underlying brane configuration.

In addition to the steepness number, there are two other numbers one can identify for a system of D3 branes sandwiched between D5 branes and NS5 branes. The first is the regime {\bf 1} ``linking number'' \cite{Gaiotto:2008sa,Gaiotto:2008ak,Hanany:1996ie}, which determines the properties of the 2+1 dimensional CFT that lives in the infrared of this system of branes. The second is the ``Page charge'' of the D5 and NS5 brane sources that one computes in regime {\bf 3}. These three numbers are gauge-dependent and in \cite{Assel:2011xz} it was proposed that  Page charges of regime {\bf 3} are to be identified with the  linking numbers of regime {\bf 1}. In this paper, we show this explicitly by identifying both quantities with the same intermediate ``steepness number'' in regime {\bf 2}.

This has several immediate implications. The first is the identification of the region of the branes where the CFT lives. As one increases the brane back-reaction to move from regime {\bf 1} to {\bf 2} to {\bf 3} the CFT does not live in the region where the branes intersect, but rather in the region where the D5 and NS5 branes exhibit a certain scaling behavior. In the absence of back-reaction, these NS5 and the D5 scaling regions are far away from each other (and also far from the region where the branes intersect), but, as we will show, the back-reaction and the scaling bring the NS5 and the D5 scaling regions into the same $AdS_4 \times S^2 \times S^2 \times \Sigma$ solution.

The second is that we can obtain a clear geometric understanding of the good-bad-ugly classification of the 2+1 dimensional quivers that give rise, or do not give rise,  to 2+1 dimensional CFT's in the infrared \cite{Gaiotto:2008sa,Gaiotto:2008ak}.

The good quivers are those in which the D5 and NS5 branes are ordered by increasing {\em linking number}, and, when the back-reaction is turned on, these branes develop scaling regions that are ordered by increasing {\em steepness number}. Furthermore, when fully back-reacted, the D5 and NS5 sources are ordered by increasing {\em Page charge}. Since these numbers are the same, the branes maintain their ordering from regime {\bf 1} to {\bf 2} to {\bf 3}.

In contrast, in the bad and ugly quivers the branes are not arranged by increasing linking numbers in regime {\bf 1}.
Since in regimes {\bf 2} and {\bf 3} the branes are always arranged by increasing {\em steepness number/Page charge}, this would suggest that the branes must cross each other. Both the Born-Infeld description of brane probes and the full supergravity back-reaction abhor intersecting branes\footnote{Strictly speaking, this is not a theorem, however we note that the Born-Infeld description does not capture the massless degrees of freedom associated with the intersection, and supergravity greatly restricts the structure of brane singularities.}. This suggests that bad and ugly quivers do not correspond to {\it complete} Born-Infeld and supergravity solutions. 

Brane intersections typically lead to new low-energy degrees of freedom coming from strings, or branes, that stretch across the intersection. These new degrees of freedom are, initially, highly localized at the intersection and cannot be tracked by supergravity or Born-Infeld actions.  However, the condensation of such degrees of freedom can drive a transition in the brane configuration that can ultimately be tracked by the supergravity or the Born-Infeld theory.   

The fact that the supergravity and  Born-Infeld descriptions of  bad, or ugly, quiver theories  in regime {\bf 1} are trying to force the branes to cross should therefore be interpreted as requiring a stringy transition that corresponds, in the field theory, to one or more Seiberg dualities \cite{Seiberg:1994pq, Giveon:1998ns,Yaakov:2013fza,Assel:2017jgo} to give rise to a good theory which can, once again, be described in supergravity or the Born-Infeld theory. 

Put differently, supergravity connects directly, and naturally, to {\it good} theories in the UV and so supergravity  prefers such UV antecedents. Some  {\it bad} and {\it ugly} theories certainly lead to IR fixed points that can be described in supergravity, but supergravity, alone, does not seem to be capable of tracking back from the IR data to arrive at  {\it bad} and {\it ugly} brane configurations. 

The third implication is that we can extend this correspondence to type-{\bf B} and type-{\bf C} solutions.
Type-{\bf B} solutions have one asymptotic $AdS_5$ region, and are expected to come from the back-reaction of semi-infinite D3 branes terminating on D5 and NS5 branes. In regime {\bf 1} the different types of boundary conditions corresponding to the ways in which the D3 branes end on D5 and NS5 branes were classified by Gaiotto and Witten in \cite{Gaiotto:2008sa,Gaiotto:2008ak}, and the D5 and NS5 branes that give rise to conformal boundary conditions are again ordered by linking number. Our identification of the linking numbers with the steepness of the D5 and NS5 spikes allows us to show that the linking-number-ordered branes in regime {\bf 1} give rise to ordered spikes in regime {\bf 2} and to brane sources with ordered Page charges in regime {\bf 3}. In contrast, if one starts in regime {\bf 1} with linking-number-unordered branes, they need to cross each other to get to the ordered branes of regime {\bf 2} and regime {\bf 3}, and hence are not in one-to-one correspondence with conformal boundary conditions. The boundaries where the D5 and NS5 branes are not ordered are expected to undergo a type of boundary-Seiberg-duality to boundaries where the branes are ordered, which then flow to conformal boundaries.

A non-trivial consequence of our analysis is that for every Gaiotto-Witten (type-{\bf B}) boundary condition in regime {\bf 1} there must exist a type-{\bf B} regime-{\bf 3} solution. While we could not prove this rigorously, we have verified it in several non-trivial examples. The challenge is that the positions of the D5 and NS5 branes in these solutions are very complicated functions of the Page charges/steepness/linking numbers of these branes.

We expect type-{\bf C} solutions to come from infinite D3 branes superposed with semi-infinite D3 branes ending on D5 branes or NS5 branes or both. Although the classification of conformal Janus defects has not been done in regime {\bf 1} with the same thoroughness as the Gaiotto-Witten  classification of conformal boundary conditions, our identification of Page charges, steepness numbers and linking numbers gives us a clear picture of what will happen: Regime-{\bf 1} brane configurations in which the D5 and NS5 are arranged by linking number will give rise to a conformal domain wall in the infrared, dual to an $AdS_4 \times S^2 \times S^2 \times \Sigma$ Janus defect solution in regime {\bf 3}. Moreover, we expect that there will be a one-to-one correspondence between linking-number-ordered brane configurations and such Janus defect solutions. Furthermore, we expect defects where the branes are not ordered will undergo a type of {\em defect-Seiberg-duality} to a defect with ordered branes, which then flows to a conformal defect.

Our brane construction also clarifies the origin of type-{\bf D} Janus interface solutions. As we will show in Section \ref{Janus-interfaces}, these solutions come from a single stack of D3 branes that are extended across a dilaton kink created by D5 and/or NS5 brane distributions that preserve an $SO(3) \times SO(3) $ symmetry in the directions transverse to the D3 branes. Using our mapping between the $AdS_4 \times S^2 \times S^2 \times \Sigma$ coordinates and the D3-D5-NS5 coordinates, we can show that there exist supersymmetry-preserving probe D3 branes in these Janus interface solutions that have six flat directions. This confirms the intuition of \cite{DHoker:2006qeo} that the six scalars of conformal interfaces of the ${\cal N}=4$ SYM theory remain massless while the fermions become massive near the interface. 

\vspace{1em}

In Section \ref{sec:General Intersection} we present the generic form of 8-supercharge supergravity solutions that have the charges and Killing spinors corresponding to D3, D5 and NS5 branes. In Section \ref{sec:Near brane intersection} we review the $AdS_4 \times S^2 \times S^2 \times \Sigma$ geometries that have sources with D5, NS5 and D3 charges, and in Section \ref{ss:mapping} we map these geometries to the D3-D5-NS5 solutions of Section \ref{sec:General Intersection}. In Section \ref{sec:Probing} we place supersymmetric D3 and D3-D5 probes in these solutions, and show that the D5 sources in the  $AdS_4 \times S^2 \times S^2 \times \Sigma$ solutions correspond to infinite D3-D5 spikes. 

In Section \ref{sec:Brane interpretation - field theory} we explore Type-{\bf A} and Type-{\bf B} solutions in the non-back-reacted regime of parameters (regime {\bf 1}) and review the relation between regime-{\bf 1} linking numbers and regime-{\bf 3} Page charges. In Section \ref{regime-2} we explore the brane systems in the partially-back-reacted regime {\bf 2} and identify the steepness numbers with linking numbers and Page charges. 

Our conclusions are made in two sections.  First, in Section \ref{sec:Brane interpretation} we pull together the threads running through the paper, unifying the three pictures of brane intersections, supergravity description and insights coming from the field theory and brane-probe analysis.  We show how the $AdS_4 \times S^2 \times S^2 \times \Sigma$ solutions capture the scaling regions of the branes and extend these ideas by discussing Janus solutions. In Section \ref{sec:Discussion} we give a broader overview of our results and discuss possible directions for future research. 

In Appendix \ref{app:IIBtoM} we describe the duality chain between Type IIB D3-D5-NS5 brane solutions and M-theory smeared M2-M5-M5' solutions. In Appendix \ref{app:susies} we show that the $AdS_4 \times S^2 \times S^2 \times \Sigma_2 $ solutions preserve the same Killing spinors as straight infinite D3-D5-NS5 branes. Finally, in Appendix \ref{appendix:Probe bulges} we illustrate the bending of D2 and D1 probe branes caused by the blast of the $C$-fields of their D6 and D7 Hanany-Witten partners. 

\section{The most general D3-D5-NS5 intersections}
\label{sec:General Intersection}

\def\zz{z}

We start by describing the most general 8-supercharge D3-D5-NS5 straight orthogonal brane solutions in asymptotically-flat space-times.  We will then show how the AdS geometries emerge from zooming-in.  

Specifically, we consider $\frac{1}{4}$-BPS supergravity solutions that describe the intersections of D3- and five-branes in type IIB supergravity. These configurations were first studied in \cite{Hanany:1996ie}. The D3-branes extend along the $(x^0 = t,x^1, x^2, x^3)$ directions, while the D5-branes have a world-volume spanning $(x^0,x^1, x^2, x^4, x^5, x^6)$. The number of preserved supersymmetries remains unchanged upon adding NS5-branes along $(x^0, x^1, x^2, x^7, x^8, x^9)$. See Table \ref{tab:conventionsBraneSystems}.   The Chern-Simons interactions  imply that whenever two of these  species of branes are present, they generically source the third one. 

Observe that all three sets of branes  share  $(x^0, x^1, x^2)$.    Orthogonal to these common directions, there are two  $\IR^3$'s, spanned by $\vec u \equiv (x^4, x^5, x^6)$ and $\vec v \equiv(x^7, x^8, x^9)$, that make up the residual directions of the  D5  and  NS5 branes respectively. The remaining direction along the D3 branes will be denoted by $\zz = x^3$.  These coordinates are motivated by the perturbative brane configurations, however the interactions between the branes induce mutual deformations, and so the metric will be deformed away from the obvious non back-reacted ``cartoon'' of the branes.  Indeed, a ``mohawk'' structure \cite{Bena:2024dre} emerges through the dependence of all the fields on $(u, v, \zz)$, where $u \equiv |\vec u|$ and $v \equiv |\vec v|$.

\begin{table}[]
    \centering
  \begin{tabular}{c|ccc|c|ccc|cccc}
  & \multicolumn{3}{c|}{Spacetime} & $z$ & \multicolumn{3}{c|}{$\vec{u}$} & \multicolumn{3}{c}{$\vec{v}$} \\
         & $t=x^0$ & $x^1$ & $x^2$ & $x^3$ & $x^4$ & $x^5$ & $x^6$ & $x^7$ & $x^8$ & $x^9$    \\ \hline 
      D3   & $-$ & $-$ & $-$ & $-$ &  &  &   & &  &  \\
      D5   & $-$ & $-$ & $-$ & &  $-$ &  $-$ & $-$  &  && \\
      NS5   & $-$ & $-$ & $-$ & & &&  & $-$ & $-$ & $-$  \\
    \end{tabular}
    
\vspace{3em}

          \begin{tabular}{c|c|c|c} 
Multiplet & Fields & NS5 Ending & D5 Ending \\  \hline 
Vector & $(A_{\mu} , \vec{v})$ & Neumann & Dirichlet \\ 
Hyper & $(A_{3} , \vec{u})$ & Dirichlet & Neumann \\ 
    \end{tabular}
    \caption{Conventions for the coordinates and brane systems, definition of the fields and their boundary conditions. }
    \label{tab:conventionsBraneSystems}
\end{table}

\subsection{Supergravity solutions of the intersections}
\label{ss:flatforms}

The solution describing  the most general intersections of M2-M5-M5' branes in M theory was derived in  \cite{Bena:2023rzm,Lunin:2007mj}. As we show in detail in Appendix \ref{app:IIBtoM}, one can dualize these solutions into D3-D5-NS5  solutions of IIB supergravity by first smearing along an M5 direction and then reducing to Type IIA String Theory along this direction to obtain a D2-D4-NS5' solution. One can then smear and  T-dualize along one of the NS5' directions, to obtain a solution corresponding to intersections of D3, D5 and NS5 branes. 

Even if the M-theory solution only had 1+1-dimensional Lorentz invariance, the smearing and the duality chain yields a solution with  2+1-dimensional Lorentz invariance along the three directions common to the D3, D5 and NS5 branes. 
Since our solution is obtained by simply imposing the appropriate Killing spinors and isometries, \emph{it is the most general solution describing intersecting D3-D5-NS5 branes}.
 We have also directly verified the eleven-dimensional supersymmetries and, in Appendix \ref{app:susies}, we describe how this was done.  Here we will simply use these results. 

The solution is entirely determined by a pre-potential $G_0 = G_0 (z , \vec{u} , \vec{v})$ satisfying a generalized version of the Monge-Ampère equation:
\begin{equation}
  \cL_v G_0  ~=~  (\partial_z^2  G_0) \,(\cL_u G_0) ~-~  (\nabla_{\vec{u}} \partial_z  G_0 )\cdot  (\nabla_{\vec{u}} \partial_z  G_0 )    \,, 
 \label{MALap}
\end{equation}
where $\cL_v\,,\cL_u$ are the Laplacian operators on the two $\IR^3$'s spanned by $\vec u$ and  $\vec v$. One introduces the auxiliary functions $w = w (z , \vec{u} , \vec{v})$ and $A_0 = A_0 (z , \vec{u} , \vec{v})$ defined by :
\begin{equation}
w ~=~  \partial_z G_0 \,, \qquad   \cL_v G_0 ~\equiv~ (-\partial_z  w)^{\frac{1}{2}} \,e^{-3   A_0} \, . 
 \label{G0defn}
 \end{equation} 
The metric, gauge fields and dilaton are   
\begin{align}
    \d s_{10}^2& = \,  e^{\frac{3}{2}A_0}\big( -\partial_z w \big)^{-\frac{1}{4}}  \bigg( \,  - \d t^2 + \d x_1^2+  \d x_2^2  + e^{-3A_0} \big( -\partial_z w \big)^{-\frac{1}{2}} \, \d\vec{ u} \cdot \d\vec{u} \nonumber \\
    &+ e^{-3A_0} \big( -\partial_z w \big)^{\frac{1}{2}} \, \d\vec{v} \cdot \d\vec{v} 
    + \big( -\partial_z w \big) \,\left( \d z + \big( \partial_z w \big)^{-1} \, \big( \nabla_{\vec{u}} w \cdot \d\vec{u})\right)^2\bigg)\, , 
     \label{eq:10met1} \\
    e^{2 \Phi}&=(-\partial_z w)^{-1}\,,\quad C_2=-\frac{\varepsilon_{ijk}}{2}(\partial_{v_i} w) \, \d v^j \wedge \d v^k\,,\quad B_2= \frac{\varepsilon_{ijk}}{2} \frac{\partial_{u_i} w}{\partial_z w} \,\d u^j \wedge \d u^k \,, \\
     C_4&=-e^{3 A_0} (-\partial_z w)^{1/2}  \d t \wedge   \d x_1 \wedge  \d x_2\wedge  \left( \d z + (\partial_z w)^{-1}  \big( \vec{\nabla} w \big)\cdot \d \vec{u}\right) \,  
    \label{eq:C4gen1}
\end{align}
and the axion is trivial. 

An important aspect of the formulation above is that it appears to introduce an asymmetry between the NS5 and D5 branes, whereas S-duality tells us that the solutions should be entirely equivalent. The asymmetry lies in the choice of the coordinate $z$, and the fibration, in (\ref{eq:10met1}), of $ \d z$ over the $u$-directions. This has a follow-on effect in the expressions for the fluxes.

As described extensively in \cite{Bena:2024dre,Bena:2023rzm,Lunin:2007mj}, the D5-NS5 symmetry (obtained by flipping coordinates and an S-duality), can be seen by doing a coordinate transformation in which $w$ is used as a coordinate along the D3 brane and $z$ becomes a non-trivial function of $(w,\vec{u},\vec{v})$.  In particular, the metric can then be rewritten as:
\begin{align}
    \d s_{10}^2 = \, & e^{\frac{3}{2}A_0}\big( -\partial_w z \big)^{\frac{1}{4}}  \bigg( \,  -  \d t^2 +  \d x_1^2+   \d x_2^2  + e^{-3A_0} \big( -\partial_w z\big)^{\frac{1}{2}} \, \d \vec{ u} \cdot \d \vec{u} \nonumber \\
    &+ e^{-3A_0} \big( -\partial_w z \big)^{-\frac{1}{2}} \, \d \vec{v} \cdot \d \vec{v} 
    + \big( -\partial_w z \big) \,\left( dw + \big( \partial_w z \big)^{-1} \, \big( \nabla_{\vec{v}} z  \cdot \d \vec{v})\right) ^2\bigg)\,.
     \label{eq:10met2}
\end{align}
We will work primarily with the first formulation using coordinates $(z , \vec{u} , \vec{v})$ and the non-trivial function, $w(z , \vec{u} , \vec{v})$.  The dual form will be important in discussing the mohawk structure.

\subsection{Spherically symmetric intersections}
\label{ss:SSInts}

 To simplify our problem, we impose spherical symmetry on the two $\IR^3$ bases. The pre-potential, $G_0$, and the function, $w$, now only depend on $z$, $u = |\vec{u}|$ and $v = |\vec{v}|$. The angular variables parametrize two 2-spheres which are called respectively $S_1^2$ and $S_2^2$. The pre-potential obeys the spherically symmetric  form of (\ref{MALap}). The string-frame metric, gauge fields and dilaton are 
\begin{equation}
 {\begin{split}
        \d s_{10}^{2} = \, & e^{\frac{3}{2}A_0} \big( -\partial_z w \big)^{-\frac{1}{4}}\bigg(   -  \d t^2 +  \d x_1^{2}+   \d x_2^2 + \big( -\partial_z w \big) \,\left(  \d z + \big( \partial_z w \big)^{-1} \, \big( \partial_u  
    w \big) \d u\right)^{2}    \\
    &  +~ e^{-3A_0} \Big(\big( -\partial_z w \big)^{-\frac{1}{2}} \big(   \d u^2+u^2  \d s_{{\rm S}_1^{2}}^2 \big) +  \big( -\partial_z w \big)^{\frac{1}{2}} \big(   d v^{2}+v^{2}  \d s_{{\rm S}_2^{2}}^2 \big)  \Big)\bigg) \,,
  \\
    e^{2 \phi}=\, & (-\partial_z w)^{-1/2}\,,\quad C_2=-v^2\partial_v w\,\text{Vol}(S_2^{2 })\,,\quad B_2=u^2(\partial_z w)^{-1} (\partial_{u} w) \text{Vol}(S_1^2)\\
     C_4= \, & -e^{3 A_0} (-\partial_z w)^{1/2}   \d t \wedge  d x_1\wedge  \d x_2 \wedge  \left( \d z + (\partial_z w)^{-1}  \big( \partial_u w \big)  \d u\right)\, . 
    \end{split}} \label{tenmet}
\end{equation} 
We have introduced the notation $\phi = \frac{\Phi}{2}$ to match with the next section.

 \section{The near-brane intersection}
\label{sec:Near brane intersection}
\def\RSz{\zeta}

In this section we present the solutions constructed in \cite{DHoker:2007zhm,DHoker:2007hhe,Assel:2011xz}. In the next section, we show how they correspond to the near-brane, decoupling limit of the general brane intersections given in Section \ref{sec:General Intersection}.

In this scaling limit, the Poincar\'e invariance in $(t, x^1, x^2)$ gets promoted to the $SO(2,3)$ conformal invariance of AdS$_4$.  Since our goal is to interpret the solution in terms of branes in flat space, we will restrict to the Poincaré section of AdS$_4$, and the radial coordinate $\mu$ will emerge from a scale invariance of $(u,v,z)$. The remaining two scale-free coordinates will then
parametrize a Riemann surface $\Sigma$ with complex coordinate $\RSz$. Since we are imposing spherical symmetry, the complete symmetry is therefore $SO(2,3) \times SO(3)_1 \times SO(3)_2$, and the geometry is $\mathrm{AdS}_4 \times S_1^2\times S_2^2 \times \Sigma$.  We take $\Sigma$ to be an infinite strip:
\begin{equation}
    0~ \le~ \text{Im}(\RSz)~\le~\frac{\pi}{2} \,.
\end{equation}
However, our results do not rely on this assumption, and similar conclusions will hold for higher-genus Riemann surfaces.

 \subsection{The near-brane solution}
\label{ss:nbi}
 
The solutions \cite{DHoker:2007zhm,DHoker:2007hhe}
are parametrized by two real harmonic functions $h_1 = h_1 (\zeta)$ and $h_2 = h_2 (\zeta)$ on $\Sigma$, which are required to be \textit{smooth} and \textit{positive} in the interior of the strip. On the boundaries, these functions are required to satisfy \cite{DHoker:2007zhm,DHoker:2007hhe}:
\begin{equation}
 \text{Im}(\RSz) = 0:  \ \ h_1 = \partial_\perp h_2 = 0 \,;   \qquad \qquad \text{Im}(\RSz) = \frac{\pi}{2}: \ \ h_2 = \partial_\perp h_1 = 0 \,,  
\label{hbcs}
\end{equation}
where $\partial_\perp$ is the normal derivative.   These functions are, however, allowed to have logarithmic singularities on the boundaries, and hence $\partial_\RSz h_j$ can have (isolated) poles on the boundary. The vanishing of $h_j$ creates non-trivial $3$-cycles out of the two-spheres while the poles in $\partial_\RSz h_j$ are the $5$-brane sources.
We will also need the harmonic conjugates, $h_j^D$ , of the $h_j$ ($j=1,2$), defined, up to an additive constant denoted $\xi_j$, by:
\begin{equation}
  \partial_\RSz \big(h_j + i\, h_j^D \big) =  0\,.  \label{eq:defHarmonicConjugate}
\end{equation}
The metric in Einstein frame is:
\begin{equation} 
    \d s^2 = f_4^2 \d s_{{\rm AdS}_4}^2 + f_1^2 \d s_{{\rm S}_1^2}^2 + f_2^2 \d s_{{\rm S}_2^{2}}^2 + 4 \rho^2 \d \RSz \d \bar \RSz\ ,
\label{Einsteinmet}
\end{equation}
where we first define:
\begin{align}
    W &= \partial  h_1 \bar\partial  h_2 + \bar\partial h_1 \partial  h_2  = \partial \bar\partial  (h_1h_2)\ ,  \nonumber \\
N_1 &= 2 h_1 h_2 |\partial h_1|^2 - h_1^2 W \ , \nonumber \\
N_2 &= 2 h_1 h_2 |\partial h_2|^2 - h_2^2 W   \ ,
\end{align}
from which one assembles the metric functions:
\begin{align}
    f_4^8 &= 16\, {N_1 N_2 \over W^2}\ ,
\qquad \qquad
\rho^8 = \frac{N_1 N_2 W^2 }{ h_1^4 h_2^4}\ ,  \nonumber \\
f_1^8 &=  16\,  h_1^8 \frac{N_2 W^2 }{ N_1^3}\ ,
\qquad  \qquad
f_2^8 = 16\,  h_2^8 \frac{N_1 W^2}{N_2^3}\,.
\end{align}
The dilaton is given by\footnote{We follow the conventions of the original paper, in these conventions, $\Phi=2 \phi$ is the dilaton. \label{footnoteDilaton}}

\begin{equation}
    e^{4\phi}=\frac{N_2}{N_1} \,,
\label{eq:dilaton1}
\end{equation}
and the gauge fields are given by:
\begin{equation}
    H_3= \d b_1\wedge \text{Vol}(S_1^2)\,,\quad F_3= \d b_2\wedge \text{Vol}(S_2^2)\,,
    \label{eq:B2flux}
\end{equation}
where $H_3$ is the NS-NS field strength sourced by the NS5 branes and $F_3$ the R-R field strength sourced by the D5 branes.  
In these expressions, the functions $b_1,\,b_2$ are given by:
\begin{equation}
 \begin{aligned}
     b_1 &= 2 i h_1 {h_1 h_2 (\partial h_1\bar  \partial h_2 -\bar \partial h_1 \partial h_2) \over N_1} ~+~ 2 \, h_2^D \ ,  \\
    b_2 &= 2 i h_2 {h_1 h_2 (\partial h_1 \bar\partial h_2 - \bar\partial h_1 \partial h_2) \over N_2} ~-~ 2\,  h_1^D \ . 
\label{3forms1}
\end{aligned}
\end{equation}
The five form field strength is given by:
\begin{align}
     F_{(5)}  =
 - 4\,  f_4^{4}\,  {\cal F}\wedge\text{Vol}({\rm AdS}_4)  ~+~  4\, f_1^{2}f_2^{2} \, \big( *_2 {\cal F}\big) \wedge \text{Vol}(S_1^2)\wedge \text{Vol}(S_2^2)\,.
 \label{F5field}
\end{align}
In this definition, $f_4^4{\cal F}$ is a closed 1-form on the Riemann surface, and $*_2 $ is the Hodge dual on $\Sigma$. Finally:
\begin{align}
    f_4^{\, 4} {\cal F} = \d j_1\   \qquad {\rm with} \qquad j_1 =
3 {\cal C} + 3 \bar  {\cal C}  - 3 {\cal D}+ i \frac{h_1 h_2}{W}\,   (\partial h_1 \bar\partial h_2 -\bar \partial h_1 \partial h_2) \, ,
\end{align}
where $\mathcal{D}=\bar{\mathcal{A}}_1 \mathcal{A}_2+\mathcal{A}_1{\bar{\mathcal{A}}}_2$ and $\mathcal{C}$ is defined by $\partial \mathcal{C}=\mathcal{A}_1 \partial \mathcal{A}_2-\mathcal{A}_2 \partial \mathcal{A}_1$.

\subsection{Most general admissible harmonic functions}

The most general forms of the admissible harmonic functions are given by \cite{Bachas:2011xa,Assel:2011xz}:
\begin{align}\begin{split}
    h_1 (\zeta) &=-i \alpha_1 \sinh(\RSz-\beta_1)-\sum_i \gamma_{1}^{(i)} \log\left(\tanh\left(-\frac{\RSz-\delta_1^{(i)}-i\frac{\pi}{2}}{2}\right)\right)~+~ \text{c.c.} \,, \\
    h_2 (\zeta) &= \alpha_2 \cosh(\RSz-\beta_2)-\sum_i \gamma_{2}^{(i)} \log\left(\tanh\left(\frac{\RSz-\delta_2^{(i)}}{2}\right)\right)~+~\text{c.c.} \,.
    \end{split}
    \label{eq: harmonic ansatz}
\end{align}
As we will discuss, the parameters, $\alpha_1, \alpha_2, \delta_1^{(i)},\delta_2^{(i)},\gamma_1^{(i)}$  and $\gamma_2^{(i)}$ encode the brane charges and do not correspond to moduli of the non-back-reacted UV brane configuration. Indeed, they define the field content of the dual CFT, and the slopes of the mohawk spikes. On the other hand, the parameters $\beta_1$ and $\beta_2$ encode continuous deformations of the solution, such as the dilaton jump of Janus interface solutions. 
The harmonic functions determining the solution can also be written in terms of $(x,y)$ of $\zeta = x + i y$ as 
\begin{align}
\label{eq:hiRealCoordinates}
\begin{split}
    h_1 (x,y) &= 2 \alpha_1 \cosh (x - \beta_1) \sin (y) + \sum\limits_i \gamma_1^{(i)} \log \left[ \frac{\cosh (x - \delta_1^{(i)}) + \sin (y)}{\cosh (x - \delta_1^{(i)}) - \sin (y)} \right] \,, \\  
    h_2 (x,y) &= 2 \alpha_2 \cosh (x - \beta_2) \cos (y) + \sum\limits_i \gamma_2^{(i)} \log \left[ \frac{\cosh (x - \delta_2^{(i)}) + \cos (y)}{\cosh (x - \delta_2^{(i)}) - \cos (y)} \right] \, . 
    \end{split}
\end{align}
The harmonic conjugates are 
\begin{align}
\label{eq:harmonic-conjs}
\begin{split}
h_1^D (\zeta) &~=~ \frac{\pi \alpha '}{2} \xi_1 + \left[   \alpha_1  \sinh(\RSz-\beta_1) - i \sum_i \gamma_{1}^{(i)} \log\left(\tanh\left(\frac{i\pi}{4}-\frac{\RSz-\delta_1^{(i)}}{2}\right)\right)  
 + \text{c.c.}  \right] \,, \cr
h_2^D (\zeta) &~=~ \frac{\pi \alpha '}{2} \xi_2 + \left[
  i \alpha_2 \cosh(\RSz-\beta_2) - i \sum_i \gamma_{2}^{(i)} \log \left(\tanh\left(\frac{ \RSz-\delta_2^{(i)}}{2}\right)\right) + \text{c.c.}  \right]  \,,
  \end{split}
\end{align}
where the constants $\xi_j$ can be assumed to be real, and, for later convenience, we have introduced the string coupling, $\alpha '$, into the normalization of the $\xi_j$. There are two sources of ambiguity in writing down the formulas \eqref{eq:harmonic-conjs}: one is the additive constants allowed by the definition \eqref{eq:defHarmonicConjugate}, the other lies in the choice of branch cuts for the complex logarithm. We make the usual choice of branch cut (on the negative real axis), so that all the ambiguity lies in the constants $\xi_j$. Using this choice, we can write\footnote{Using the standard branch cut to define the complex logarithm, the relations  $$-i \log \tanh \left(\frac{i \pi}{4} - \frac{x + i y}{2} \right) + i \log \tanh \left(\frac{-i \pi}{4} - \frac{x - i y}{2} \right) = \pi + 2 \arctan \left( \frac{\sinh x}{\cos y} \right)\,,$$ $$-i \log \tanh \left( \frac{x + i y}{2} \right) + i \log \tanh \left( \frac{x - i y}{2} \right) = \pi - 2 \arctan \left( \frac{\sinh x}{\sin y} \right)$$ hold for every $x \in \mathbb{R}$ and $0 < y < \frac{\pi}{2}$. } 
\begin{align}
\begin{split}
    h_1^D (x,y) &=  \frac{\pi \alpha '}{2}\xi_1 + 2 \alpha_1 \sinh (x - \beta_1) \cos (y) + \sum\limits_i   \gamma_1^{(i)} \left( \pi + 2 \arctan \left[ \frac{\sinh (x - \delta_1^{(i)}) }{ \cos (y)} \right] \right) \,,\\   
    h_2^D (x,y) &= \frac{\pi \alpha '}{2}\xi_2  - 2 \alpha_2 \sinh (x - \beta_2) \sin (y) + \sum\limits_i   \gamma_2^{(i)} \left( \pi - 2 \arctan \left[ \frac{\sinh (x - \delta_2^{(i)}) }{  \sin (y)} \right] \right) \, . 
    \end{split}
    \label{eq:hiDrealcoordinates}
\end{align}

\begin{figure}
    \centering
\begin{tabular}{cc}
    $h_1(x,y)$ & \raisebox{-0.5\height}{\includegraphics[scale=.5]{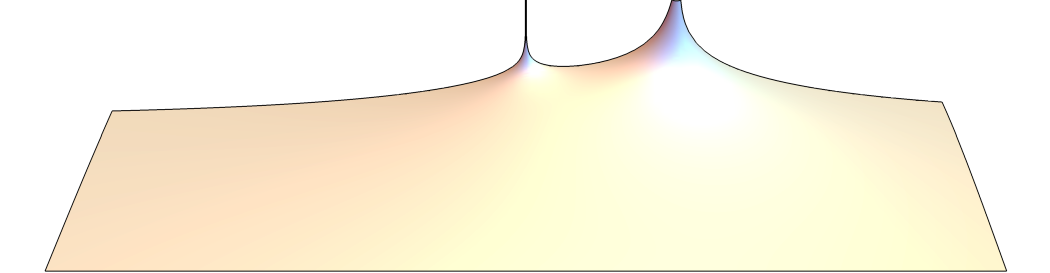}} \\[40pt]
    $h_2(x,y)$ & \raisebox{-0.5\height}{\includegraphics[scale=.5]{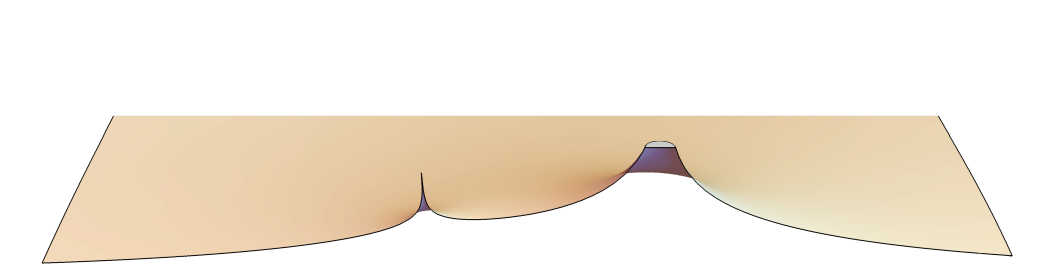}} \\[40pt]
    $h_1^D(x,y)$ & \raisebox{-0.5\height}{\includegraphics[scale=.5]{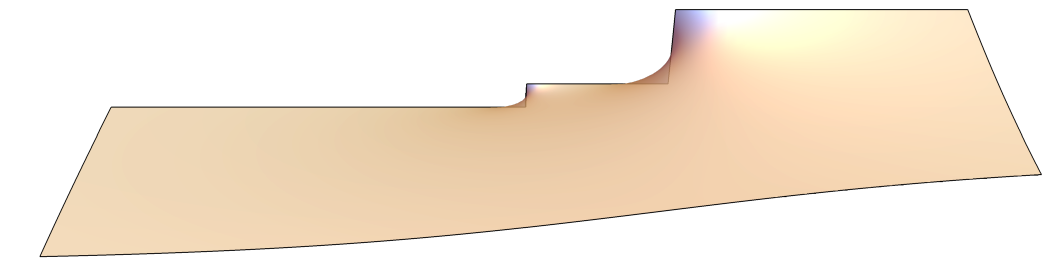}} \\[40pt]
    $h_2^D (x,y)$ & \raisebox{-0.5\height}{\includegraphics[scale=.5]{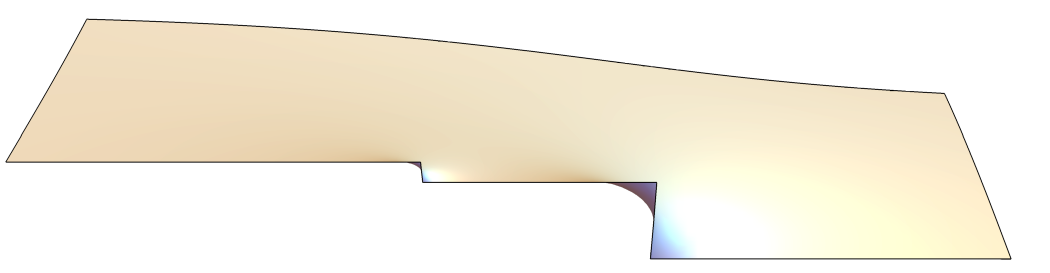}} \\[40pt]
    $f_1^8 (x,y)$ & \raisebox{-0.5\height}{\includegraphics[scale=.5]{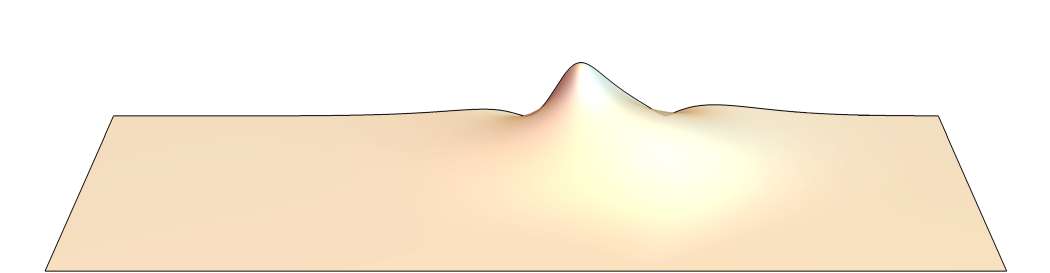}} \\[40pt]
    $f_2^8 (x,y)$ & \raisebox{-0.5\height}{\includegraphics[scale=.5]{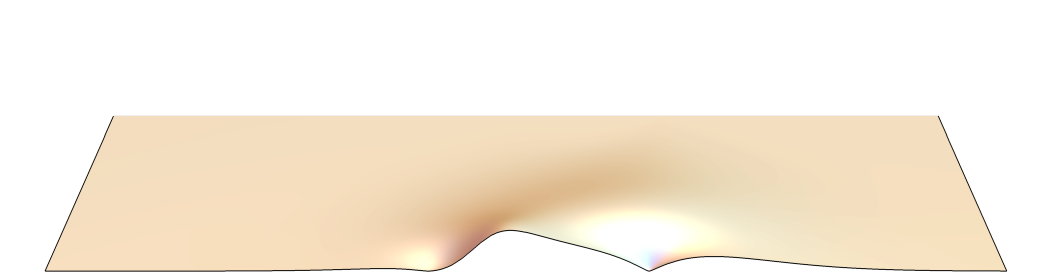}} \\[40pt]
    $f_4^8 (x,y)$ & \raisebox{-0.5\height}{\includegraphics[scale=.5]{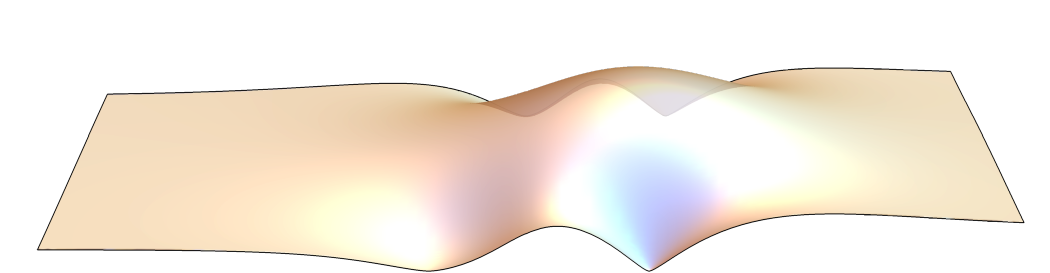}} 
\end{tabular}
    \caption{Plot of the various functions for the example of the $T^{\sigma = [3,2,2,2]}_{\rho = [2,2,2,2,1]} [SU(9)]$ theory (introduced in Section \ref{sec:Brane interpretation - field theory}). In each plot, the horizontal axis is $x$ and the depth axis is $y$ (with $y=0$ in front and $y=\frac{\pi}{2}$ in the back).  }
    \label{fig:stripFunctionsExampleTRhoSigma}
\end{figure} 

\paragraph{Behavior at the bottom and top of the strip. }
\begin{itemize}
    \item \underline{Bottom of the strip}. Here, $h_1 = 0$ while $h_2$ has a divergence at $\zeta = \delta_2^{(i)}$ corresponding to the presence of $\gamma_2^{(i)}$ NS5 branes. The function,  $h_2^D (\zeta = x)$, is locally constant and jumps at $x = \delta_2^{(i)}$ by an amount $2 \pi \gamma_2^{(i)}$: 
\begin{equation}
    h_2^D \big(\RSz = x \big) = \frac{\pi \alpha '}{2} \xi_2 + 2 \pi  \sum_i \gamma_{2}^{(i)} \theta ( \delta_2^{(i)} - x ) \,, \label{eq:loc-const2}
\end{equation}
where $\theta$ is the Heaviside step function.  
    \item \underline{Top of the strip}. Similarly, $h_2 = 0$, while $h_1$ has a divergence at $\zeta = \delta_1^{(i)} + \frac{i\pi}{2}$ corresponding to the presence of $\gamma_1^{(i)}$ D5 branes. The function, $h_1^D$, is locally constant and jumps at $x = \delta_1^{(i)}$ by an amount $2 \pi \gamma_1^{(i)}$: 
    \begin{equation}
        h_1^D \bigg(\RSz = x + \frac{i\pi}{2} \bigg) =\frac{\pi \alpha '}{2} \xi_1 +  2 \pi \, \sum_i \gamma_{1}^{(i)} \, \theta (x -\delta_1^{(i)})  \label{eq:loc-const1} \,.
    \end{equation}
\end{itemize} 
Positivity of $h_1$ and $h_2$ in the interior of the stripe requires that $\alpha_{1,2}\gamma_{1,2}^{(i)}>0$, and we can choose the orientations of our charges without loss of generality to give
$\alpha_{1,2} \geq 0$ and $\gamma_{1,2}^{(i)} \geq 0$ for all $i$ \cite{DHoker:2007hhe}.  Physically this means that we only want one orientation of D5 and NS5 branes: branes without anti-branes. The situation where $\alpha_1 = \alpha_2 = 0$ is depicted in Fig. \ref{fig:stripFunctionsExampleTRhoSigma}, in which the divergences of $h_{1,2}$ and the local constancy of $h^D_{1,2}$ are apparent. 

\subsection{Brane charges}
\label{ss:Bcharges}

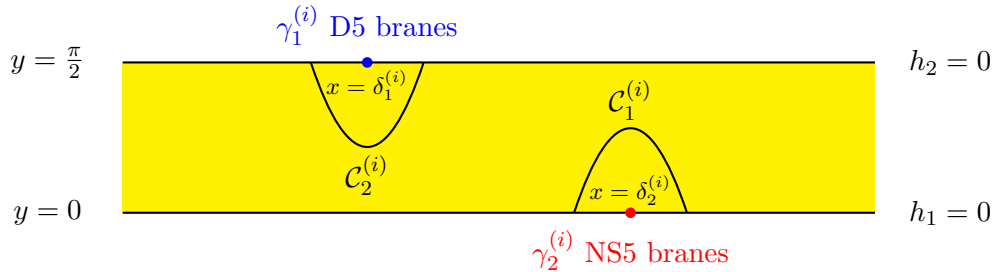
\begin{figure}[b]
\centering
\raisebox{-.5\height}{\begin{tikzpicture}
    \fill[yellow] (-5,-1) rectangle (5,1);
    \draw[thick] (-5,-1) -- (5,-1);
    \draw[thick] (-5,1) -- (5,1);
    \draw[thick] (-2.5,1) .. controls (-2,-0.5) and (-1.5,-0.5) .. (-1,1); 
    \draw[thick] (2.5,-1) .. controls (2,0.5) and (1.5,0.5) .. (1,-1); 
    \node at (-1.75,-.5) {$\mathcal{C}_2^{(i)}$};
    \node at (1.75,.50) {$\mathcal{C}_1^{(i)}$};
    \fill[blue] (-1.75,1) circle (2pt);
    \node at (-1.75,0.7) {\footnotesize $x=\delta_1^{(i)}$};
    \node at (-1.75,1.5) {\color{blue}  $\gamma_1^{(i)}$ D5 branes};
    \fill[red] (1.75,-1) circle (2pt);
    \node at (1.75,-0.7) {\footnotesize $x=\delta_2^{(i)}$};
    \node at (1.75,-1.5) {\color{red} $\gamma_2^{(i)}$   NS5 branes};
    \node at (-6,1) {$y=\frac{\pi}{2}$};
    \node at (-6,-1) {$y=0$};
    \node at (6,1) {$h_2=0$};
    \node at (6,-1) {$h_1=0$};
\end{tikzpicture}}
    \caption{ The infinite strip with D5 and NS5 sources. The $3$-cycles around D5 and NS5 sources are defined by contours, $\mathcal{C}_a^{(i)}$, on the Riemann surface. }
    \label{fig:infinite strip with sources}
\end{figure}

\paragraph{NS5 and D5 charges. }
On the upper boundary of the strip, $h_2$ vanishes, and as a result, the two-sphere $S^2_2$ smoothly shrinks to zero size. Fibering $S^2_2$ along a path going around a pole of $h_1$ and ending on $\text{Im}(\RSz) = \frac{\pi}{2}$ defines a homological $3$-cycle over which $F_3$ has a non-trivial period integral. Similarly, on the lower boundary, where $\text{Im}(\RSz) = 0$, the function $h_1$ vanishes, causing $S_1^2$ to pinch off.  This defines another homological $3$-cycle over which $H_3$ has a non-trivial period integral.
Indeed, one finds:
\begin{equation}
Q^{(i)}_{\rm D5}=\int_{\mathcal{C}_2^{(i)} \times S_2^2} F_{3}=(4\pi)^2 \gamma_1^{(i)}\,,\qquad Q^{(i)}_{\rm NS5}=\int_{\mathcal{C}_1^{(i)} \times S_1^2} H_{3}=-(4\pi)^2\gamma_2^{(i)}\,,
\label{eq:5charges}
\end{equation}
where the path $\mathcal{C}^{(i)} \subset \Sigma$ encircles the $i^{\text{th}}$ pole, as depicted in Fig. \ref{fig:infinite strip with sources}. Hence the parameters $\gamma_1^{(i)}$ and $\gamma_2^{(i)}$ determine the five-brane supergravity charges.

\paragraph{D3 charges. }
In the presence of five-branes, the definition of D3-brane charge requires one to distinguish between two notions of charge: the Maxwell charge and the Page charge.  Specifically, the equations of motion of the $5$-form flux are:
\begin{equation}
\d F_5 = H_3 \wedge F_3\,,
\end{equation}
which can be written as:
\begin{equation}
\d(F_5-B_2\wedge F_3) = 0\,, \qquad \text{or} \qquad \d(F_5+H_3\wedge C_2)=0 \,.
\end{equation}
Since D3 branes source $F_5$, it is natural to define the gauge-invariant Maxwell charge as:
\begin{equation}
\label{eq:D3BraneCharge}
Q_{\text{Maxwell}}^{\rm D3} = \int_{\mathcal{C}_5} F_5 \,,
\end{equation}
where $\mathcal{C}_5$ is a closed five-cycle. This charge, however, is neither localized nor quantized. The \emph{Page charge}, by contrast, is both and can be defined using either of the following two choices of closed form, each adapted to a different region of the strips, as we explain below:
\begin{align}
\begin{split}
Q_{\text{Page} , \, \rm D5}^{{\rm D3}\,(i)} &= \int_{S_1^2 \times S_2^2 \times \mathcal{C}_2^{(i)}}   F_5 - F_3 \wedge B_2 \,   \qquad \qquad  \text{(Upper part of strip)} \\
Q_{\text{Page} , \, \rm NS5}^{{\rm D3}\,(i)} &= \int_{S_1^2 \times S_2^2 \times \mathcal{C}_1^{(i)}}  F_5 + C_2 \wedge H_3 \,  \qquad \qquad    \text{(Lower part of strip)}
\end{split}
\label{eq:D3PageCharges}
\end{align}
The two choices are labeled with a subscript NS5 or D5 depending on which 3-form is used in the definition, respectively $H_3$ and $F_3$. 
The choice here depends on which integrand is globally well-defined on the five-cycle on which the integral is computed. On the upper part of the strip,  $B_2$ (and not $C_{2}$) is well defined\footnote{These forms are proportional to the volumes of $S^2_1$ and $S^2_2$ and for these forms to be well-defined the corresponding sphere must {\it not} pinch off.} and therefore we use $Q_{\text{Page} , \,\rm D5}^{\rm D3}$.  One integrates $F_3$ over the $3$-cycle defined by the sphere $S^2_2$ that pinches off along $\cC_2^{(i)}$, and this gives $Q_{\rm D5}$.  On the upper part of the strip, the only non-vanishing contribution to $B_2$  comes from $h_2^D$ and so integrating it over $S^2_1$ gives $8 \pi h_2^D$.  {\it Mutatis mutandis} for $Q_{\text{Page} , \, \rm NS5}^{\rm D3}$ on the lower part of the strip.
We thus obtain:
\begin{align}
\label{eq: Page charge D3 brane}
\begin{split}
 Q_{\text{Page} , \, \text{D5}}^{\text{D3}\,(i)} & =-8\pi \, Q_{\rm D5}^{(i)} \, h_2^{D}\bigg(\RSz=\frac{i\pi}{2}+\delta_1^{(i)}\bigg)\, ,     \\  Q_{\text{Page} ,  \, \text{NS5}}^{\text{D3}\,(i)} & =-8\pi \,Q_{\rm NS5}^{(i)} \, h_1^{D}\big(\RSz=\delta_2^{(i)}\big) \, . 
 \end{split}
\end{align}

\noindent Hence the parameters $\delta_1^{(i)}$ and $\delta_2^{(i)}$ are directly related to the three-brane charges carried by the five-branes.

These Page charges are conserved, they are independent of the Gaussian surface and can therefore be localized and quantized.  However, they are not gauge-invariant under large gauge transformations of the two-form potentials $C_2$ and $B_2$. 
Physically, this means that the presence of five-branes induces a non-trivial D3-brane charge, which can change discontinuously across different regions of space that might require different  gauge choices for $B_2$ and $C_2$. This effect is directly related to the Hanany-Witten transition, where the crossing of NS5- and D5-branes results in the creation or annihilation of D3-branes.
Thus, while the brane charge \eqref{eq:D3BraneCharge} is invariant under gauge transformations, it is not necessarily quantized because of flux contributions. Conversely, the Page charges \eqref{eq:D3PageCharges} are quantized but shift under large gauge transformations. This distinction plays a crucial role in defining and interpreting D3-brane charge in supergravity backgrounds containing five-branes.
Note that $h_a^D$ is parametrized by the $\gamma_a^{(i)}$, which determine the 5-brane charges (\ref{eq:5charges}).  This means that the D3 charges are proportional to sums of combinations of $Q_{NS5}^{(i)}Q_{D5}^{(j)}$, as one would expect from the Chern-Simons interaction. 
Explicitly, equation \eqref{eq: Page charge D3 brane} combined with \eqref{eq:hiDrealcoordinates}-\eqref{eq:5charges} gives\footnote{We use the identities $ \pi \pm 2 \arctan \sinh x = 4 \arctan e^{\pm x}$ valid for all $x \in \mathbb{R}$. }
\begin{equation}
\begin{split}
    Q_{\text{Page} ,  \, \text{D5}}^{\text{D3}\,(i)} & = 2^8 \pi^3 \gamma_1^{(i)} \left[ - \frac{\pi \alpha '}{4}  \xi_2  + \alpha_2 \sinh \left( \delta_1^{(i)} - \beta_2 \right) - 2 \sum\limits_j \gamma_2^{(j)} \arctan \left( e^{\delta_2^{(j)} - \delta_1^{(i)}} \right) \right] \, , \\
    Q_{\text{Page} ,  \, \text{NS5}}^{\text{D3}\,(i)} & = 2^8 \pi^3 \gamma_2^{(i)} \left[ \frac{\pi \alpha '}{4}  \xi_1 + \alpha_1 \sinh \left( \delta_2^{(i)} - \beta_1 \right) + 2 \sum\limits_j \gamma_1^{(j)} \arctan \left( e^{\delta_2^{(i)} - \delta_1^{(j)}}  \right)\right] \, .  
      \end{split}
      \label{eq:3-chargesPage}
\end{equation}

\section{Mapping the AdS$_4$ solution to D3-D5-NS5 intersections}
\label{ss:mapping}

The ${\rm AdS}_4$ solution emerges as a near-brane limit of the general intersection. One can establish how this comes about by matching the symmetry-invariant pieces of the solution in Section \ref{ss:SSInts} with those in Section \ref{ss:nbi}. 
In this section, we find an explicit mapping of coordinates $$(\mu , x,y) \rightarrow (u,v,z)$$ realizing this matching. The map is given in \eqref{uvzmap1}. 

\subsection{Brane Coordinates}

We choose the Poincaré metric on ${\rm AdS}_4$:
\begin{equation}
\label{eq:Poincare metric}
    \d s_{{\rm AdS}_4}^2=\left(\frac{\d\mu}{\mu}\right)^2+\mu^2\left(- \d t^2+ \d x_1^2+ \d x_2^2\right)\,.
\end{equation}
We start by matching the coefficients of $\d s_{{\rm S}^2_i}^2$ and $\d t^2$ between \eqref{tenmet} and \eqref{Einsteinmet}: 
\begin{align}
    e^{\frac{3}{2}A_0}\big( -\partial_z w \big)^{-1/4}& ~=~ \mu^2\, f_4^2\, e^{\phi}\,,  \label{ident2}\\
    u^2\,e^{-\frac{3}{2}A_0} \big( -\partial_z w \big)^{-3/4} &~=~ f_1^2 \, e^{\phi}  \,, \\
    v^2\, e^{-\frac{3}{2}A_0} \big( -\partial_z w \big)^{1/4} &~=~ f_2^2 \, e^{\phi}  \label{ident3}\,, 
\end{align}
where the factor of $e^{\phi}$ arises because \eqref{tenmet} is in string frame and \eqref{Einsteinmet} is in Einstein frame. 

Using the expression for the dilaton $e^{2\phi}=(-\partial_z w)^{-1/2}$ (see footnote \ref{footnoteDilaton}), we can identify the coordinates $u,v$ with:  
\begin{equation}
    u^2~=~\mu^2e^{-2\phi}f_1^2f_4^2 ~=~ 4 \,\mu^2 \, h_1^2\,,\qquad v^2~=~ \mu^2e^{2\phi}f_2^2f_4^2 ~=~ 4\, \mu^2\, h_2^2\,.
\label{ident5}
\end{equation}
Matching $B_2$ between \eqref{eq:B2flux} and \eqref{tenmet} gives 
\begin{equation}
    b_1 =  u^2(\partial_z w)^{-1} (\partial_{u} w) \, . 
\end{equation}
Finally, the entire metric (\ref{tenmet}) must be scale invariant, which justifies the ansatz\footnote{The argument goes as follows. Define a scaling action on $(u,v,z,w)$ with scale factors $(\lambda_u ,\lambda_v ,\lambda_z ,\lambda_w )$. The fact that the dilaton is scale invariant implies $\lambda_z = \lambda_w$. Then the metric being invariant implies $\lambda_u = \lambda_v$, and furthermore $e^{+ \frac{3}{2} A_0} \d z^2$ scales like $e^{- \frac{3}{2} A_0} \d u^2$, which proves $\lambda_z = - \lambda_u$. The ansatz then follows from \eqref{ident5} which shows that $z$ is proportional to $\mu^{-1}$, the proportionality constant being $\Sigma$-dependent only.  } 
\begin{equation}
    z = \frac{F(\zeta , \bar{\zeta})}{\mu} \, . 
\end{equation}
Plugging this in the metric \eqref{tenmet} and identifying the coefficient of $\left(\frac{\d\mu}{\mu}\right)^2$ with \eqref{Einsteinmet} gives an equation for $F(\zeta , \bar{\zeta})$: 
\begin{equation}
    \left( \frac{b_1}{2 h_1} - F \right)^2 = (f_4^2 - f_1^2 - f_2^2) e^{2 \phi} \frac{f_1^2}{4 h_1^2} \, , 
\end{equation}
which admits the following solution 
\begin{equation}
\label{Fmessy}
    F = \frac{1}{2 h_1} \left[ b_1 \pm e^{\phi} f_1 f_4 \sqrt{1-\left(\frac{f_1}{f_4}\right)^2-\left(\frac{f_2}{f_4}\right)^2} \right] \, . 
\end{equation}
Matching the last terms in the metric picks the sign $\pm = -$ and, using the explicit expressions \eqref{3forms1}, equation \eqref{Fmessy} simplifies massively\footnote{Using the identity $\sqrt{1-\left(\frac{f_1}{f_4}\right)^2-\left(\frac{f_2}{f_4}\right)^2 }=-i \frac{h_1 h_2\,\left(\partial h_1 \bar\partial h_2 - \bar\partial h_1 \partial h_2)\right)}{(N_1 N_2)^{\frac{1}{2}}}$. } to $F = \frac{h_2^D}{h_1}$. Hence we have $z = \frac{h_2^D}{\mu \, h_1}$. A similar argument, beginning with the matching of the $C_2$ fields $b_2 = - v^2 \partial_v w$ yields $w = \frac{h_1^D}{\mu \, h_2}$. We have thus identified the ``brane coordinates'' as functions of the $AdS_4$ radius and the Riemann-surface harmonic functions:
\begin{equation}
\boxed{
    u~=~ 2 \,\mu \, h_1\,,\qquad v~=~2\, \mu\, h_2\,,\qquad z~=~\frac{h_2^D}{\mu \,h_1}  \,, \qquad w~=~\frac{h_1^D}{\mu \,h_2} \,, }  
\label{uvzmap1}
\end{equation}
which is completely analogous to the result of the M2-M5-M5' system \cite{Bena:2023rzm}. Observe that $
z \sim \frac{c_1}{u}$ and $z \sim \frac{c_2}{v}$ where $c_1$ and $c_2$ are functions on $\Sigma$.  These describe the natural ``harmonic spike'' sourced by D3 branes stretching the $\IR^3$ directions of the 
5-brane world-volume along the $z$ direction.  These spikes define the mohawk structure -- this will be expanded upon in Section \ref{sec:Brane interpretation}, see, in particular, Fig. \ref{fig:D3 D5 spike one side}.
Note that $u$ vanishes at $ \Im(\RSz) =0$ and so the discussion in Section \ref{ss:Bcharges} means that $u$ defines the radius of a Gaussian surface around the NS5 branes, and thus the coordinate $v$ runs along the NS5 branes.  Similarly, $v$ vanishes at $ \Im(\RSz) =\frac{\pi}{2} $   and thus the coordinate $u$ runs along the D5 branes. This is precisely the coordinate choice outlined at the beginning of Section 
\ref{sec:General Intersection}.

\subsection{A simple geometry: D3 branes}
\label{ss:D3s}

It is very instructive to see how all of the foregoing discussion applies to the simplest example:  pure D3 branes. This  solution is given by \eqref{eq:hiRealCoordinates}, \eqref{eq:hiDrealcoordinates} with $\gamma_{1}^{(i)} = \gamma_{2}^{(i)} = 0$ and, to keep things simple, $\beta_1 = \beta_2 =0$: 
\begin{equation}
\label{eq:D3form1}
    h_1~=~  2\, \alpha_1\, \cosh(x) \, \sin(y) \,,\qquad  
   h_2~=~  2\,  \alpha_2\, \cosh(x) \, \cos(y) \,,
\end{equation}
\begin{equation}
\label{eq:D3harmonic-conjs}
h_1^D ~=~ 2\,  \alpha_1  \sinh(x) \, \cos(y)\,,
 \qquad h_2^D ~=~ - 2\,   \alpha_2  \sinh(x) \, \sin(y)  \,.
\end{equation}
With these choices, the metric (\ref{Einsteinmet}) becomes 
\begin{equation} 
\label{eq:AdS5 S5}
    \d s^2 = 4 \sqrt{ |\alpha_1\alpha_2| }\left(\d x^2 ~+~ \cosh^2 x \, \d s_{{\rm AdS}_4}^2 ~+~  \d y^2 ~+~ \sin^2 y \, \d s_{{\rm S}_1^2}^2 ~+~ \cos^2 y \, \d s_{{\rm S}_2^{2}}^2 \right)\,,
\end{equation}
which is the metric on AdS$_5$ $\times S^5$.
Indeed, if one takes the Poincar\'e form of AdS$_4$ (\ref{eq:Poincare metric}), then one can obtain the Poincar\'e form of AdS$_5$:
\begin{align}
\label{eq:Poincare5}
    \d s_{{\rm AdS}_5}^2 & =\left(\frac{d\nu}{\nu}\right)^2+\nu^2\left(- \d t^2+ \d x_1^2+ \d x_2^2+ \d x_3^2\right) \\ & =  \d x^2  + \cosh^2 x \left[ \left( \frac{d \mu}{\mu} \right)^2 +\mu^2  \left( - \d t^2+ \d x_1^2+ \d x_2^2 \right) \right]
\end{align}
by taking
\begin{equation}
\label{eq:AdS5coords}
    \nu ~=~ \mu \, \cosh x \,, \qquad x_3 ~=~  \frac{1}{\mu} \, \tanh x\,.
\end{equation}

Note that from \eqref{uvzmap1}, the brane coordinates are 
\begin{equation}
\label{eq:D3uvzw}
\begin{aligned}
u  ~=~& 4 \, \alpha_1 \, \mu \,   \cosh(x) \, \sin(y)\,,
 \qquad v  ~=~ 4 \, \alpha_2 \, \mu  \, \cosh(x) \, \cos(y)  \,,\\
 z  ~=~& - \frac{\alpha_2}{\alpha_1} \, \frac{1}{\mu}  \,   \tanh(x)  \,,
 \qquad  \qquad w  ~=~  \frac{\alpha_1}{\alpha_2} \, \frac{1}{\mu}  \,   \tanh(x) \,.
 \end{aligned}
\end{equation}

From this one sees that 
\begin{equation}
\label{eq:uvzwinterp}
\nu   ~=~  \frac{1}{4}\, \sqrt{\bigg(\frac{u}{\alpha_1 }\bigg)^2  ~+~  \bigg(\frac{v}{\alpha_2 }\bigg)^2}   \,, \qquad 
x_3 ~=~ - \frac{\alpha_1}{ \alpha_2}\,  z ~=~   \frac{\alpha_2}{ \alpha_1}\,  w \,. 
\end{equation}
As one would expect from our earlier discussion, $z$ or $w$, are coordinates that follow the world-volume of the D3 branes and, up to scaling, ``$u^2 + v^2$'' is a radial coordinate transverse to the D3's.

This simple solution  can also be exhibited as a solution to the generalized Monge-Ampère system \eqref{MALap}. We find:
\begin{equation}
\begin{aligned}
G_0~=~& -\left(\frac{\alpha_1}{\alpha_2}\right)^2\frac{z^2}{2}-16~\alpha_2^2~\frac{\arctan\left(\frac{v}{u} \frac{\alpha_1}{\alpha_2}\right)}{2 ~u~v} \,,\\
    w~=~&-z~\left(\frac{\alpha_1}{\alpha_2}\right)^2\,,\qquad e^{3 A_0}~=~ \frac{(\alpha_2^2u^2+\alpha_1^2v^2)}{16|\alpha_1^2\alpha_2^4|}\,,
\end{aligned}
\end{equation}
which gives AdS$_5\times S^5$ as in \eqref{eq:AdS5 S5}.

\subsection{The mohawk}
\label{ss:mohawk1}

To understand these geometries  one must first appreciate that in the ``core'' of the solution, away from the singularities, the geometry is dominated by D3 fluxes, even when there are no asymptotic D3 branes present. Indeed, as we noted earlier,  the Chern-Simons interaction gives rise to an $F_5$ flux proportional to $Q_{D5} Q_{NS5}$, which overwhelms the individual $F_3$ and $H_3$ away from their singular sources. The fact that these geometries are dominated by D3 fluxes is implicit in (\ref{eq:dilaton1}), and the fact that the dilaton is {\it independent} of the scale parameter $\mu$ (if one were creating the AdS factor by zooming in on the five branes, one would expect the dilaton to scale with $\mu$). The five-branes then deform the geometry, creating spikes, as one approaches their singular sources.

Whenever a brane ends on another brane, the intersection is distorted as a result of their relative tensions.  This was one of the core results of brane probe analyses like \cite{Callan:1997kz}, and must also be a feature of the fully back-reacted solution. If one zooms in on an  intersection, one focuses on a spike created by one brane pulling on another  and a scale invariance can emerge from the spike profile.  It is possible that the same zooming in can capture multiple brane intersections, and then the spikes nest inside one another creating a self-similar ``mohawk'' \cite{Bena:2024dre}.  The AdS coordinate, $\mu$, becomes the scale parameter of the self-similar structure.

One way to see this structure is to use the brane coordinates $(u,v,z)$ or $(u,v,w)$ defined above.   The $\mu$-scaling behavior in (\ref{uvzmap1})   reveals the mohawk spikes.  In particular, the functions
\begin{equation}
    \hat{z} ~\equiv~  u \, z ~=~ 2 \, h_2^D ~\equiv~ b_1^c\,, \qquad \hat{w} ~\equiv~ v\, w ~=~ 2 \, h_1^D ~\equiv~ -b_2^c\,,
\end{equation}
are scale invariant, and their values define the steepness of the spikes.  At the D5 sources on the top of the strip, the coordinate, $u$, diverges and $z$ vanishes, and the steepness is determined by $h_2^D$. At the NS5 sources on the bottom of the strip, the coordinate, $v$, diverges and $w$ vanishes, and now the steepness is determined by $h_1^D$. 

At the bottom of the strip, $\hat z$ is locally constant and the discrete jumps are determined by the NS5 charges (\ref{eq:5charges}), see Fig.~\ref{fig:stripFunctionsExampleTRhoSigma}.  At the top of the strip, $\hat z$ is continuous but carries the knowledge of the NS5 charges into the environment of the D5 charges and the interaction determines the contribution to the D3 charge from both sources (\ref{eq: Page charge D3 brane}).  Specifically, we have:
\begin{equation}
 \hat z\big(x= \delta_1^{(i)} \big)~=~    \frac{1}{4\pi}\frac{Q_{\text{Page} ,\, \text{D5}}^{D3\,(i)}}{Q^{(i)}_{D5}} \qquad \Leftrightarrow \qquad   z\big(x= \delta_1^{(i)} \big)~=~ \frac{1}{4\pi}   \frac{Q_{\text{Page},\, \text{D5}}^{D3\,(i)}}{Q^{(i)}_{D5}} \, \frac{1}{u} \,.
\end{equation}
Thus, exactly as in \cite{Bena:2024dre}, the steepness of the spike is determined by the charge ratio.  

It is also important to note that $h_2^D$ is monotonically decreasing in $x$ (see Fig.~\ref{fig:stripFunctionsExampleTRhoSigma}), which means that the spike steepness is monotonic, and so the spikes themselves are nested inside one another, defining the layers of the mohawk. In particular, this is how the D5 branes avoid crossing one another. We will discuss in Section \ref{sec:3dN=4} how this corresponds to the dual theory being a good quiver. 

At the top of the strip, the mohawk spikes are defined by the D5 branes deforming the D3 background, with $z \to 0$ and $u \to \infty$.  There is a
parallel story  at the bottom of the strip with NS5 spikes, at which $w \to 0$ and $v \to \infty$, but with $\hat w$ defining the steepness of the layered NS5 mohawk.

Interestingly, if one looks at the bottom of the strip using the $u$ and $z$ coordinates, one finds $u \to 0$, $z \to \infty$, with the steepness function $\hat z$ being locally constant. However, as $u \to 0$, one is approaching the NS5 regime, and one flips to the $w$ and $v$ coordinates to reveal the NS5 mohawk.

The important conclusion here is that  each singular D5 and NS5 source of the \AdSSS solution is an infinite D3-D5 or D3-NS5 spike when written in the ``brane coordinates''. Collectively, these spikes form self-similar mohawks, exactly as in \cite{Bena:2024dre}: Every five-brane bends in exactly the same way, nested inside each other, and there are no other scales associated with their shapes. 

\subsection{Different kinds of solutions depending on the number of asymptotic AdS$_5$ regions}

Depending on the values of $\alpha_k,\,\beta_k$ we can have \AdSSS   
solutions with different types of asymptotic behavior, which we summarize below. We will give the precise physical interpretation and dictionary between these solutions and branes in flat space in Section \ref{sec:Brane interpretation}.

\subsubsection*{Two AdS$_5$ regions}

If $\alpha_1$ and $\alpha_2$ are both non-zero in \eqref{eq:hiRealCoordinates}, 
the AdS$_4$ combines with the Riemann-surface coordinate $x$ to give rise to asymptotic AdS$_5$ ($\times S^5$) geometries:
\begin{equation}
    \d s_{10}^2= L_{\pm}^2 \left[ \left( \d x^2+\cosh(x)^2 \d s^2_{{\rm AdS}_4}  \right)+  \left(  \d y^2+\sin(y)^2 \d \Omega_1^2+\cos(y)^2d\Omega_2^2\right) \right]
\end{equation}
with the radius $L_{\pm}$ given by:
\begin{equation}
    L_{\pm}^4=16\left( \vert\alpha_1 \alpha_2\vert\cosh(\beta_1-\beta_2)+ 2\vert\alpha_2\vert e^{\mp \beta_2}\sum_i\vert \gamma_1^{(i)}\vert e^{ \pm \delta_1^{(i)}}+ 2\vert\alpha_1\vert e^{\mp \beta_1}\sum_j \vert \gamma_2\vert e^{ \pm\delta_2^{(j)}}\right)\,.
    \label{eq:AdSradius}
\end{equation}

The AdS radius gives the number of asymptotic D3 branes in each region by integrating the RR five form field strength: $\int_{S^5} F_5=-4\pi^3 L^4_{\pm}\,.$ It is interesting to note that, except in some peculiar limit, all the five-branes give a contribution to the number of three branes at infinity. 

The number of D3-branes that go all the way through the geometry is 
$\frac{4}{\pi (\alpha')^2} \vert\alpha_1 \alpha_2\vert\cosh(\beta_1-\beta_2)$, where we used \eqref{eq:ND3} to relate the charges in \eqref{eq:D3PageCharges} to the corresponding quantized number of D3-branes.
One can compute the difference between the number of D3s at the two asymptotic regions and obtain:
\begin{equation}
    N_{\textrm{D3}\, ,+\infty}-N_{\textrm{D3}\, ,-\infty}=\frac{16}{\pi (\alpha ')^2}\left( \sum_i\vert \alpha_2\, \gamma_1^{(i)}\vert \sinh(\delta_1^{(i)}-\beta_2)+ \sum_j \vert \alpha_1 \,\gamma_2^{(j)}\vert \sinh(\delta_2^{(j)}-\beta_1)\right)\,. \label{eq:diff ND3}
\end{equation} 
Depending on the sign of $\sinh(\delta_1^{(i)}-\beta_2)$, we either interpret $\vert\alpha_2\, \gamma_1^{(i)}\vert \sinh(\delta_1^{(i)}-\beta_2)$ as the net number of asymptotic D3 branes which arrive from $-\infty$ and end on the D5 branes, or as the number of D3 branes which emanate up from the D5 branes and reach $+\infty$. Similarly, the second term of \eqref{eq:diff ND3} concerns the D3 branes ending on NS5s branes.

\subsubsection*{One asymptotic AdS$_5$ region}
\label{sec:One AdS5 region}
It is possible to close one of the asymptotic regions by taking a particular limit of the parameters (see for example \cite{GarciaEtxebarria:2024jfv}). Indeed, by taking the limit 
\begin{equation}
    \alpha_i\rightarrow 0,\, \beta_i\rightarrow \infty \qquad \textrm{with} \qquad \alpha_i\, e^{\beta_i}=2\theta_i \, \, \textrm{fixed}
\end{equation}
we obtain the harmonic functions:
\begin{align}
    h_1&=i~ \theta_1 \exp(-\RSz)-\sum_i \gamma_{1}^{(i)} \log\left(\tanh\left(-\frac{\RSz-\delta_1^{(i)}-i\frac{\pi}{2}}{2}\right)\right)~+~ \text{c.c} \,, \\
    h_2&= \theta_2 \exp(-\RSz)-\sum_i \gamma_{2}^{(i)} \log\left(\tanh\left(\frac{\RSz-\delta_2^{(i)}}{2}\right)\right)~+~\text{c.c} \,.
\end{align}

The corresponding solution has $L_{+}=0$ while $L_{-}$ is non trivial and given by:
\begin{equation}
    L_{-}^4=16 \left(\frac{\vert\alpha_2\vert e^{\beta_2}}{2}\sum_i\vert \gamma_1^{(i)}\vert e^{- \delta_1^{(i)}}+ \frac{\vert\alpha_1\vert e^{ \beta_1}}{2}\sum_j \vert \gamma_2^{(j)}\vert e^{ \delta_2^{(j)}}\right)\,.
\end{equation}
We see that the D3 charge contribution that we interpreted above as counting the number of D3 branes that go all the way from $-\infty$ to $\infty$ now vanishes. To make sense of this solution we need at least one stack of five branes (one type is enough). Since there is only one asymptotic AdS$_5\times S^5$ region, it is clear that one must have at least one type of five branes on which the three branes can end. 

As we will see in Section \ref{sec:Brane interpretation}, these solutions come from taking a near-horizon limit of a combination of D3-D5 and D3-NS5 spikes.

\subsubsection*{No asymptotic AdS$_5$ region}
It is also possible to close both asymptotic AdS$_5\times S^5$ regions by taking the limit $\alpha_1,\,\alpha_2 \rightarrow 0$ in \eqref{eq: harmonic ansatz}. 
These solutions require both species of branes to be present, otherwise either $h_1$ or $h_2$ is everywhere trivial and one cannot satisfy the regularity condition that $h_i>0$ inside the Riemann surface. 

As we will explain later, this requirement makes intuitive sense: these solutions are expected to capture scaling regions of a system of D3 branes suspended between D5 and NS5 branes, and the field theory living on these D3 branes only flows to an infrared CFT when both D5 and NS5 branes are present. The field theory on D3 branes suspended {\em only} between NS5 branes, with no D5 branes present, is confining, and hence does not flow to a CFT in the infrared\cite{Affleck:1982as,Aharony:1997bx,deBoer:1996mp}.

The most general such solution is given by:
\begin{align}
\label{eq:most general sol with two closing}
\begin{split}
    h_1 (x,y) &= \sum\limits_i \gamma_1^{(i)} \log \left[ \frac{\cosh (x - \delta_1^{(i)}) + \sin (y)}{\cosh (x - \delta_1^{(i)}) - \sin (y)} \right] \,, \\  
    h_2 (x,y) &=  \sum\limits_i \gamma_2^{(i)} \log \left[ \frac{\cosh (x - \delta_2^{(i)}) + \cos (y)}{\cosh (x - \delta_2^{(i)}) - \cos (y)} \right] \, . 
    \end{split}
\end{align}

\section{Born-Infeld meets supergravity }
\label{sec:Probing}

In this section, we probe the \AdSSS solutions first by D3 branes and then by D5-D3 brane spikes. The coordinates in which the near brane limit is expressed, $(x,y,\mu)$ are well-adapted to D3-D5 brane spikes, but not to pure D3 branes. In contrast, the D3-brane probes are completely trivial in the $(u,v,z)$ coordinates.

\subsection{D3 probes}
\label{D3probe}

The Dirac-Born-Infeld (DBI) and Wess-Zumino (WZ) actions for a D3-brane probe are: 
\begin{equation}
    S_{DBI}=-T_{D3}\int d^4\sigma e^{-\phi} \sqrt{-\rm{det}\left( \tilde{g}_{\alpha \beta}+\mathcal{F}_{\alpha\beta}\right)}-T_{D3} \sum_p \int d^4\sigma \,e^{\mathcal{F}} \wedge \tilde{C}_{p}\,,
\end{equation}
where $\mathcal{F}_{\alpha \beta}=\tilde{B}_{\alpha \beta}+2\pi \alpha' F_{\alpha \beta}$ with $F_{\alpha \beta}$ the world-volume gauge field, and $\tilde{g}_{\alpha \beta}$, $\tilde{B}_{\alpha \beta}$ and $\tilde{C}_n$ are the metric, NS-NS and RR fields pulled back to the D3-brane world-volume. We choose this world-volume to span $(t, x_1, x_2, z)$, while sitting at fixed $(u,v)$. With this embedding, the on-shell action vanishes identically, signaling a supersymmetric floating probe.

We therefore study trajectories with $u$ and $v$ held constant in the coordinates adapted to the near-brane \AdSSS limits: 
\begin{equation}
u=2\mu\,h_1=U_0\,,\qquad v=2\mu\,h_2=V_0\,.
\end{equation}
The behavior of this probe depends on the number of asymptotic AdS$_5$ regions. Recall that an AdS$_5$ region is characterized by the divergence of both $h_1$ and $h_2$, which, for the probe at constant $u,~v$ ,  corresponds  to the limit $\mu\to 0$. In contrast, when one of the ends of the \AdSSS solution is closed (or capped) by an AdS$_4\times$ B$_6$ region, both $h_1$ and $h_2$ must vanish, and hence $\mu$ goes to infinity. In these regimes, the D3-branes exit the near-horizon \AdSSS solution. It is important to note that the brane can sit at an arbitrary point in the $\mathbb{R}^6$ spanned by $(u, S_1^2, v, S_2^2)$. This implies that the D3 probe has six flat directions, and that the scalar fields on its world-volume remain massless in this background. 

This is consistent with the fact that the \AdSSS solutions preserve the Killing spinors of flat orthogonal D3, D5, and NS5 branes and probe flat D3 branes are compatible with these Killing spinors.

Note that this makes the \AdSSS solutions qualitatively different from other solutions where D3 branes interact with transverse RR and NSNS 3-form field strengths, like the Polchinski-Strassler solution \cite{Polchinski:2000uf}. Even if these fields do induce position-dependent fermion mass terms on a D3 brane probe \cite{DHoker:2006qeo, Grana:2000jj}, the six scalars remain massless, and this can be ascertained by the fact that our probe D3 branes can be placed at any position in the six-dimensional transverse space spanned by $\vec u$ and $\vec v$.

In the next subsection we probe different regions of the solutions \eqref{Einsteinmet} using flat D3 brane probes, with a particular focus on the asymptotic regions.

\subsubsection*{Probing the AdS$_5$ regions}

We take the Riemann surface to be the strip, and suppose that towards one of its ends both $h_1$ and $h_2$ diverge. Setting $\beta_i=0$ and writing $\zeta=x+iy$, at $x\to\infty$ we obtain:
\begin{equation}
h_1=\alpha_1\,e^{x}\sin y\,,\qquad h_2=\alpha_2\,e^{x}\cos y\,.
\end{equation}
Integrating $u=U_0$ and $v=V_0$ is then straightforward and gives
\begin{equation}
\mu^2 e^{2x}=\left(\frac{U_0}{\alpha_1}\right)^2+\left(\frac{V_0}{\alpha_2}\right)^2\,,\qquad
\tan y=\frac{U_0\,\alpha_2}{V_0\,\alpha_1}\,.
\end{equation}
Thus, asymptotically, the probe D3 brane approaches a straight horizontal line in the Riemann surface: the coordinate $y$ is constant along the brane world-volume, and $\mu\sim e^{-x}$ as $x\to\infty$.

If the geometry has two AdS$_5$ regions, one at each end of the strip, the two harmonic functions never vanish simultaneously. Consequently, solutions exist only for $\mu$ ranging from $0$ up to some finite maximal value, so the probe always remains within the near-horizon region. This makes perfect sense, since having two asymptotic AdS$_5\times S^5$ regions corresponds to having some asymptotic D3-branes at both ends of the strip.

\subsubsection*{Probing the AdS$_4\times B^6$ regions}

The other limiting region of the \AdSSS solutions arises when the spacetime closes off, with both $h_1$ and $h_2$ tending to zero. This corresponds to setting $\alpha_1=\alpha_2=0$, and writing $\zeta=x+iy$. The $x\to\infty$ asymptotics become
\begin{align}
h_1 &\approx 4 e^{-x}\,\sin y \,\sum_i \gamma_1^{(i)} e^{\delta_1^{(i)}}\,,\\
h_2 &\approx 4 e^{-x}\,\cos y \,\sum_i \gamma_2^{(i)} e^{\delta_2^{(i)}}\,.
\end{align}
The trajectory $u=U_0$ and $v=V_0$ can again be integrated and yields
\begin{equation}
\mu^2 \sim e^{2x}\,,\qquad \sin y = \text{const}\,.
\end{equation}
Similarly, $\mu^2\sim e^{-2x}$ as $x\to -\infty$ and hence we see that the D3-brane leaves the near-brane region at the two ends of the strip. It is also worth noting that, along the brane, the coordinate
\begin{equation}
z=\frac{2 h_2^D}{u} \,,
\end{equation}
over which the brane extends, has a range:
\begin{equation}
z \in \left[0,\ \frac{2\pi \sum_i \gamma_2^{(i)}}{U_0}\right]\,.
\end{equation}
This range scales like $1/U_0$ and therefore diverges as $U_0\to 0$. The D3-branes ending on the five-branes form a spike-like configuration, $z\sim 1/u$. While such a geometry can be probed by a D3-brane, the fact that the space closes at both ends of the strip (and hence carries no asymptotic D3-brane charge) implies that, in this limit, the probe necessarily exits the geometry.

Even though the range of $z$ is finite, the proper length of the brane segment inside AdS is still infinite. Indeed, along a trajectory, $(u=U_0$, $v=V_0)$, we have $dz = \frac{1}{u}\,dh_2^D$, and in the simple example of a single stack of D5- and NS5-branes one finds
\begin{equation}
e^{\frac{3}{2}A_0}\bigl(-\partial_z\omega\bigr)^{\frac{3}{4}}\,dz^2
\sim dx^2 \sim \left(\frac{d \mu}{\mu}\right)^2
\end{equation}
as $x\to\infty$. Thus, the proper length of the brane diverges.

\subsection{D3-D5 probes}
\subsubsection*{The probe D3-D5 spike in flat space}

In order to obtain a better intuition for the computation of the next subsection, we first consider a probe D5-spike with D3 charge, placed in a flat ten-dimensional space with the metric
\begin{equation}
    ds^2 = -dt^2 + dx_1^2+ dx_2^2+  dz^2 +du^2+   u^2 d\Omega_2^2 + dv^2 + v^2 d\Omega_2'^2 \,.
\end{equation}

We will choose the static gauge and the world-volume  coordinates $(t,x_1,x_2,u,\theta,\phi)$, where we parametrized an $\mathbb{R}^3$ inside the D5 world-volume by $(u,\theta,\phi)$. Moreover, we will turn on a scalar field, $\Xi(u)$, that describes a direction, $z=\Xi(u)$, transverse to the world-volume, and a field strength, $F \equiv u^2\mathcal{F}(u)\sin(\theta)d \theta d \phi $, along the $S^2$ parametrized by $\theta$ and $\phi$, whose purpose is to induce a D3-brane charge on the D5-brane world-volume. The action of this D5-D3 configuration is: 

\begin{equation}
    S=-T_{D5}\int  d t \,d x_1 \, d x_2 \,  d u \,\text{Vol}(S^2)\,u^2\sqrt{\left(1+(\partial_u\Xi)^2\right)\left(1+\mathcal{F}^2\right)} \,.
\end{equation}

 Solving the equations of motion for $\Xi$ and $\mathcal{F}$ is straightforward and one obtains:
    \begin{equation}
    \mathcal{F}(u)=\frac{q_0}{u^2}~,~~~ \Xi(u)=\frac{q_0}{u}\,,
\label{D5flat}
\end{equation}
where $q_0$ is a constant. The interpretation of this solution is that the D3-brane extended along the $z$ direction is captured in the D5 world-volume action by $ F_{\theta\phi}$, and the   pulling that this D3 brane exerts on the D5 brane is encoded by the scalar $\Xi(u)$ \cite{Callan:1997kz,Constable:1999ac}. The shape of the D5-D3 configuration is a spike.

We can view the same result in a different way, by parameterizing the $(u,z)$ plane with $(u, q \equiv u z)$:
\begin{equation}
    du^2+ dz^2 = du^2 + \left(\frac{dq}{u}-\frac{q}{u^2}du\right)^2 \,.
\end{equation}
If we now place a D5-brane probe again along $(t,x_1,x_2,u,\theta,\phi)$, the solution \eqref{D5flat} corresponds simply to turning on the same world-volume field strength $\mathcal{F}=\frac{q_0 }{u^2} $, and the scalar field is constant and equal to $\Xi=q_0$. 

In the next subsection, we show that similar D5–D3 spikes probing the \AdSSS solution extend in fact along AdS$_4\times$S$^2$ and are localized at a point on the Riemann surface. The D3 Page charge of the spike is related to the location of this point, exactly like the D3 Page charge of the flat-space D3-D5 spike is related to $q=q_0$. This shows that \AdSSS supergravity solutions are naturally adapted to spike configurations, and that the AdS factor arises from the near-brane limit of fully back-reacted D3–D5–NS5 spike(-like) regions, in a sense that will be made precise later.

\subsubsection*{D5-D3 probe spikes in the \AdSSS solutions}

It is now time to probe the solutions of Section \ref{ss:nbi}, using a D5-brane probe extending along AdS$_4 \times S_1^2$ with world-volume D3 flux on it, in order to further elucidate their brane interpretation. We will work in the string frame, in which the solution \eqref{Einsteinmet} is: 
\begin{equation}
    \d s^2=e^{\phi}f_4^2 \d s^2_{AdS_4}+e^{\phi}f_1^2 \d s_{S_1^2}^2 + e^{\phi}f_2^2 \d s_{S_2^2}^2+4e^{\phi}\rho^2 (\d x^2+\d y^2) \,.
\label{stringframemet}
\end{equation}
As before, $\phi$ is related to the actual dilaton $\Phi$ via $\phi=\Phi/2$, and like in the rest of the paper, we write AdS$_4$ in Poincar\'e coordinates.

We choose a static gauge for the world-volume coordinates of the D5 probe, which means that we identify the first four coordinates with the AdS$_4$ coordinates $(t,\mu,x_1,x_2)$, and the last two with the coordinates on $S_1^2$. As $(t,x_1,x_2)$ and the $S_1^2$ represent isometry directions, we assume that the Riemann-surface coordinates, $x$ and $y$, which become scalar fields on the world-volume of the probe-brane, depend only on $\mu$. The induced metric on the probe brane is therefore:
\begin{equation}
\begin{aligned}
    \d \tilde{s}_6^2=e^{\phi}\mu^2 f_4^2 (- \d t^2 + \d x_1^2+ \d x_2^2) +e^{\phi}\left(\frac{f_4^2}{\mu^2}+4\rho^2\left( \left( \frac{\partial x}{\partial \mu} \right)^2+ \left(\frac{\partial y}{\partial \mu} \right)^2 \right) \right) \d \mu^2 +e^{\phi}f_1^2 \d s^2_{S_1^2}\,.
\end{aligned}
\end{equation}
The induced NS-NS and RR fields are trivially pulled back on the D5-brane world-volume and are given by
\begin{equation}
    B_2=b_1 \text{Vol}(S_1^2)\,, \quad \quad \quad C_4=-4j_1 \text{Vol}(\text{AdS}_4) \,.
\end{equation}
In order to account for the world-volume D3 charge, we need to turn on a world-volume two-form field of the form: 
\begin{equation}
    F_2=\Pi ~ \text{Vol}(S_1^2)\,,
\end{equation}
where $\Pi$ is proportional to the ratio of the D3 and D5 charges of the probe brane.

It is now straightforward to compute the DBI and WZ actions:
\begin{equation}
\begin{aligned}
    S_{DBI}&=-T_{D5} \int \d^6 \sigma \,   e^{-2\phi} \sqrt{-\det \left( \tilde{G}_{\alpha \beta}+F_{\alpha \beta} + \tilde{B}_{\alpha \beta} \right)} \\ 
    &=-T_{D5}\int \text{Vol}({\rm{AdS}}_4)\text{Vol}(S_1^2) ~f_4^3~ \sqrt{\left(e^{2\phi}\,f_1^4+(b_1+\Pi)^2 \right)\left(f_4^2+4\mu^2 \rho^2 \left( \left(\partial_{\mu} x\right)^2+ \left(\partial_{\mu} y \right)^2\right)\right)} \,,\\
    S_{WZ}&=T_{D5} \int\, e^{\tilde{B}_2+F_2}\wedge \sum_n \tilde{C}_n=T_{D5} \int \text{Vol}({\rm{AdS}}_4)\text{Vol}(S_1^2) \left(-4j_1 (b_1+\Pi) \right)+T_{D5} \int \, \tilde{C}_6\,. 
\end{aligned}
\end{equation}
In order to determine $\tilde{C}_6$ we need the following relations:
\begin{equation}
\begin{aligned}
\label{conventions}
    F_7&=\d C_6 + H_3\wedge C_4 \,,  \\
    F_7&= \star F_3 \,. 
\end{aligned}
\end{equation}

Performing the Hodge star operation and using \eqref{conventions} we find:
\begin{equation}
\begin{aligned}
    \d C_6=&\Bigg(\left(-\frac{e^{2\phi}f_1^2f_4^4}{f_2^2}\partial_y b_2+4j_1 \partial_x b_1 \right)\, \d x+ \left( \frac{e^{2\phi}f_1^2f_4^4}{f_2^2}\partial_x b_2 + 4j_1 \partial_y b_1  \right)\, \d y \Bigg)\wedge \text{Vol}(AdS_4)\wedge \text{Vol}(S_2) \,,
\end{aligned}
\end{equation}
where we kept only the first term of $F_5$ in \eqref{F5field}, keeping in mind that we will eventually take the pullback of this expression to the D5 world-volume. 
We do not need to find $C_6$, since we are going to solve the equations of motion, where only $\d C_6$ appears. To see this, we write the part of $C_6$ that concerns us as:
\begin{equation}
    C_6=g_1 \omega^{012345}+g_2 \omega^{02345}\wedge \d x + g_3 \omega^{02345} \wedge  \d y \,,
\label{dC6}
\end{equation}
where $(g_1,g_2,g_3)$ are functions of $(\mu,x,y)$. In \eqref{dC6}, 
we denoted $\text{Vol}(AdS_4)\wedge \text{Vol}(S_1^2)$ by $\omega^{012345}$, and the part of $\text{Vol}(AdS_4)\wedge \text{Vol}(S_1^2)$ whithout the $\frac{d\mu}{\mu}$ factor by $\omega^{02345}$. Then 
\begin{equation}
\label{dC6an}
    \d C_6=\left( \partial_x g_1 -\partial_{\mu} g_2\right)\omega^{012345}\wedge \d x + \left( \partial_y g_1 -\partial_{\mu} g_3\right)\omega^{012345}\wedge  \d y \,,
\end{equation}
and the pullback $\tilde{C}_6$ is given by:
\begin{equation}
    \tilde{C}_6=(g_1+g_2 \partial_{\mu} x+g_3 \partial_{\mu} y) \,\omega^{012345} \,.
\end{equation}
The equations of motion for $x$ and $y$ will now give:
\begin{equation}
\begin{aligned}
    &\frac{\delta \mathcal{L}_{DBI}}{\delta x}+T_5\partial_x \left( (\tilde{B}_2+F_2)\wedge \tilde{C}_4\right)+T_5\left(\partial_x g_1 -\partial_{\mu}g_2 \right)=0 \,, \\
     &\frac{\delta \mathcal{L}_{DBI}}{\delta y} +T_5\partial_y \left( (\tilde{B}_2+F_2)\wedge \tilde{C}_4\right)+T_5\left(\partial_y g_1 -\partial_{\mu}g_3 \right)=0 \,.
\end{aligned}
\label{eom1}
\end{equation}
We see, therefore, that for the equations of motion we only need the two components of $dC_6$ in \eqref{dC6an}.  

Since we are interested in putting D5-D3 spike probes at a fixed point at the upper boundary of the strip, we will take the $y\rightarrow \pi/2$ limit of $\delta \mathcal{L}_{DBI+WZ}/\delta x$ and set $\partial x/\partial \mu$ to zero to arrive at: 
\begin{equation}
    \frac{\delta \mathcal{L}_{DBI+WZ}}{\delta x}=-\partial_x \left(f_4^4 \sqrt{e^{2\phi}f_1^4+(b_1+\Pi)^2} \right)-4(b_1+\Pi)\partial_x j_1-\frac{e^{2\phi}f_1^2f_4^4}{f_2^2}\partial_y b_2 \,.  
\end{equation}

\begin{figure}
    \centering
\begin{tabular}{cc}
\includegraphics[scale=.5]{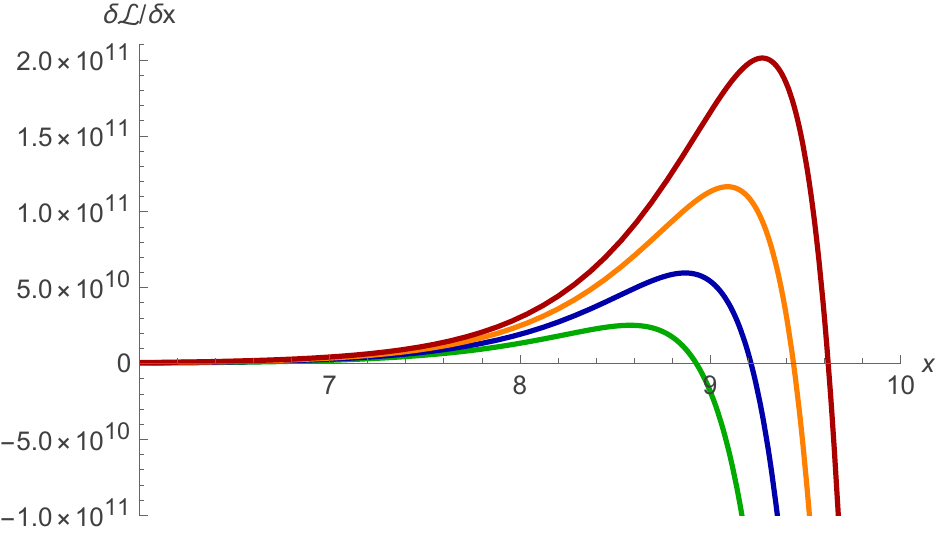} & \includegraphics[scale=.5]{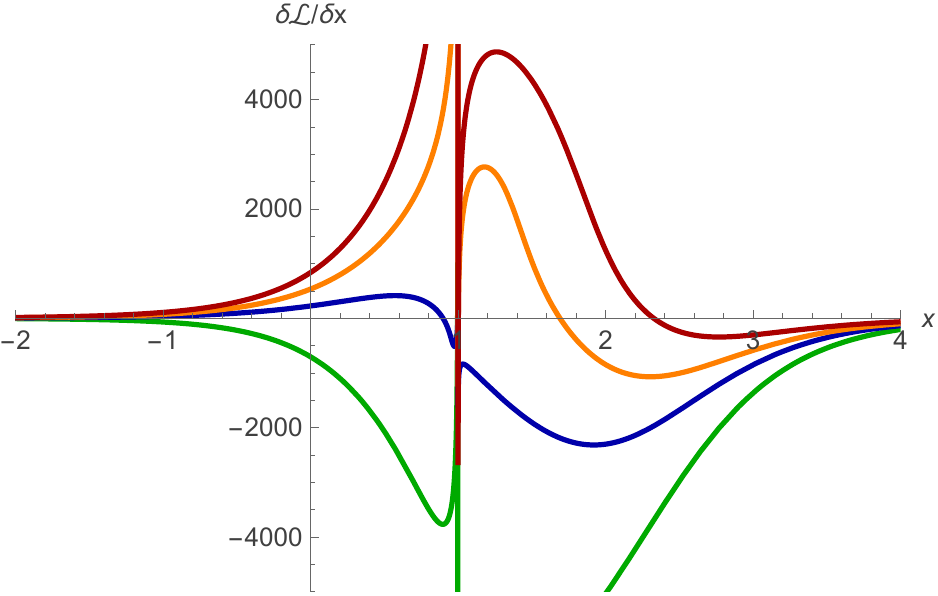}
\end{tabular}
    \caption{The $x$ equation of motion for various values of $\Pi$ for a Janus solution (a), and a solution with no asymptotic AdS$_5 \times S^5$ regions (b). The intersection of  a curve with the $x$ gives the position where the D5-D3-brane probe feels no force. \\[0.3em]
    (a): Janus solution with parameters $(\alpha_1, \alpha_2,\beta_1,\beta_2)=(1,0.2,-5,3)$. The D3-D5 charge ratio, $\Pi$, takes the values $(150,200,250,300)$ from left to right. There are solutions for all values of $\Pi$. \\[0.3em]
    (b): Solution with parameters $(\gamma_1, \gamma_2,\delta_1, \delta_2)=(1,4,1,2)$. Here $\Pi$ takes the values $(\text{Green, Blue, Orange, Red})=(-60,-40,-30,-20)$. There are solutions only for $\Pi \in [-4\pi\,\gamma_2,0]$ as one can easily deduce from \eqref{noAdS5Pi}.}
    \label{probeplots}
\end{figure}

In Fig.~\ref{probeplots} we depict the above function for various values of $\Pi$ for the Janus solution, where $\gamma_1^{(i)}= \gamma_2^{(i)}=0$, and 
for the solution with no asymptotic AdS$_5 \times S^5$ regions, where $\alpha_1=\alpha_2=0$, with a single stack of D5-branes and a single stack of NS5-branes. By explicitly computing the zeros, $x_0$, of this function we observe that for the Janus solution, $\Pi$ obeys:
\begin{equation}
    \Pi=4 \alpha_2 \sinh (x_0- \beta_2) \,,
\end{equation}
whereas for the solutions with no asymptotic AdS$_5\times S^5$ regions we find 
\begin{equation}
    \Pi=-8 \gamma_2 \arctan (e^{\delta_2 -x_0}) \,.
\label{noAdS5Pi}
\end{equation}

As was reviewed in Section \ref{ss:Bcharges}, the D3-brane charge associated with a singularity located at $\delta_1^{(a)}=x_0$ in a solution with a single stack of NS5-branes at $\delta_2$ is given by 
\begin{equation}
    Q_{D3}=2^8\pi^3 \left( \alpha_2 \gamma_1 \sinh (x_0-\beta_2)-2\gamma_1 ~ \gamma_2 \arctan (e^{\delta_2 -x_0}) \right) \,,
\label{QD3brane}
\end{equation}
while its D5-brane charge is $Q_{D5}=(4\pi)^2\gamma_1^{(a)}$. Note that we set $\xi_2=0$ in \eqref{QD3brane}. Since by definition $\Pi$ is given by 
\begin{equation}
    \Pi=\frac{1}{4\pi}\frac{Q_{D3}}{Q_{D5}} \,,
\end{equation}
we see that in both examples our probe calculation matches the D3-brane charge of a solution with an extra singularity at $(x=x_0,y=\pi/2)$. The fact that we find a perfect agreement means that all the \AdSSS  solutions describe a zoom-in of a system of D3 branes ending on or suspended between D5 and NS5 branes that focuses on the region where the D5 and NS5 branes have a scaling symmetry. Even if these regions are finite, the zoom-in transforms all the five-branes into infinite spikes\footnote{This analysis makes the probe computation of \cite{Coccia:2021lpp} clear: the fundamental-string and D1-brane probes take a spiking form, such that their world-volumes respectively follow the profiles of the NS5- and D5-branes. A similar argument holds for the D5 probes.}.

\section{Brane systems and field theory}
\label{sec:Brane interpretation - field theory}

In this section we turn off back-reaction, and we review the physics of 8-supercharge D3-D5-NS5 systems and the three-dimensional $\mathcal{N}=4$ SCFTs that are realized on their world-volume.

\subsection{Linking numbers from supergravity}

\paragraph{Brane charges and brane numbers.} The starting point are the charges computed at the end of Section \ref{sec:Near brane intersection}, which can be converted to integer quantized five-brane numbers, $N_{\rm D5}^{(i)}$ and $N_{\rm NS5}^{(i)}$,  via  
\begin{equation}
    Q_{\rm D5}^{(i)} = \left( 2 \pi \sqrt{\alpha '} \right)^2 N_{\rm D5}^{(i)} \, , \qquad Q_{\rm NS5}^{(i)} = - \left( 2 \pi \sqrt{\alpha '} \right)^2 N_{\rm NS5}^{(i)} 
\end{equation}
and to the integer quantized D3 Page-charge numbers on the five-branes, $N_{\rm D5}^{{\rm D3}\,(i)}$ and $N_{\rm NS5}^{{\rm D3}\,(i)}$, via
\begin{equation}
    Q_{\text{Page} ,  \, \text{D5}}^{\text{D3}\,(i)}  = \left( 2 \pi \sqrt{\alpha '} \right)^4 N^{{\rm D3} \, (i)}_{\rm D5} \, , \qquad Q_{\text{Page} ,  \, \text{NS5}}^{\text{D3}\,(i)}  = \left( 2 \pi \sqrt{\alpha '} \right)^4 N^{{\rm D3} \, (i)}_{\rm NS5} \,.
     \label{eq:ND3}
\end{equation}
Replacing the charges by integers, Equation \eqref{eq:3-chargesPage} allows us to define the linking numbers
\begin{equation}
   \begin{split}
        \ell^{(i)}_1 = \frac{ N_{\rm D5}^{\text{D3}\,(i)} }{N_{\rm D5}^{(i)}} &=   -\xi_2   + \frac{4 \alpha_2}{\pi \alpha '}    \sinh \left( \delta_1^{(i)} - \beta_2 \right) - \frac{2}{\pi}  \sum\limits_j N_{\rm NS5}^{(j)}  \arctan \left( e^{\delta_2^{(j)} - \delta_1^{(i)}} \right) \,, \\ \ell^{(i)}_2 = \frac{ N_{\rm NS5}^{\text{D3}\,(i)} }{N_{\rm NS5}^{(i)}} &=  \xi_1  + \frac{4 \alpha_1}{\pi \alpha '} \sinh \left( \delta_2^{(i)} - \beta_1 \right) +  \frac{2}{\pi} \sum\limits_j  N_{\rm D5}^{(j)}  \arctan \left( e^{\delta_2^{(i)} - \delta_1^{(j)}}  \right) \, . 
    \end{split}
\label{LinkingCharges}
\end{equation}
This equation imposes a quantization condition on $\delta_1^{i}$ and $\delta_2^{i}$ \cite{Assel:2011xz}. Note that, unlike some earlier  work, we do not introduce any minus sign in \eqref{LinkingCharges}.

\paragraph{Linking numbers in brane systems} 
Consider the situation where $\alpha_1 = \alpha_2 = 0$, and all the $\delta^{(i)}_{1,2}$ are far apart. In this situation, the functions $h_{1,2}$, $h^D_{1,2}$ and $f_i$ are depicted in Fig. \ref{fig:stripFunctionsExampleTRhoSigma}. Then 
\begin{equation}
    \frac{2}{\pi} \arctan (e^{\delta_2 - \delta_1}) \sim \begin{cases} 0 & \text{ if } \delta_2 < \delta_1 \\ 1 & \text{ if } \delta_2 > \delta_1\end{cases}
\end{equation}
so 
\begin{equation}
   \begin{split}
        \ell^{(i)}_1 &\sim  -\xi_2  -   \sum\limits_{j | \delta_2^{(j)} > \delta_1^{(i)}} N_{\rm NS5}^{(j)}  \,, \\ \ell^{(i)}_2 &\sim \xi_1 + \sum\limits_{j | \delta_2^{(i)} > \delta_1^{(j)}} N_{\rm D5}^{(j)}    \, . 
    \end{split}
\end{equation}
Naively, the sum $\sum_{j | \delta_2^{(j)} > \delta_1^{(i)}} N_{\rm NS5}^{(j)}$ is the number of NS5 branes located on the right of the D5 stack labeled by $i$ in the $z$ direction.
 However, the physical location of the D5 stack can be changed as long as the linking is not modified. The interaction between the D5 and NS5 branes elongates the five-brane world-volumes, and can create very long spikes that can be described as  effective 3+1-dimensional world-volumes, which can be interpreted as D3 branes. This is another incarnation of the fact that D3 brane charges are given by formulas \eqref{eq:D3PageCharges} and not \eqref{eq:D3BraneCharge}, and this is the Hanany-Witten effect.
Note that one can pick a ``symmetric gauge" by adding a background gauge field with
\begin{equation}
    \xi_1 = - \frac{1}{2} \sum_j  N_{\rm D5}^{(j)} \, , \qquad \xi_2 = - \frac{1}{2} \sum_j  N_{\rm NS5}^{(j)} \, . 
    \label{eq:symmetricGauge}
\end{equation}
In the symmetric gauge \eqref{eq:symmetricGauge} we have 
\begin{equation}
   \begin{split}
        \ell^{(i)}_1 &=     \frac{1}{2} \left( N_{\text{NS5,left}} -  N_{\text{NS5,right}} \right) + N_{\text{D3,right}} - N_{\text{D3,left}}  \,,  \\ \ell^{(i)}_2 &=   \frac{1}{2} \left( N_{\text{D5,left}} -  N_{\text{D5,right}} \right)    + N_{\text{D3,right}} - N_{\text{D3,left}}   \, . 
    \end{split}
    \label{eq:symmetricGaugeLinkingNumbers}
\end{equation}
This uniquely defines brane systems up to Hanany-Witten moves, as we now review.  

\paragraph{Holographic duals of \AdSSS solutions:}
These holographic duals involve the 4-dimensional $\mathcal{N}=4$ SYM theory with $U(N)$ gauge group, its supersymmetric 3-dimensional boundary conditions and its domain walls. The half-BPS boundary conditions of 4-dimensional $\mathcal{N}=4$ SYM have been analyzed thoroughly in \cite{Gaiotto:2008sa,Gaiotto:2008ak}. 
 
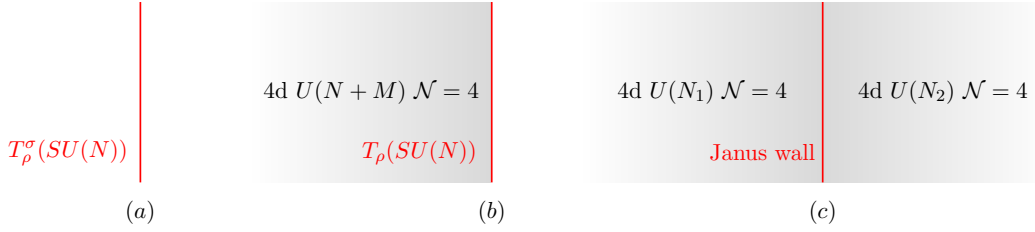
\begin{figure}
\begin{center}
\scalebox{.8}{\begin{tikzpicture}
\node at (0,0) {\begin{tikzpicture}
  \shade[left color=white, right color=white] 
    (0,0) rectangle (-1,3);
  \shade[right color=white, left color=white] 
    (0,0) rectangle (1,3);
 \draw[thick,red] (0,0)--(0,3);
 \node at (-1.2,.5) {\color{red} $T_{\rho}^{\sigma}(SU(N))$};
 \node at (0,-0.5) {$(a)$};
\end{tikzpicture}};
\node at (5,0) {\begin{tikzpicture}
  \shade[left color=white, right color=gray!30] 
    (0,0) rectangle (-4,3);
  \shade[right color=white, left color=white] 
    (0,0) rectangle (1,3);
 \draw[thick,red] (0,0)--(0,3);
 \node at (-1.2,.5) {\color{red} $T_{\rho}(SU(N))$};
 \node at (-2,1.5) {4d $U(N+M)$ $\mathcal{N}=4$};
 \node at (0,-0.5) {$(b)$};
\end{tikzpicture}};
\node at (12,0) {\begin{tikzpicture}
  \shade[left color=white, right color=gray!30] 
    (0,0) rectangle (-4,3);
  \shade[right color=white, left color=gray!30] 
    (0,0) rectangle (4,3);
 \draw[thick,red] (0,0)--(0,3);
 \node at (-1,.5) {\color{red} Janus wall};
 \node at (-2,1.5) {4d $U(N_1)$ $\mathcal{N}=4$};
 \node at (2,1.5) {4d $U(N_2)$ $\mathcal{N}=4$}; 
 \node at (0,-0.5) {$(c)$};
\end{tikzpicture}};
\end{tikzpicture}}
\end{center}
    \caption{The red line denotes a $1+2$-dimensional $\mathcal{N}=4$ theory embedded into $1+3$-dimensional space (the added dimension being the horizontal direction). The three dimensional theory can be coupled to a four-dimensional theory on one side (panel $(b)$) or two sides (panel $(c)$). }
    \label{fig:N=4boundaries}
\end{figure}

There are three kinds of holographically dual theories,  illustrated in Fig. \ref{fig:N=4boundaries}, each of which comes from the near-horizon limit of a certain brane system, reviewed below. When the solution can be written as AdS$_4$ times a compact space, it is dual to a 2+1 dimensional CFT (Fig.~\ref{fig:N=4boundaries}$(a)$). When the solution has one  AdS$_5$ asymptotic region it is dual to $\mathcal{N}=4$ SYM theory with a conformal boundary (Fig.~\ref{fig:N=4boundaries}$(b)$). When the solution has two  AdS$_5$ asymptotic regions it is dual to a Janus domain wall (Fig.~\ref{fig:N=4boundaries}$(c)$). We describe in turn each of these three kinds of solutions, and leave the discussion of multi-Janus solutions with more than two AdS$_5$ asymptotic regions to Section \ref{Multi-Janus}.

\subsection{Three-dimensional  $\mathcal{N}=4$ SCFTs}
\label{sec:3dN=4}

When there is no asymptotic AdS$_5$  region (Fig.~\ref{fig:N=4boundaries}$(a)$), we are dealing with a purely three-dimensional field theory, dual to an AdS$_4$ geometry in the near horizon limit.  Here, we recall some basic facts about such 3-dimensional $\mathcal{N}=4$ superconformal field theory (SCFT).  

\paragraph{Good 3-dimensional $\mathcal{N}=4$ theories. }
In four spacetime dimensions, some supersymmetric gauge theories are conformally invariant. Examples include $\mathcal{N}=4$ SYM and $\mathcal{N}=2$ $SU(N_c)$ gauge theories with $N_f = 2 N_c$ hypermultiplets in the fundamental representation. 

In three spacetime dimensions, on the contrary, the Yang-Mills coupling constant always runs. A simple dimensional analysis shows that $g_{YM} \rightarrow 0$ in the UV and $g_{YM} \rightarrow \infty$ in the IR: in other words, gauge theories are UV complete, but can flow to non-trivial strongly interacting fixed point in the IR. Depending on the situation, the flow can be more or less easily tractable. This led Gaiotto and Witten to introduce a classification of 3-dimensional $\mathcal{N}=4$ theories\footnote{3-dimensional $\mathcal{N}=4$ theories have the same number of supercharges as 4-dimensional $\mathcal{N}=2$  theories, that is 8 supercharges. At the superconformal point, there are 8 additional fermionic generators in the superconformal algebra. } into \emph{good}, \emph{bad} and \emph{ugly} theories. The good theories are, roughly speaking, those for which the flow is most tractable. 
The precise definition involves the charge of monopole operators\footnote{The supersymmetry algebra implies an $R$-symmetry $\mathfrak{so}(4) = \mathfrak{su}(2)_C \oplus \mathfrak{su}(2)_H$. Monopole operators are uncharged under the $\mathfrak{su}(2)_H$ summand, and transform in a spin $q \in \frac{1}{2} \mathbb{N}$ representation of $\mathfrak{su}(2)_C$.  We say that the theory is good if all monopole operators have charge $q \geq 1$.}, but for our purposes, it is enough to state the following properties of good theories: 
\begin{itemize}
    \item The moduli space of vacua of good 3-dimensional $\mathcal{N}=4$ SCFTs contains a \emph{unique} most singular point, located generically at the intersections between the Higgs and Coulomb branch.\footnote{Recall that at a generic (equivalently, smooth) point on the moduli space of vacua, there is an effective theory, which can be in various phases (Coulomb phase on the Coulomb branch, Higgs phase on the Higgs branch). Singularities signal a breaking of the validity of the effective theory corresponding to having integrated out massless degrees of freedom. In other words, singularities correspond to interacting fixed points. The singular locus can always be stratified into smooth strata, which are partially ordered under closure \cite{Assel:2017jgo,Bourget:2019aer}. This allows to define the \emph{most singular} locus, which may or may not be a point. For good theories, it is a point, and this point is the intersection of the Higgs and Coulomb branches. } This defines a unique SCFT, which is 
    \begin{enumerate}
        \item[$(i)$] entirely determined by the UV Lagrangian, which encodes the gauge group and matter content;\footnote{For simplicity, we assume throughout this work that masses and Fayet-Iliopoulos terms are set to zero. }
        \item[$(ii)$] contains all the degrees of freedom of the UV theory, and additional ones -- typically, massless monopole operators. 
    \end{enumerate}
    \item A theory with gauge group $U(N_c)$ and $N_f$ hypermultiplets transforming in the fundamental representation is good if (and only if) 
    \begin{equation}
    \label{eq:condition_good}
        N_f \geq 2 N_c \, . 
    \end{equation} 
\end{itemize}

\paragraph{The class of $T^{\sigma}_{\rho} [SU(N)]$ theories. } 
Using this, we can construct a large class of good 3-dimensional $\mathcal{N}=4$ SCFTs as framed linear unitary quivers. This means that we have a chain of gauge groups of the form $U(N_c)$, in which the bound \eqref{eq:condition_good} is satisfied locally at each gauge node. The whole quiver (hence the whole  good UV theory, hence the whole IR SCFT) can be encoded in a triple $(N,\rho,\sigma)$ where $N \geq 2$ is an integer, and $\rho = [\rho_1 , \rho_2 , \dots , \rho_N]$, $\sigma = [\sigma_1 , \sigma_2 , \dots , \sigma_N]$ are two partitions\footnote{A partition $\rho$ of $N$ is a non-increasing sequence of non-negative integers which sums to $N$. This means $\rho_1 \geq \rho_2 \geq \dots \geq \rho_N \geq 0$ with $\rho_1 + \dots + \rho_N = N$. If needed, we can assume the sequence is infinite, with $\rho_{N+1} = \dots = 0$. The number of $i$ such that $\rho_i >0$ is denoted $|\rho|$.  Each partition can be represented as a Young diagram. The transpose of a partition $\rho$ is a partition that we denote $\hat{\rho}$, and it corresponds to the transposed Young diagram. } of $N$ satisfying 
\begin{equation}
\label{eq:orderPartition}
    \hat{\rho} \geq \sigma \, , 
\end{equation}
where the hat denotes the transpose (see footnotes) and we use the \emph{dominance order} on partitions: 
\begin{equation}
     \hat{\rho} \geq \sigma \qquad \Leftrightarrow \qquad \forall \,i = 1 , \dots , N \, , \quad \sum\limits_{j=1}^i \hat{\rho}_j \geq \sum\limits_{j=1}^i \sigma_j \, .  \end{equation}
The corresponding theory is called $T^{\sigma}_{\rho} [SU(N)]$. 
From these data, we define for all $j \geq 1$
\begin{equation}
\label{eq:formulaMN}
    M_j \equiv \hat{\sigma}_j - \hat{\sigma}_{j+1}  \, , \qquad N_j \equiv \sum\limits_{k \geq j+1} (\rho_k - \hat{\sigma}_k) \, . 
\end{equation}
 Note that $M_j \geq 0$ because the partitions are ordered, and $N_j \geq 0$ as a consequence of \eqref{eq:orderPartition}. In what follows, we make an assumption which is stronger than \eqref{eq:orderPartition}, namely we assume
 \begin{equation}
    \forall j \in \{ 1 , \dots , |\rho|\} \, , \quad  N_j > 0 \, . 
 \end{equation}
This ensures that no NS5 brane can decouple from the brane system.  
From there, one can build a linear quiver 
 \begin{equation}
    \raisebox{-.5\height}{\begin{tikzpicture}[xscale=0.75]
    \node[gauge,label=below:{$\mathrm{U}(N_1)$}] (1) at (0,0) {};
    \node[gauge,label=below:{$\mathrm{U}(N_2)$}] (2) at (2,0) {};
    \node (d) at (3,0) {$\cdots$};
    \node[flavor,label=left:{$M_1$}] (3) at (0,1) {};
    \node[flavor,label=left:{$M_2$}] (4) at (2,1) {};
    \draw (1)--(2)--(d);
    \draw (1)--(3);
    \draw (2)--(4);
    \end{tikzpicture}} \;  
    \label{eq:quiverGeneraln}
\end{equation}
The (electric) flavor symmetry is $H_{\sigma} = S (U(M_1) \times U(M_2) \times \dots)$. There is also a similarly defined magnetic symmetry $H_{\rho}$ (which can also be read from the magnetic quiver introduced below).

\begin{figure}
    \centering
 \newcommand{\branesystempelc}{\begin{tikzpicture}
\node at (0,0) {\curvebellD{3}{.7}};
\node at (.3,0) {\curvebellD{1}{-.7}};
\node at (1.15,0) {\curvebellNS{1}{-1.5}};
\draw [thick,red] (0,-3)--(0,3);
\draw [thick,red] (0.1,-3)--(0.1,3);
\draw [thick,red] (0.2,-3)--(0.2,3);
\draw [thick,red] (0.3,-3)--(0.3,3);
\draw[thick] (0,0)--(-.8,0);
\draw[thick] (.3,0)--(1.9,0);
\end{tikzpicture}}
   \newcommand{\branesystempartitions}{
\begin{tikzpicture}[xscale=.75]
\draw [NS5](0,-1)--(0,1);
\draw [NS5](2,-1)--(2,1);
\draw [NS5](4,-1)--(4,1);
\draw [NS5](6,-1)--(6,1);
\draw [NS5](8,-1)--(8,1);
\draw[thick] (-8,0)--(8,0);
\node[D5] at (-2,0) {};
\node[D5] at (-4,0) {};
\node[D5] at (-6,0) {};
\node[D5] at (-8,0) {};
\node at (-7,-.3) {2};
\node at (-5,-.3) {4};
\node at (-3,-.3) {6};
\node at (-1,-.3) {9};
\node at (1,-.3) {7};
\node at (3,-.3) {5};
\node at (5,-.3) {3};
\node at (7,-.3) {1};
\node at (-8,.4) {\textcolor{blue} {$-\frac{1}{2}$}};
\node at (-6,.4) {\textcolor{blue} {$-\frac{1}{2}$}};
\node at (-4,.4) {\textcolor{blue} {$-\frac{1}{2}$}};
\node at (-2,.4) {\textcolor{blue} {$+\frac{1}{2}$}};
\node at (0.3,.7) {\textcolor{red} {$0$}};
\node at (2.3,.7) {\textcolor{red} {$0$}};
\node at (4.3,.7) {\textcolor{red} {$0$}};
\node at (6.3,.7) {\textcolor{red} {$0$}};
\node at (8.3,.7) {\textcolor{red} {$1$}};
\end{tikzpicture}}
\newcommand{\branesystemHiggs}{\begin{tikzpicture}[xscale=.75]
\draw [NS5](0,-1)--(0,2);
\draw [NS5](2,-1)--(2,2);
\draw [NS5](4,-1)--(4,2);
\draw [NS5](6,-1)--(6,2);
\draw [NS5](8,-1)--(8,2);
\node at (0.3,1.7) {\textcolor{red} 0};
\node at (2.3,1.7) {\textcolor{red} 0};
\node at (4.3,1.7) {\textcolor{red} 0};
\node at (6.3,1.7) {\textcolor{red} 0};
\node at (8.3,1.7) {\textcolor{red} 1};
\node at (3.5,.5) {\textcolor{blue} {$-\frac{1}{2}$}};
\node at (3.5,1) {\textcolor{blue} {$-\frac{1}{2}$}};
\node at (3.5,1.5) {\textcolor{blue} {$-\frac{1}{2}$}};
\node at (5.5,1) {\textcolor{blue} {$+\frac{1}{2}$}};
\node[D5] at (3,.5) {};
\node[D5] at (3,1) {};
\node[D5] at (3,1.5) {};
\node[D5] at (5,1) {};
\draw[D3] (0,0)--(8,0);
\draw[D3] (0,-.1)--(6,-.1);
\draw[D3] (2,-.2)--(6,-.2);
\draw[D3] (2,-.3)--(4,-.3); 
\node[gauge,label=below:{$\mathrm{U}(2)$}] (1) at (1,-2) {};
\node[gauge,label=below:{$\mathrm{U}(4)$}] (2) at (3,-2) {};
\node[gauge,label=below:{$\mathrm{U}(3)$}] (3) at (5,-2) {};
\node[gauge,label=below:{$\mathrm{U}(1)$}] (4) at (7,-2) {};
\node[flavor,label=left:{$3$}] (12) at (3,-1) {};
\node[flavor,label=left:{$1$}] (13) at (5,-1) {};
\draw[red] (1)--(2)--(3)--(4);
\draw[blue] (2)--(12);
\draw[blue] (3)--(13);
\end{tikzpicture}}
\newcommand{\branesystemCoulomb}{\begin{tikzpicture}[xscale=.75]
\draw[D3] (-8,.05)--(-2,.05);
\draw[D3] (-8,-.05)--(-2,-.05);
\draw[D3] (-6,-.15)--(-4,-.15);
\draw[D3] (-6,.15)--(-4,.15);
\node[D5] at (-2,0) {};
\node[D5] at (-4,0) {};
\node[D5] at (-6,0) {};
\node[D5] at (-8,0) {};
\draw [NS5](-5.3,-1)--(-5.3,1);
\draw [NS5](-5.1,-1)--(-5.1,1);
\draw [NS5](-4.9,-1)--(-4.9,1);
\draw [NS5](-4.7,-1)--(-4.7,1);
\draw [NS5](-3,-1)--(-3,1);
\node at (-5.3,1.3) {\textcolor{red} 0};
\node at (-5.1,1.3) {\textcolor{red} 0};
\node at (-4.9,1.3) {\textcolor{red} 0};
\node at (-4.7,1.3) {\textcolor{red} 0};
\node at (-3,1.3) {\textcolor{red} 1};
\node at (-8,.4) {\textcolor{blue} {$-\frac{1}{2}$}};
\node at (-6,.4) {\textcolor{blue} {$-\frac{1}{2}$}};
\node at (-4,.4) {\textcolor{blue} {$-\frac{1}{2}$}};
\node at (-2,.4) {\textcolor{blue} {$+\frac{1}{2}$}};
\node[gauge,label=below:{$\mathrm{U}(2)$}] (1) at (-7,-2.5) {};
\node[gauge,label=below:{$\mathrm{U}(4)$}] (2) at (-5,-2.5) {};
\node[gauge,label=below:{$\mathrm{U}(3)$}] (3) at (-3,-2.5) {};
\node[flavor,label=left:{$4$}] (12) at (-5,-1.5) {};
\node[flavor,label=left:{$1$}] (13) at (-3,-1.5) {};
\draw[blue] (1)--(2)--(3);
\draw[red] (2)--(12);
\draw[red] (3)--(13);
\end{tikzpicture}}
\newcommand{\branesystemSugra}{
\begin{tikzpicture}[xscale=1.2]
\draw [NS5](0,-2)--(0,2);
\draw [NS5](.1,-2)--(.1,2);
\draw [NS5](.2,-2)--(.2,2);
\draw [NS5](.3,-2)--(.3,2);
\draw [D5h](-1-.5,-1)--(-1+.5,1);
\draw [D5h](-1.1-.5,-1)--(-1.1+.5,1);
\draw [D5h](-1.2-.5,-1)--(-1.2+.5,1);
\draw [D5h](+1-.5,-1)--(+1+.5,1);
\draw [NS5](2,-2)--(2,2);
\draw[D3] (0,0)--(-1,0);
\draw[D3] (.3,0)--(2,0);
\node at (-.5,.2) {6};
\node at (.7,.2) {2};
\node at (1.6,.2) {1};
\node at (-1.6,-1.5) {$\ell_1^{(2,3,4)}=-\frac{1}{2}$};
\node at (-1.6,-2) {$\gamma_1^{(2,3,4)}=3$};
\node at (1.1,-1.5) {$\ell_1^{(1)}=+\frac{1}{2}$};
\node at (1.1,-2) {$\gamma_1^{(1)}=1$};
\node at (0.15,3) {$\ell_2^{(1,2,3,4)}=0$};
\node at (0.15,2.4) {$\gamma_2^{(1,2,3,4)}=4$};
\node at (2,3) {$\ell_2^{(5)}=1$};
\node at (2,2.4) {$\gamma_2^{(5)}=1$};
\end{tikzpicture}}
\begin{tikzpicture}
\node at (0,13) {\branesystempartitions};
\node at (-5,7) {\branesystemHiggs};
\node at (5,7) {\branesystemCoulomb};
\node at (-4,-1) {\branesystemSugra};
\node at (4,-1) {\branesystempelc};
\draw[->] (-1,11.5)--(-3,10);
\draw[->] (1,11.5)--(3,10);
\node at (-5,11) {\begin{tabular}{c}
Electric frame \\ Electric quiver
\end{tabular}};
\node at (5,11) {\begin{tabular}{c}
Magnetic frame \\ Magnetic quiver
\end{tabular}};
\node at (-4,2.5) {Non-back-reacted frame};
\node at (4.5,2.5) {Partially-back-reacted frame};
\draw[->] (-5,4)--(-4.5,3);
\draw[->] (5,4)--(-1.5,2.5);
\draw[->] (-1,-1)--(1,-1);
\end{tikzpicture}
    \caption{Brane systems and quivers for the 3-dimensional $\mathcal{N}=4$ theory $T^{\sigma = [3,2,2,2]}_{\rho = [2,2,2,2,1]} [SU(9)]$. We depict the D5 branes in blue, the NS5 branes in red, the D3 branes in black. The linking numbers for D5 / NS5 branes are also indicated in blue / red. The black numbers denote the multiplicity of the D3 branes. The top panel shows the partition frame, from which one can reach the Higgs/electric (respectively Coulomb/magnetic frame) by Hanany-Witten moves in such a way that no D3 ends on a D5 brane (respectively an NS5 brane). The bottom left panel shows the stacks of branes ordered by their linking number. This is the frame which reflects the supergravity description. The bottom right panel shows the same system with the asymptotic back-reaction taken into account. Whenever we quote numerical values for the $\gamma$'s we set $\alpha' = 4$, so that the $\gamma$’s take integer values.}
    \label{fig:summaryBranes}
\end{figure}
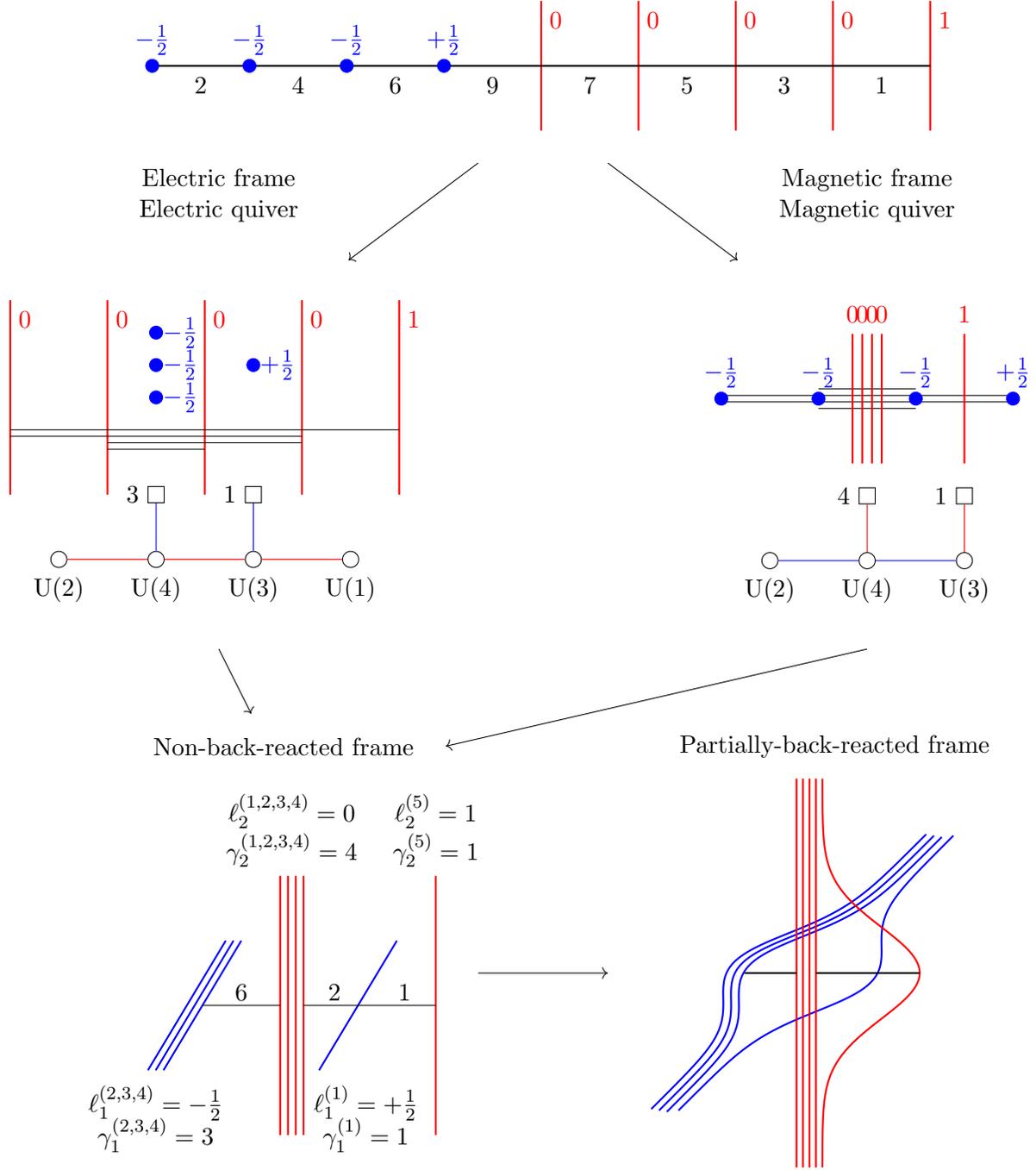

\paragraph{Brane system. } 
Linear unitary quiver gauge theories can be realized as the low-energy limit of non-back-reacted  D3-D5-NS5 Hanany-Witten brane systems, in which D3 branes stretch between NS5 branes, and do not end on D5 branes. In this frame, there are $N_i$ D3 branes stretching between the $i$-th and $(i+1)$-th NS5 branes, and in addition there are $M_i$ D5 branes between these same NS5 branes. The Hanany-Witten moves consist of permutation of D5 and NS5 branes with D3 brane creation in such a way that the linking numbers, defined in equation \eqref{eq:symmetricGaugeLinkingNumbers}, are preserved. For the $T^{\sigma}_{\rho} [SU(N)]$, the linking numbers \eqref{eq:symmetricGaugeLinkingNumbers} can be expressed as  
\begin{equation}
    \ell_1^{(i)} = \sigma_i - \frac{1}{2} |\rho| \,  , \qquad \ell_2^{(i)} = - \rho_i + \frac{1}{2} |\sigma| \, . 
\label{LinkingPartition}
\end{equation}

In order to make the discussion more concrete, and illustrate the discussion with  pictures,  from now on until the end of Section \ref{sec:3dN=4}, we focus on a particular example, the $T^{\sigma = [3,2,2,2]}_{\rho = [2,2,2,2,1]} [SU(9)]$ theory. The generic theory is an obvious generalization. 
Using Hanany-Witten transitions, one can obtain several equivalent brane systems, as shown in Fig. \ref{fig:summaryBranes}. 
\begin{itemize}
    \item The simplest starting point is the \emph{partition frame}, in which the partition, $\sigma$, indicates the net number of D3 branes ending on each D5 brane, and similarly with partition $\rho$ for D3 branes ending on NS5 branes. This is depicted at the top of Fig. \ref{fig:summaryBranes}. Note that the partitions are read from the center towards each end of the brane system. 
    \item From there, one can perform Hanany-Witten moves in such a way that \emph{no D3 brane ends on a D5 brane}. This is the \emph{electric phase}, shown on the middle left of Fig. \ref{fig:summaryBranes}. The D3 brane can break into segments moving along the $v$ direction, parametrizing the Higgs branch. The D5 branes can also move in the $v$ direction, corresponding to mass deformations. One can read the quiver gauge theory (and the quiver is usually called the \emph{electric quiver}). 
    \item Similarly, one can go into a frame where \emph{no D3 brane ends on an NS5 brane}. This is the \emph{magnetic phase}, shown in the middle right. The D3 branes and NS5 branes can move along the $u$ direction. The associated quiver is the 3-dimensional mirror of the previous one, and is usually called the \emph{magnetic quiver}. Explicitly, one finds that the 3-dimensional mirror of $T^{\sigma}_{\rho} [SU(N)]$ is $T_{\sigma}^{\rho} [SU(N)]$. Note that the inequality \eqref{eq:orderPartition} is invariant under swapping $\sigma$ and $\rho$. 
    \item Finally, the most relevant frame for our purposes is the \emph{non-back-reacted frame} in which all the five-branes are arranged by order of increasing linking number, as shown in the bottom part of Fig. \ref{fig:summaryBranes}. In this frame the $z$ coordinate of each stack of five-branes is an increasing function of their corresponding $\delta$. 
\end{itemize}

\begin{figure}
    \centering
\hspace*{-0.3cm}\begin{tabular}{cc}
\includegraphics[scale=.45]{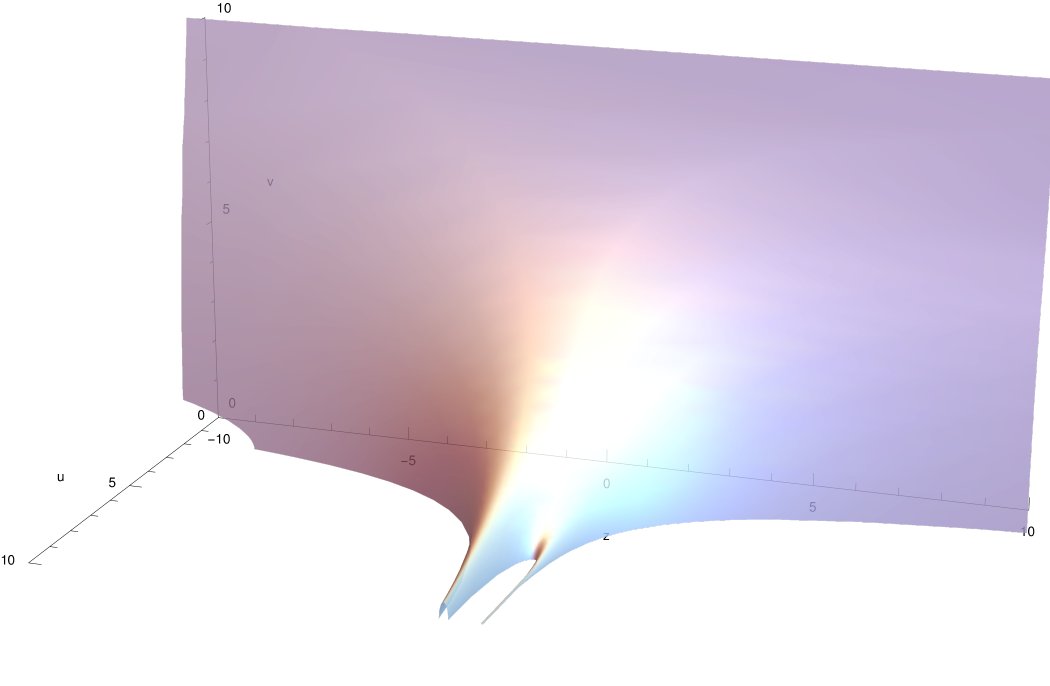} & \includegraphics[scale=.45]{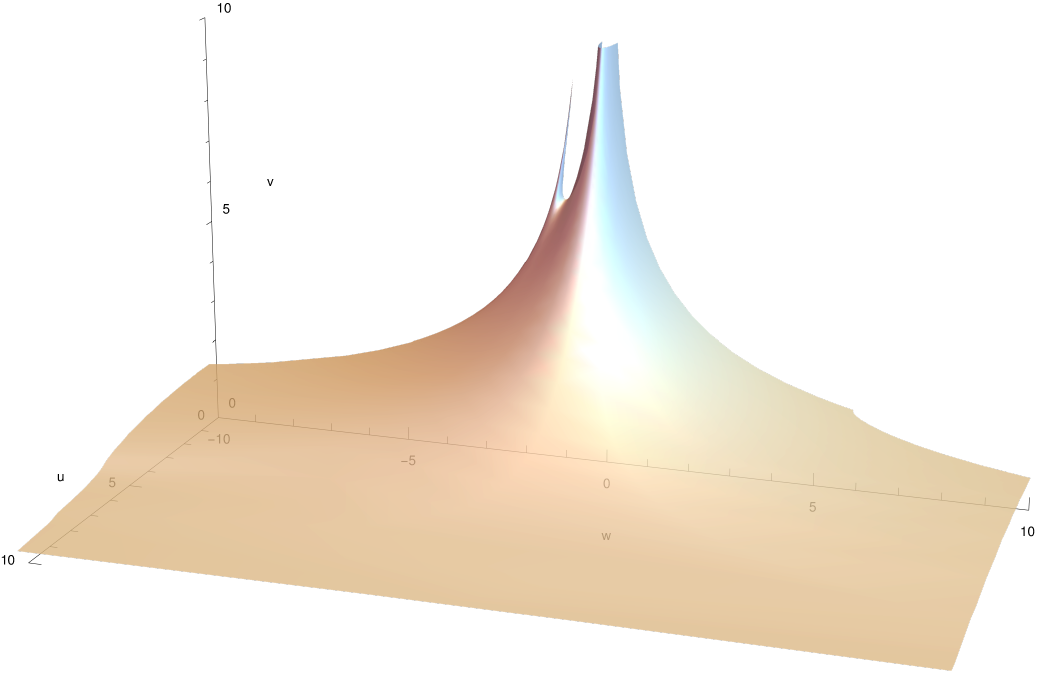}
\end{tabular}
    \caption{Image of the strip $\{(x,y) \in \mathbb{R} \times (0 , \frac{\pi}{2}) \}$ under the map \eqref{uvzmap1} at fixed $\mu$, for the $T^{\sigma = [3,2,2,2]}_{\rho = [2,2,2,2,1]} [SU(9)]$ example. The left panel shows the $(u,v,z)$ image ($u$ comes out of the paper, $z$ horizontally, $v$ vertically) and the right panel shows the $(u,v,w)$ image ($u$ comes out of the paper, $w$ horizontally, $v$ vertically). The spikes correspond to the asymptotic, scaling regions of the stacks of five-branes, which appear in the partially back-reacted frame at the bottom of Fig. \ref{fig:summaryBranes}: In the large-$u$ region of the left panel we can see the two D5 scaling regions, and in the large-$v$ of the right panel we can see the two NS5 scaling regions. Because of the near-brane limit, both these scaling regions appear as infinite spikes.}
    \label{fig:StripImage}
\end{figure}

\begin{figure}[t]
    \centering
    \begin{tabular}{cc}
        $\mu$ large & \raisebox{-.5\height}{\includegraphics[scale=.4]{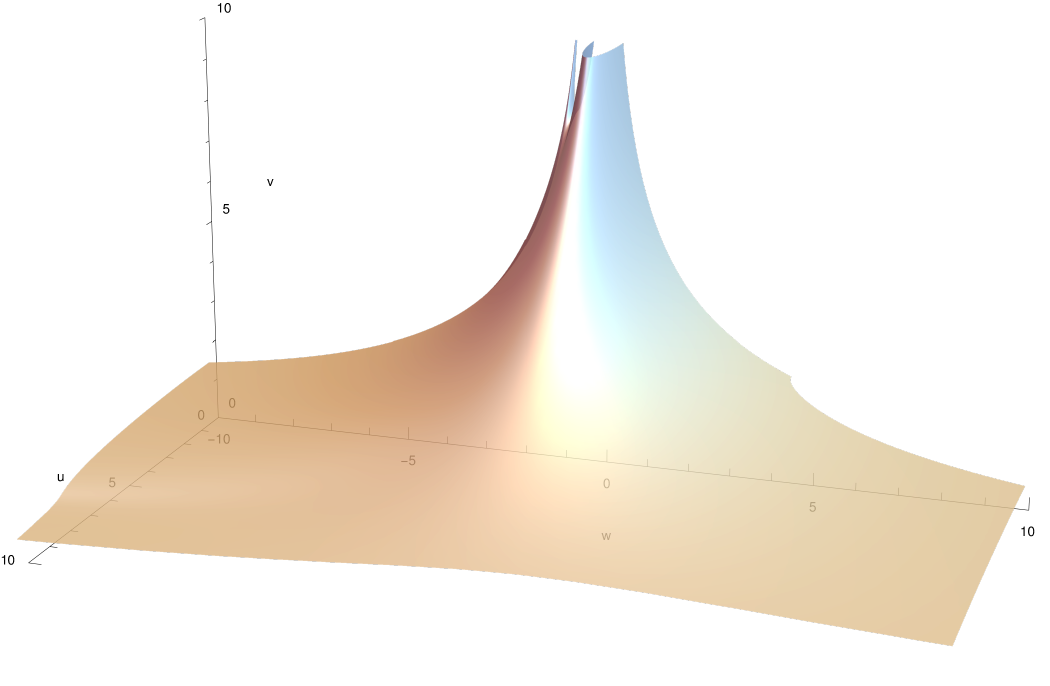}} \\
        $\mu$ medium & \raisebox{-.5\height}{\includegraphics[scale=.4]{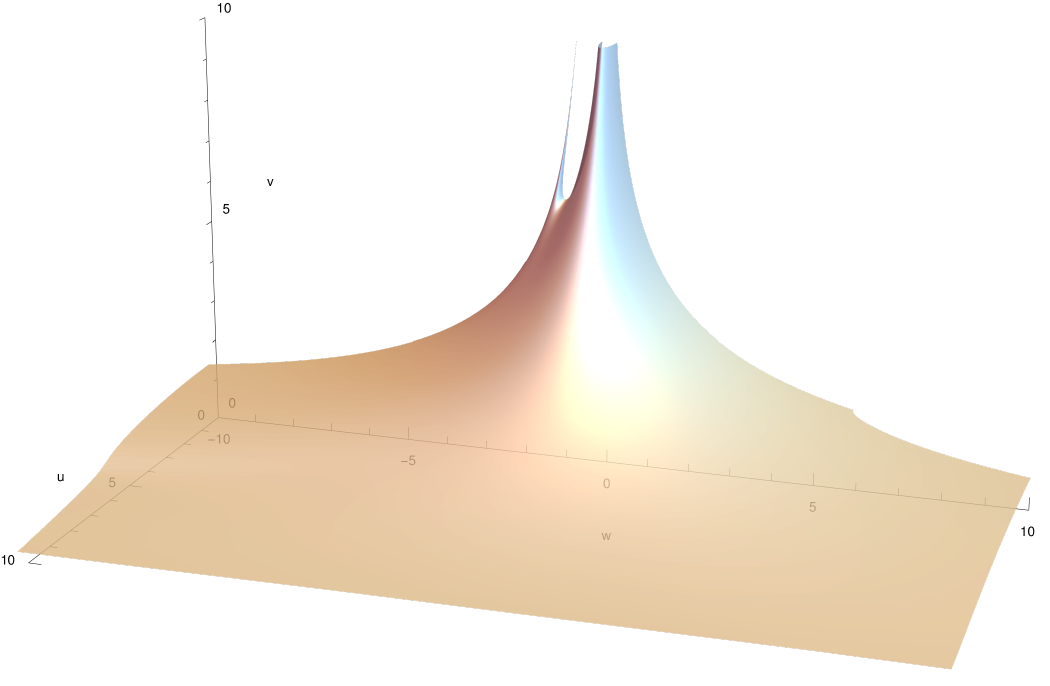}} \\
        $\mu$ small  & \raisebox{-.5\height}{\includegraphics[scale=.4]{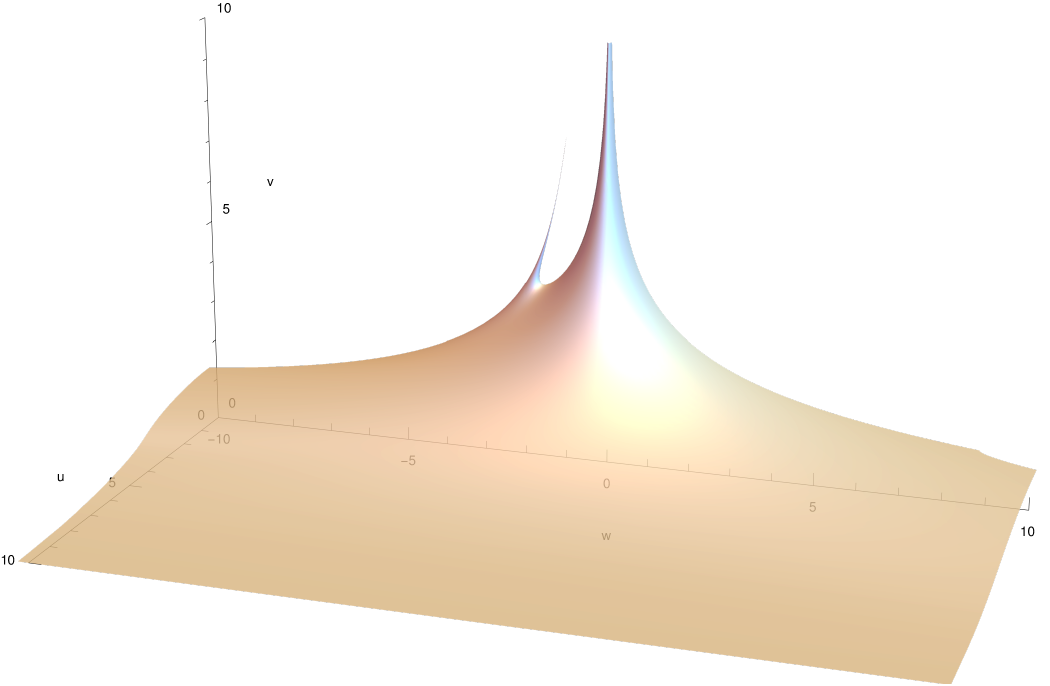}} 
    \end{tabular}
    \caption{Impact of varying $\mu$ in the plots shown in Fig. \ref{fig:StripImage}. We look only at the $(u,v,w)$ plots. When $\mu$ gets small, the location of the branes becomes sharper and sharper, reproducing the asymptotic region at the bottom of Fig. \ref{fig:summaryBranes}. }
    \label{fig:StripImageMu}
\end{figure}

\begin{figure}
    \centering
    \includegraphics[width=0.75\linewidth]{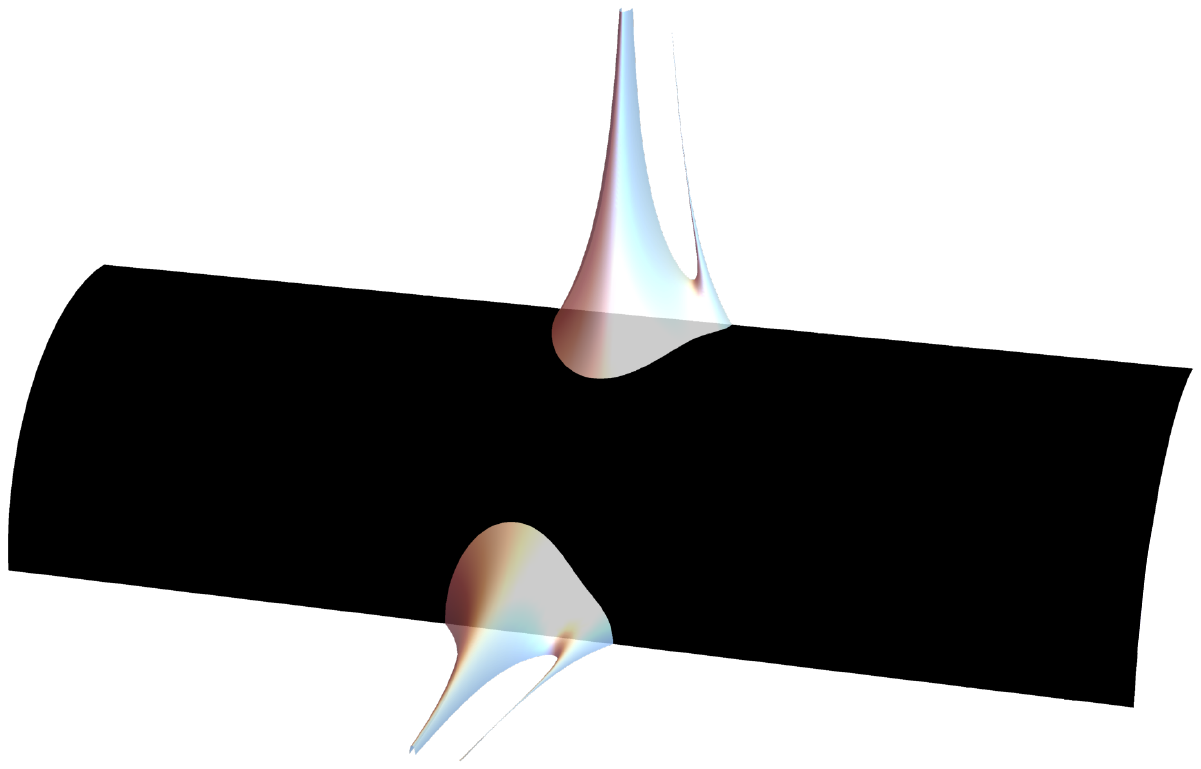}
    \caption{Combined plot showing the two charts $(u,v,z)$ and $(u,v,w)$. The horizontal axis is $z$ for the bottom part, and $-w$ for the top part. The black cylinder hides schematically the part of spacetime which is not described by the near-brane scaling limit. This is only schematic, as in principle the cylinder should hide almost all of the plot, leaving only the asymptotic spikes visible.}
    \label{fig:StripImageCylinder}
\end{figure}

\subsection{Comments on Bad and Ugly theories}

In this subsection, we explain why it can be assumed that the theories that are holographically dual to AdS$_4\times S^2\times S^2$  are only good theories, by showing the geometric underpinning of the good / bad criterion.  

Consider for simplicity a three-dimensional $\mathcal{N}=4$ theory with gauge group $U(N_c)$ and $N_f$ fundamental hypermultiplets. 
As mentioned above, the inequality \eqref{eq:condition_good} characterizes \emph{good} theories; when it is not satisfied, the theory can be \emph{ugly} ($N_f = 2 N_c -1$)  or \emph{bad} ($N_f \leq 2 N_c -2$), following the terminology of \cite{Gaiotto:2008ak}. The theories with $N_f < 2 N_c$ have been studied in detail from the point of view of the moduli space of supersymmetric vacua (MSV) in \cite{Assel:2017jgo,Bourget:2021jwo,Bourget:2023cgs}. We insist that bad and ugly theories are perfectly well-defined quantum field theories. 
The qualifiers only reflect how easy it is to track the IR properties from the UV gauge-theoretic description. This also translates into geometrical properties of the moduli space \cite[A.1]{Bourget:2023cgs}: good theories have a Coulomb branch that is a conical symplectic singularity, with a maximally singular locus consisting of a point (fixed by the canonical $\mathbb{C}^{\ast}$-action). In contrast, ugly and bad theories have a most singular locus which has dimension $>0$.

In terms of $T^{\sigma}_{\rho} [SU(N)]$ theories, we have $N = N_f$ and the partitions are $\sigma = [1^{N_f}]$, $\rho = [N_f - N_c , N_c]$\cite{Yaakov:2013fza,Assel:2017jgo}. Hence the fact that the theory is good is reflected in the correct non-increasing ordering of the entries of the partition $\rho$: $N_f - N_c \geq N_c$. This is a general feature, and it shows that the duals of the solutions of Section \ref{sec:Near brane intersection} are good by construction if we label the singular points $\delta_1^{(i)}$ and $\delta_2^{(i)}$ monotonically. Geometrically, as discussed in Section \ref{ss:mohawk1}, this corresponds to spikes which are nested inside each other, forming a mohawk.

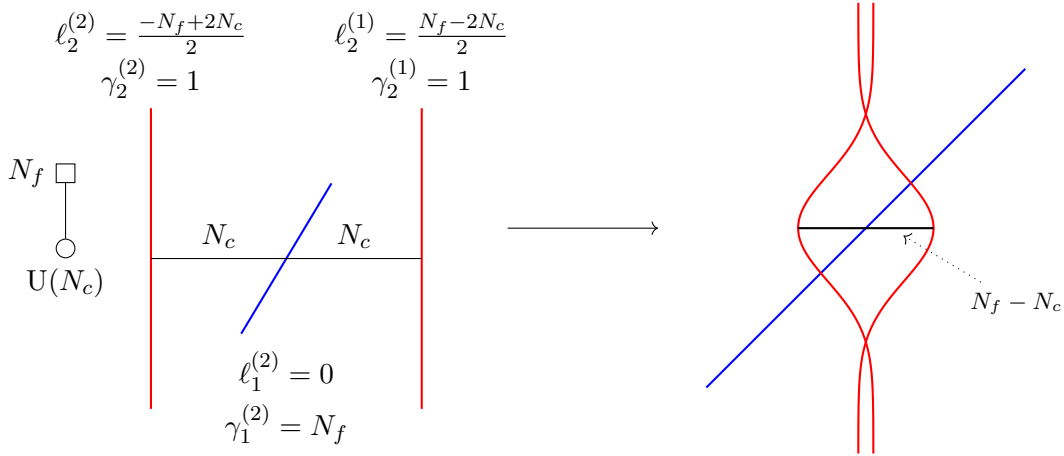
\begin{figure}
    \centering
 \newcommand{\branesystempelcx}{\begin{tikzpicture}
\node at (0,0) {\curvebellD{1}{0}};
\node at (.4,0) {\curvebellNS{1}{-1}};
\node at (-.4,0) {\curvebellNS{1}{1}};
\draw[thick] (.9,0)--(-.9,0);
\node (1) at (2,-1) {\footnotesize $N_f-N_c$};
\draw [->,dotted] (1)--(.5,-.1);
\end{tikzpicture}}
\newcommand{\branesystemSugrax}{
\begin{tikzpicture}[xscale=1.2]
\draw [NS5](0,-2)--(0,2);
\draw [NS5](3,-2)--(3,2);
\draw [D5h](1.5-.5,-1)--(1.5+.5,1);
\draw[D3] (0,0)--(3,0);
\node at (1.5,-1.5) {$\ell_1^{(2)}=0$};
\node at (1.5,-2.2) {$\gamma_1^{(2)}=N_f$};
\node at (0,3) {$\ell_2^{(2)}=\frac{-N_f+2N_c}{2}$};
\node at (0,2.4) {$\gamma_2^{(2)}=1$};
\node at (3,3) {$\ell_2^{(1)}=\frac{N_f-2N_c}{2}$};
\node at (3,2.4) {$\gamma_2^{(1)}=1$};
\node at (.75,.3) {$N_c$};
\node at (2.25,.3) {$N_c$};
\end{tikzpicture}}
\begin{tikzpicture}
\node at (-7,-1) {\begin{tikzpicture}[xscale=0.75]
    \node[gauge,label=below:{$\mathrm{U}(N_c)$}] (1) at (0,0) {};
    \node[flavor,label=left:{$N_f$}] (3) at (0,1) {};
    \draw (1)--(3);
    \end{tikzpicture}};
\node at (-4,-1) {\branesystemSugrax};
\node at (4,-1) {\branesystempelcx};
\draw[->] (-1,-1)--(1,-1);
\end{tikzpicture}
    \caption{On the left, the non-back-reacted brane diagram corresponding to the depicted gauge theory. When $N_f - 2 N_c < 0$, the branes are not arranged in increasing linking number from left to right. Thus the back-reaction leads to crossing, depicted on the right. The remaining number of D3 branes, $N_f - N_c$ (assuming this number is nonnegative) is obtained from the linking numbers.  }
    \label{fig:branesBadtheory}
\end{figure}

If we insist on labeling the poles in the wrong order, one can see heuristically that the naive back-reaction would ultimately lead to branes crossing each other. This crossing corresponds to Seiberg duality \cite{Seiberg:1994pq, Giveon:1998ns,Yaakov:2013fza,Assel:2017jgo}.
For instance, considering the example above with $N_c < N_f < 2 N_c$ and using the conventions of Fig. \ref{fig:summaryBranes}, we get the brane configurations shown in Fig. \ref{fig:branesBadtheory}. In supergravity such a crossing solution does not appear to exist, which suggests that the field theory must explore some other phases.

In the example above, this allows us to identify the (Seiberg-dual) theory which is holographically dual to the supergravity solution. It is the $U(N_f-N_c)$ gauge theory with $N_f$ fundamentals, and this is indeed what would have been obtained by labeling the partition $\rho$ in the correct non-increasing order $[N_c,N_f-N_c]$. Note that for this to work, we need to assume $N_c < N_f$. This makes sense from the $T^{\sigma}_{\rho} [SU(N)]$ perspective as the entries of the partitions should be positive. In our heuristic brane diagram on the right of Fig. \ref{fig:branesBadtheory}, this condition is also required: the NS5 branes cannot cross if they lie in the same plane, so the crossing requires moving them slightly in the transverse direction, which corresponds in the gauge theory to turning on Fayet-Iliopoulos parameters. However this is not possible without breaking supersymmetry for these theories (see \cite[App. B]{Bourget:2021jwo}), so everything is consistent. 

\subsection{Four-dimennsional $\mathcal{N}=4$ SYM with a boundary}
 
We now describe the middle panel, $(b)$ of Fig. \ref{fig:N=4boundaries}. Consider first theories with $M=0$, for which there is a natural construction whereby the flavor symmetry of a theory $T_{\rho} [SU(N)]$ as described above is gauged in a 4-dimensional bulk. In terms of branes, this corresponds to pulling all the D5 branes to infinity, yielding for instance the configuration (here $N=9$ and $\rho=[2,2,2,2,1]$ as before)
\begin{equation}
    \raisebox{-.5\height}{\begin{tikzpicture}[xscale=.75]
\draw [NS5](0,-1)--(0,1);
\draw [NS5](2,-1)--(2,1);
\draw [NS5](4,-1)--(4,1);
\draw [NS5](6,-1)--(6,1);
\draw [NS5](8,-1)--(8,1);
\draw[D3] (-6,0)--(8,0);
\draw[D3] (-6,-.1)--(6,-.1);
\draw[D3] (-6,-.2)--(6,-.2);
\draw[D3] (-6,-.3)--(4,-.3);
\draw[D3] (-6,-.4)--(4,-.4);
\draw[D3] (-6,-.5)--(2,-.5);
\draw[D3] (-6,-.6)--(2,-.6);
\draw[D3] (-6,-.7)--(0,-.7);
\draw[D3] (-6,-.8)--(0,-.8);
\node at (0.3,0.7) {\textcolor{red} 2};
\node at (2.3,0.7) {\textcolor{red} 2};
\node at (4.3,0.7) {\textcolor{red} 2};
\node at (6.3,0.7) {\textcolor{red} 2};
\node at (8.3,0.7) {\textcolor{red} 1};
\end{tikzpicture}}
\label{eq:braneSystemTrho4d}
\end{equation}
This corresponds to 4-dimensional $\mathcal{N}=4$ $U(N)$ theory coupled to a 3-dimensional $\mathcal{N}=4$ SCFT with $U(N)$ flavor symmetry living on its boundary. It is also possible to consider a more general boundary condition, as explored in \cite{Gaiotto:2008sa}. 
Recall that the bosonic field content of 4-dimensional $\mathcal{N}=4$ SYM is a gauge field, $A_{\mu}$, and six scalars which can be decomposed into two triplets, $\vec{u}$ and $\vec{v}$, which coincide with the coordinates introduced in Table \ref{tab:conventionsBraneSystems}. The boundary conditions illustrated in \eqref{eq:braneSystemTrho4d} correspond in field theory to Neumann boundary conditions for the 3-dimensional vector multiplets. It is also possible to impose Dirichlet conditions to the 3-dimensional vector multiplets. This is interpreted in the brane systems as ending the D3 branes on D5 branes, and in the field theory as $\vec{u}$ having a Nahm pole when approaching the boundary (while $\vec{v}$ remains constant up to gauge transformation). 

Combining the two constructions, we obtain a 4-dimensional $\mathcal{N}=4$ SYM theory with gauge group $U(N+M)$, with a Nahm pole reducing the gauge group $U(N+M) \rightarrow U(N)$, specified by a partition $\lambda$ of $M$, and then the coupling to a 3-dimensional theory $T_{\rho} [SU(N)]$. In terms of branes, the picture is 
\begin{equation}
    \raisebox{-.5\height}{\begin{tikzpicture}[xscale=.75]
\draw [NS5](6,-1)--(6,1);
\draw [NS5](7,-1)--(7,1);
\draw [NS5](9,-1)--(9,1);
\draw [NS5](10,-1)--(10,1);
\draw[line width=.10cm] (-1,0) -- (1,0);
\draw[line width=.09cm] (1,0) -- (2,0);
\draw[line width=.08cm] (2,0) -- (2.5,0);
\draw[line width=.07cm] (3.5,0) -- (4,0);
\draw[line width=.05cm] (4,0) -- (6,0);
\draw[line width=.04cm] (6,0) -- (7,0);
\draw[line width=.03cm] (7,0) -- (7.5,0);
\draw[line width=.02cm] (8.5,0) -- (9,0);
\draw[line width=.01cm] (9,0) -- (10,0);
\draw[dotted] (7.5,0)--(8.5,0);
\draw[dotted] (2.5,0)--(3.5,0);
\node[D5] at (1,0) {};
\node[D5] at (2,0) {};
\node[D5] at (4,0) {};
\node at (2.5,.5) {\color{blue} $\lambda$ Nahm pole};
\node at (8,-1.5) {\color{red} $T_{\rho}[SU(N)]$};
\node at (-1,-0.5) {\color{black} $N+M$ D3};
\node at (5,-0.5) {\color{black} $N$ D3};
\end{tikzpicture}}
\label{eq:braneSystemTrho4dNahm}
\end{equation} 
The linking number for the $i$-th D5 brane (counting from the left) is $\ell^{(i)}_1 = - \lambda_i - \frac{1}{2} |\rho| < 0$. 

Finally, one can consider the most general configuration, where the 4-dimensional theory of the picture above is coupled to an arbitrary good 3-dimensional  $T_{\rho}^{\sigma}[SU(N)]$ theory. Although we will not focus on this configuration in the rest of the paper, it is worth noting that these configurations are also captured by the supergravity solutions of Section \ref{sec:Near brane intersection}. Indeed, for any choice of gauge group $U(N+M)$, Nahm pole data, and partitions $\rho$ and $\sigma$, one can verify this directly by solving the system of equations \eqref{LinkingCharges}.

\begin{figure}
    \centering
\hspace*{-1.5cm}\begin{tabular}{cc}
\includegraphics[scale=.5]{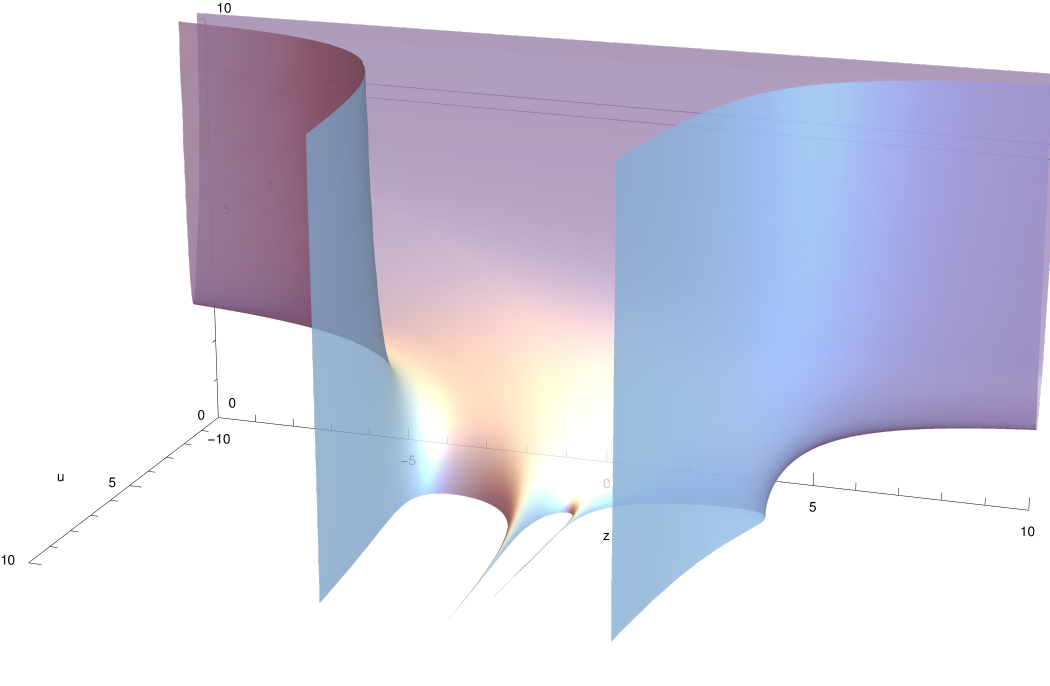} & \includegraphics[scale=.5]{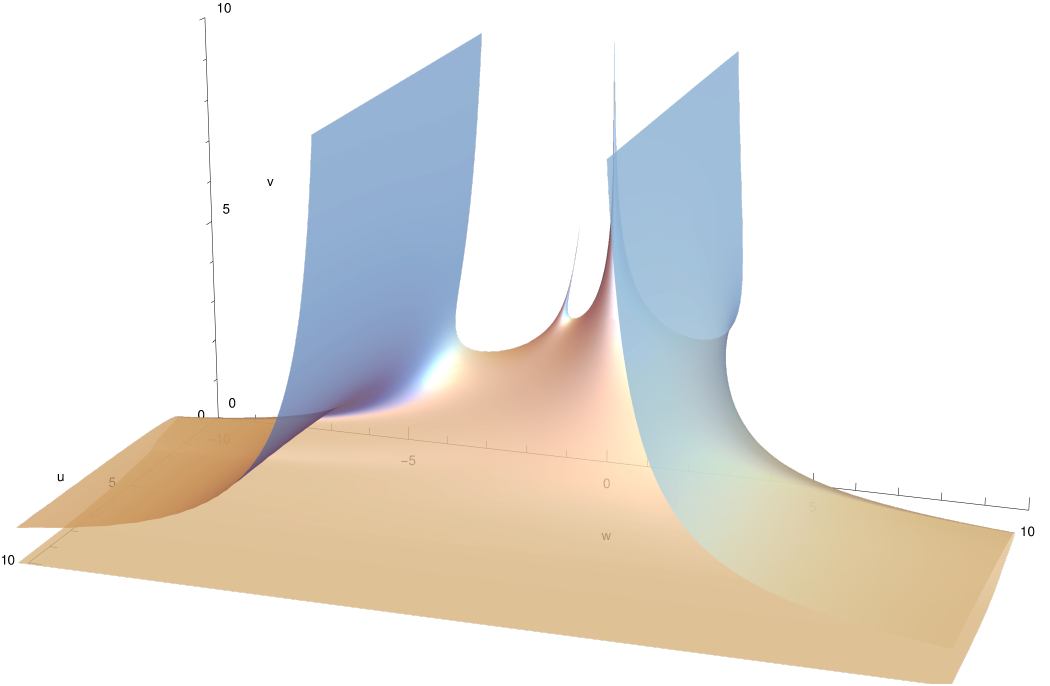}
\end{tabular}
    \caption{ Same as Fig. \ref{fig:StripImage} but with $\alpha_{1,2} \neq 0$. The horizontal plane at the bottom is the $v=0$ plane; the vertical plane at the back is the $u=0$ plane.  }
    \label{fig:StripImageALPHA}
\end{figure}

\section{The partially-back-reacted picture}

\label{regime-2}

The second regime in which we can analyze our system is when some of the branes back-react and some are treated as probes. Since the NS5 branes can never be treated as probes (there is no regime of parameters where their back-reaction can be ignored), we will take the NS5 branes to be fully back-reacted and consider D5 brane probes with possible D3-brane world-volume charges.

In Appendix \ref{appendix:Probe bulges} we demonstrate how different, and simpler, incarnations of the same essential physics emerge from different species of branes.

\subsection{Probe D5-brane in an NS5 background: induced D3 flux and the bulge}
\label{sec:Probe D5 in NS5 background}
Following the conventions of Section \ref{sec:General Intersection}, we take the NS5-brane world-volume coordinates to be $(x_1, x_2, \vec{v})$. The space transverse to the NS5-brane is parametrized by cylindrical coordinates $(\vec{u}, z)$, and we express $\vec{u}$ in spherically symmetric coordinates to fit with the solutions used in this paper. The directions $(x_1, x_2)$ are shared by all branes. The geometry for a stack of $n_5$ coincident NS5-branes located at $u = z = 0$ is:  
\begin{align}
    \d s_{10}^2&=- \d t^2+ \d x_1^2+ \d x_2^2+ \d v^2+v^2 \d s_{S_2^2}^2+ H_5\left(\d u^2+u^2 \d s_{S_1^2}^2+ \d z^2\right) \,, \\
    H_5&=1+\frac{n_5\alpha'}{u^2+z^2} \,,\\
    e^{2\phi}&=g_s^2 H_5 \,, \\
    B_2&=n_5\alpha'\left(\frac{-u z}{u^2+z^2}+\left(\arctan\left(\frac{u}{z}\right)-\psi_1\right)\right) \text{Vol}(S_1^2) \equiv B\, \text{Vol}(S_1^2)\,,
\end{align}
where $\psi_1$ is an integration constant. This parameter has no direct physical interpretation, it corresponds to a gauge choice. Much like in \cite{Pelc:2000kb}, we now consider a probe D5-brane in static gauge, with asymptotic world-volume coordinates chosen to be $(t, x_1, x_2, \vec{u})$. At finite $u$, the shape of the D5-brane is specified by a function $z(u)$. The gauge-invariant field strength on the D5-brane is given by $\mathcal{F} = B_2 + F$, where $F$ denotes the field strength of the world-volume $U(1)$ gauge field. In what follows we use $\mathcal{F} = B_2$ up to a meaningful (gauge invariant) shift of the constant $\psi_1$. Using this expression, one can write down the $\kappa$-symmetry projection condition \cite{Simon:2011rw,Bena:2024oeq} and obtain\footnote{We use flat indices here so that all $\Gamma$ matrices square to one, for example $(\Gamma^{\theta})^2=1$.}:
\begin{equation}
    \label{eq:kappa D3 NS5}
    \frac{\left(u^2H_5\,\Gamma^{012u\theta\phi}\sigma_1+B\Gamma^{012u}i\sigma_2+u^2H_5 \partial_uz\Gamma^{0 12z\theta\phi}\sigma_1+ \partial_uz B \Gamma^{012z}i\sigma_2\right)}{\sqrt{\left(u^4H_5^2+B^2\right)\!\left(1+(\partial_uz)^2\right)}}~\varepsilon=\varepsilon\,.
\end{equation}
The globally preserved supersymmetries should coincide with those of a D5-brane extending along $(x_1, x_2, \vec{u})$, while also satisfying the projection condition imposed by the background NS5-branes. As shown in \cite{Hanany:1996ie}, D3-branes extended along $(x_1, x_2, z)$ preserve the same set of supersymmetries. We impose these supersymmetry conditions in order to solve the $\kappa$-symmetry projection equations:
\begin{align}
    &\Gamma^{012u\theta\phi}\sigma_1\varepsilon=\varepsilon \,,\\
    &\Gamma^{012z}i\sigma_2\varepsilon=-\varepsilon\,.
\end{align}
From these one obtains the following condition for preserved supersymmetry:
\begin{equation}
\label{eq:differentialEqB2}
    \partial_uz=-\frac{B}{u^2 H_5}\,,
\end{equation}
which has the implicit solution:
\begin{equation}
    z-\frac{n_5\alpha'}{u}\left(\arctan(\frac{u}{z})-\psi_0\right)=z_{\infty}\,.
    \label{z_inf}
\end{equation}
We define $z_{\infty}$ as the asymptotic value of $z$ in the limit $u \rightarrow \infty$, and $\psi_0$ as a constant that determines the induced flux on the D5-brane. The fully back-reacted AdS$_4$ geometries described in Section \ref{sec:Near brane intersection} correspond to the infrared limit of the brane intersection, and our discussion will focus on this regime. As we will explain, these geometries come from regions where the branes exhibit a scaling symmetry, which is easiest to illustrate when $z_{\infty} = 0$, but whose existence and properties are independent of the value of $z_{\infty}$. We will therefore set for now $z_{\infty} = 0$ and obtain the D3-D5 brane shape:
\begin{equation}
\label{eq:probe D5 in NS5 background}
    z~u=n_5 \alpha'\left( \arctan\left(\frac{u}{z}\right)-\psi_0\right)\,.
\end{equation}
Recall that by definition $u>0$ while $z \in \mathbb{R}$.

\begin{figure}
    \centering
\begin{tabular}{cc}
   \includegraphics[scale=0.8]{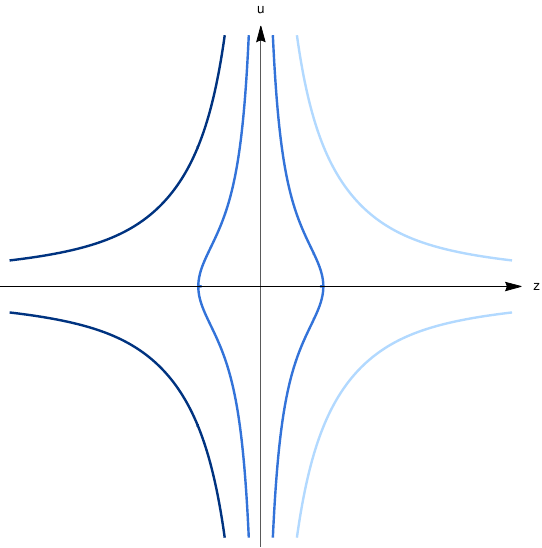}  & \includegraphics[scale=0.8]{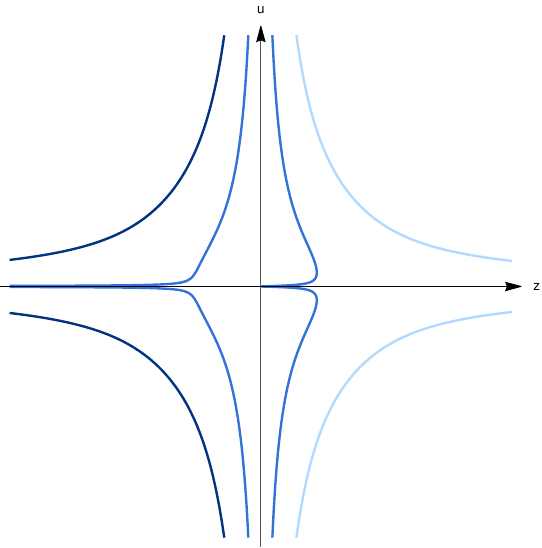} \\
    $\widetilde\psi_0=0$  &  $\widetilde\psi_0=0.02 \pi $ \\ & \\ 
   \includegraphics[scale=0.8]{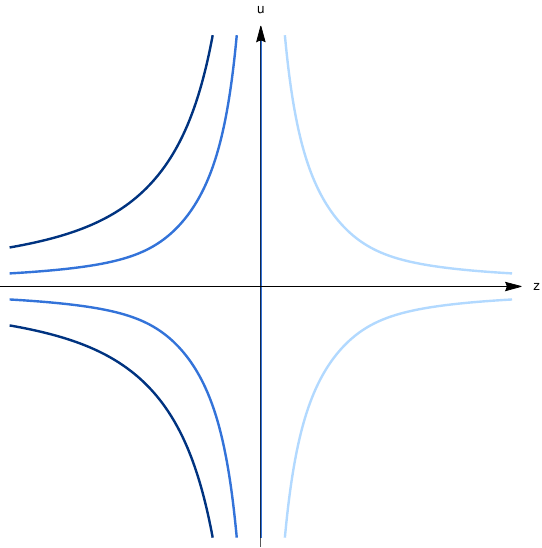}  & \includegraphics[scale=0.8]{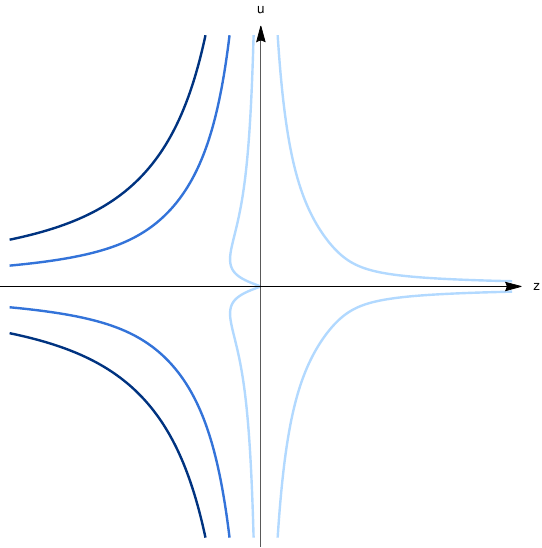} \\
    $\widetilde\psi_0=\frac{\pi}{2}$  &  $\widetilde\psi_0=0.8 \pi$ \\
\end{tabular}
    \caption{Plots of equation \eqref{eq:probe D5 in NS5 background} with $\psi_0 = \widetilde\psi_0 + k \pi$ showing the shape of a D5 brane with D3 world-volume density in an NS5 background, for various values of $\widetilde\psi_0$ and $k = -1, 0, +1$ (light, normal and dark blue respectively).  }
    \label{fig:pelc}
\end{figure}

We review how the brane shapes given by \eqref{eq:probe D5 in NS5 background} change as functions of $\psi_0$ in Fig. \ref{fig:pelc}. There are several families of solutions: 
\begin{itemize}
    \item \underline{ $\psi_0 \leq - \frac{\pi}{2}$}. There is a unique solution, $z_+(u)$, with $z_+(u) >0$. When $\psi_0 = -\frac{\pi}{2}$, we can also find the ``limit'' solution $z_- (u) = 0$.
    \item \underline{$- \frac{\pi}{2} < \psi_0 < \frac{\pi}{2}$}. There are two solutions, $z_{\pm}(u)$, with $z_-(u) < 0 < z_+ (u)$. For $\psi_0 >0$, the solution $z_+ (u)$ is bounded while the solution $z_- (u)$ diverges when $u \rightarrow 0$ as $z_- (u) \rightarrow - \infty$. For  $\psi_0 <0$, the opposite happens: $z_- (u)$ is bounded while $z_+ (u) \rightarrow + \infty$ as $u \rightarrow 0$. For $\psi_0 = 0$, $z_+ (u) = - z_- (u)$ and both solutions are bounded.  This is the only solution where $z_{\pm} (u)$ reaches a finite limit as $u \rightarrow 0$, namely $z_{\pm} (0) = \pm \sqrt{n_5 \alpha '}$. 
    \item \underline{$\psi_0 \geq \frac{\pi}{2}$}. There is a unique solution $z_-(u)$, satisfying $z_-(u) <0$. When $\psi_0 = \frac{\pi}{2}$, we find the ``limit solution" $z_+ (u) = 0$. 
\end{itemize}

In order to get a solution which varies continuously as $\psi_0$ varies, we focus on the solution 
\begin{equation}
    z (u ; \psi_0) = \begin{cases}
        z_+ (u ; \psi_0) & \psi_0 < \frac{\pi}{2} \\  
        0 & \psi_0 = \frac{\pi}{2} \\  
        z_- (u ; \psi_0-\pi) & \psi_0 > \frac{\pi}{2} 
    \end{cases}
    \label{eq:choiceD5}
\end{equation} 

For $0 < \psi_0 < \pi$, the solution is bounded and $\psi_0$ is the angle at which the D5 brane emerges off the NS5 stack. We can give an interpretation of this fact by realizing that this solution comes from the back-reaction of a solution with $N_{D3}$
D3 branes stretched between the NS5 stack on the left and the D5 brane on the right, where
\begin{equation}
    N_{D3} = n_5 \frac{\psi_0}{\pi}\,.
    \label{eq:ND3Pelc}
\end{equation}
The choice of signs made in \eqref{eq:choiceD5} corresponds to choosing the D5 brane that avoids colliding with the NS5 stack through the $z>0$ region, when no D3 branes are stretched between the five-branes ($\psi_0 = 0$). The D3 branes always end on the D5 from the left (negative $z$ -- again this reflects our (arbitrary) choice of signs in \eqref{eq:choiceD5}), but when $N_{D3} > \frac{1}{2} n_5$, the back-reacted D5 brane is actually pulled on the left of the NS5 stack. 

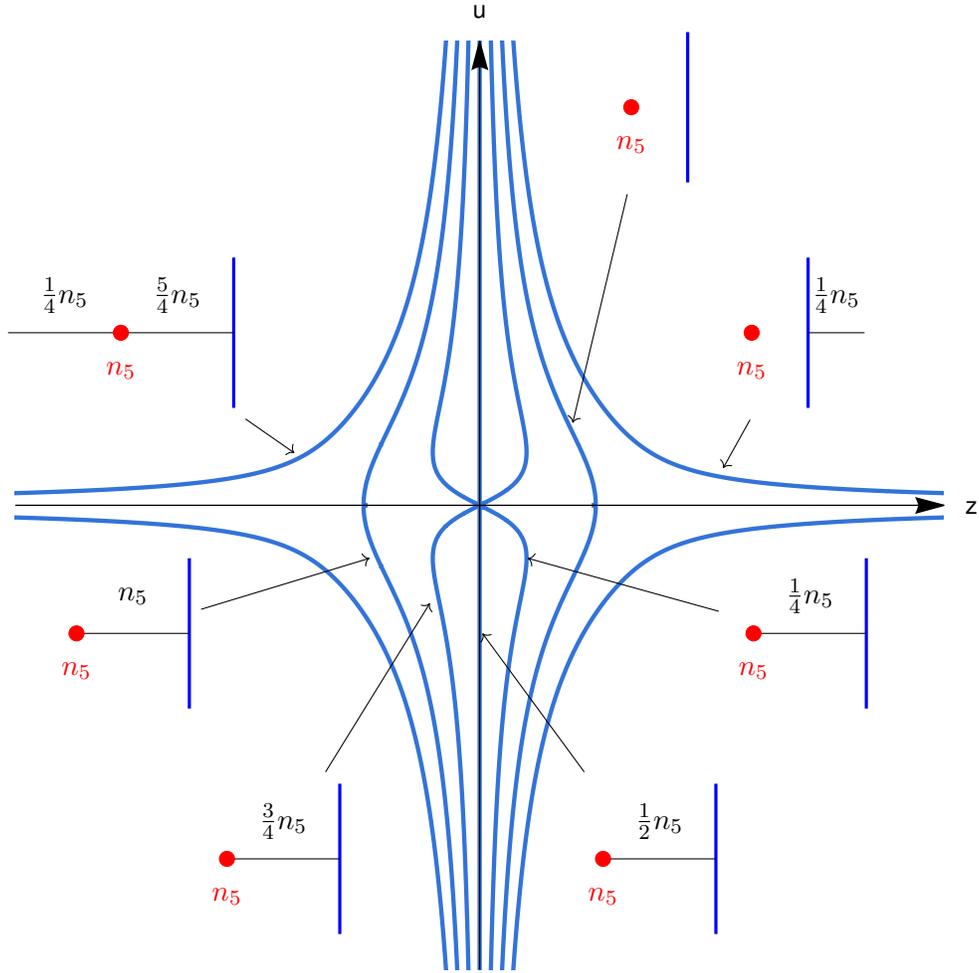
\begin{figure}
    \centering
\begin{tikzpicture}
\node at (0,0){\includegraphics[scale=1.4]{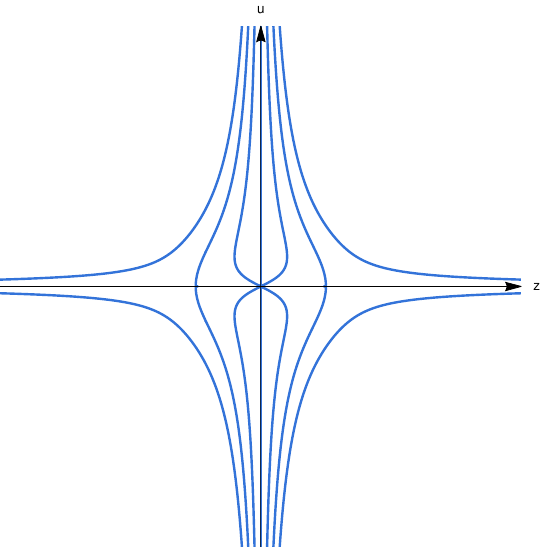}};
\node (a) at (2,5) {\begin{tikzpicture}[xscale=.75]
\draw[blue, very thick] (2,1)--(2,-1);
\node[style={circle, draw=red, inner sep=0pt, fill=red, minimum size=2mm}] at (1,0) {}; 
\node at (1,-.5) {${\Red n_5}$};
\end{tikzpicture}};
\draw[->] (a)--(1,.8);
\node (b) at (4,2) {\begin{tikzpicture}[xscale=.75]
\draw[blue, very thick] (2,1)--(2,-1);
\draw (2,0)--(3,0);
\node[style={circle, draw=red, inner sep=0pt, fill=red, minimum size=2mm}] at (1,0) {}; 
\node at (2.5,.5) {$\frac{1}{4} n_5$};
\node at (1,-.5) {${\Red n_5}$};
\end{tikzpicture}};
\draw[->] (b)--(3,.2);
\node (c) at (4,-2) {\begin{tikzpicture}[xscale=.75]
\draw[blue, very thick] (2,1)--(2,-1);
\draw (0,0)--(2,0);
\node[style={circle, draw=red, inner sep=0pt, fill=red, minimum size=2mm}] at (0,0) {}; 
\node at (1,.5) {$\frac{1}{4} n_5$};
\node at (0,-.5) {${\Red n_5}$};
\end{tikzpicture}};
\draw[->] (c)--(.4,-1);
\node (d) at (2,-5) {\begin{tikzpicture}[xscale=.75]
\draw[blue, very thick] (2,1)--(2,-1);
\draw (0,0)--(2,0);
\node[style={circle, draw=red, inner sep=0pt, fill=red, minimum size=2mm}] at (0,0) {}; 
\node at (1,.5) {$\frac{1}{2} n_5$};
\node at (0,-.5) {${\Red n_5}$};
\end{tikzpicture}};
\draw[->] (d)--(-.2,-2);
\node (e) at (-3,-5) {\begin{tikzpicture}[xscale=.75]
\draw[blue, very thick] (2,1)--(2,-1);
\draw (0,0)--(2,0);
\node[style={circle, draw=red, inner sep=0pt, fill=red, minimum size=2mm}] at (0,0) {}; 
\node at (1,.5) {$\frac{3}{4} n_5$};
\node at (0,-.5) {${\Red n_5}$};
\end{tikzpicture}};
\draw[->] (e)--(-.9,-1.6);
\node (f) at (-5,-2) {\begin{tikzpicture}[xscale=.75]
\draw[blue, very thick] (2,1)--(2,-1);
\draw (0,0)--(2,0);
\node[style={circle, draw=red, inner sep=0pt, fill=red, minimum size=2mm}] at (0,0) {}; 
\node at (1,.5) {$  n_5$};
\node at (0,-.5) {${\Red n_5}$};
\end{tikzpicture}};
\draw[->] (f)--(-1.7,-1);
\node (g) at (-5,2) {\begin{tikzpicture}[xscale=.75]
\draw[blue, very thick] (2,1)--(2,-1);
\draw (-2,0)--(2,0);
\node[style={circle, draw=red, inner sep=0pt, fill=red, minimum size=2mm}] at (0,0) {}; 
\node at (1,.5) {$\frac{5}{4} n_5$};
\node at (-1,.5) {$\frac{1}{4} n_5$};
\node at (0,-.5) {${\Red n_5}$};
\end{tikzpicture}};
\draw[->] (g)--(-2.7,.4);
\end{tikzpicture}
    \caption{The main plot shows the shape of the D5 brane in the background of $n_5$ NS5 branes located at $u=z=0$. This corresponds to the graph of the function $z(u ; \psi_0)$ defined in \eqref{eq:choiceD5} for $\psi_0 = \kappa \frac{\pi}{4}$ with $\kappa = -1,0,1,2,3,4,5$ (we assume here that $n_5$ is a multiple of 4). For each curve we also draw the non-back-reacted brane system, where the red dot denotes the NS5 stack, the blue line is the D5 brane and the black line is a stack of D3 branes. Note that for $\psi_0=\frac{\pi}{2}$, the D5-brane is completely straight, since exactly $\frac{n_5}{2}$ D3 branes are stretched between the five-branes. }
    \label{fig:PelcVariousPsi}
\end{figure}

From \eqref{eq:ND3Pelc} one sees that the angle, $\psi_0$, is quantized. More importantly, the fact that $N_{D3} \leq n_5$ for bounded solutions is a manifestation of the S-rule. Indeed, for $\psi_0$ outside of the $[0  , \pi]$ interval, the solution is not bounded, meaning that asymptotic D3 branes are present. Consider for instance $\psi_0 > \pi$ (the solutions with $\psi_0 <0$ are analogous). We can write $\psi_0 = \widetilde\psi_0 + k \pi$ for some positive integer $k$ and $0 < \widetilde\psi_0 < \pi$. This solution represents the partial back-reaction of a brane configuration in which $k n_5$ D3 branes extend from $z = - \infty$ to the D5 brane, and $n_5 \frac{\widetilde\psi_0}{\pi}$ additional D3 branes are stretched between the NS5 stack and the D5 branes. This is illustrated in Fig. \ref{fig:PelcVariousPsi}.

\subsection{Brane Bending and Linking Numbers}

For all values of $\psi_0$, equation \eqref{eq:ND3Pelc} gives the net number of D3s which end from the negative-$z$ region on the D5 brane. As one can see from equation \ref{z_inf}, the $u \rightarrow \infty$ behavior of $z(u ; \psi_0)$ is   
\begin{equation}
        z(u ; \psi_0) \sim \frac{n_5 \alpha'}{u}\left( \frac{\pi}{2}-\psi_0\right) \, , 
    \end{equation}
and this scaling behavior is independent of the value of $z_{\infty}$. The coefficient in front of $\frac{1}{u}$ can be rewritten as  
\begin{equation}
     n_5 \alpha' \left( \frac{\pi}{2}-\psi_0\right) = \pi \alpha ' \left( \frac{n_5}{2} - N_{D3} \right) = \pi \alpha ' \left(  \frac{1}{2} \left( N_{\text{NS5,left}} -  N_{\text{NS5,right}} \right) + N_{\text{D3,right}} - N_{\text{D3,left}}  \right) \, .
     \label{linkingSUGR}
\end{equation}

Hence, when the D3-D5 branes are considered as probes in the background sourced by the NS5 branes, their world-volume is bent and has a universal scaling region that is controlled by a quantized  ``steepness number'' \eqref{linkingSUGR} and is independent of the value of the continuous parameter $z_{\infty}$.\footnote{Plots illustrating the D5-brane shape with fixed $\psi_0$ as on changes $z_{\infty}$ can be found in Figures 3 and 4 of \cite{Pelc:2000kb}.}

Of course, once the back-reaction of the D5 branes is also taken into account, we expect the democracy between five-branes to be restored, and the NS5 branes to exhibit a similar scaling region, controlled by the corresponding steepness number. These steepness numbers are the same as the (democratic definition of) non-back-reacted linking numbers \eqref{eq:symmetricGaugeLinkingNumbers}.

Note that these steepness numbers do not capture whether the D5 brane is on the right of the NS5 branes ($z_{\infty}>0$) or on the left ($z_{\infty}<0$). This understanding is only meaningful when none of the branes are back-reacted. The only information transmitted from the non-back-reacted regime to the partially-back-reacted regime are the linking numbers (or the quiver data), which now control the steepness of the asymptotic scaling regions of the five-branes, at $z \rightarrow 0$, $u \rightarrow \infty$ for the D5-branes, and at $w \rightarrow 0$, $v \rightarrow \infty$ for the NS5-branes.
\section{Brane interpretation in supergravity}
\label{sec:Brane interpretation}

We now turn to the underlying brane structure of the \AdSSS solutions.  We start simply with D3-branes ending on D5-brane walls and proceed by adding walls of both types of five-branes, keeping a finite non-trivial number of asymptotic D3-branes. Then, we look at configurations that have no AdS$_5$ throats, and finally we discuss the brane content of the Janus solutions. 


\subsection{D3 branes ending on D5 branes and no NS5 branes}

We start with the simplest non trivial solution which corresponds to D3-branes ending on D5-branes without any NS5-branes. Obviously, solutions with D3-branes ending on NS5-branes without D5-branes can be equivalently described. What makes these solutions more tractable is that there is only one kind of five brane. We consider solutions without Janus-interface-type deformations (which we discuss in subsection \ref{Janus-interfaces}), and hence set  $\beta_1 = \beta_2$ in the general solution \eqref{eq:hiRealCoordinates}. We also set to zero the free parameter $\beta_1+\beta_2$. This leads to the most general solution without NS5-branes and without Janus-interface-type deformations:
\begin{align}
\begin{split}
    h_1 (x,y) &= 2 \alpha_1 \cosh (x) \sin (y) + \sum\limits_i \gamma_1^{(i)} \log \left[ \frac{\cosh (x - \delta_1^{(i)}) + \sin (y)}{\cosh (x - \delta_1^{(i)}) - \sin (y)} \right] \,, \\  
    h_2 (x,y) &= 2 \alpha_2 \cosh (x) \cos (y)\, . 
    \end{split}
\end{align}

These solutions have two asymptotic AdS$_5 \times S^5$ regions. We return to solutions with only one asymptotic region at the end of this subsection. We now recall the relation between brane charges and supergravity parameters:
\begin{equation}
    Q_{\rm D5}^{(i)}=(4\pi)^2 \gamma_1^{(i)}\,,\quad
    Q_{\text{Page} ,  \, \text{D5}}^{\text{D3}\,(i)}  = 2^8 \pi^3 \gamma_1^{(i)} \alpha_2 \sinh  \delta_1^{(i)} \,.
\end{equation}
The difference between the asymptotic numbers of D3-branes can be computed from the difference between the radii of the AdS$_5\times S^5$ geometries in the two asymptotic regions of the Riemann-surface infinite strip
and is given by (recall \eqref{eq:ND3}):
\begin{equation}
    N_{\textrm{D3}\, ,+\infty}-N_{\textrm{D3}\, ,-\infty}=\frac{16}{\pi (\alpha ')^2}  \sum_i\vert \alpha_2\, \gamma_1^{(i)}\vert \sinh \delta_1^{(i)}   =\sum_i N^{\textrm{D3} \, (i)}_{\textrm{D5}}\,.
\end{equation}
At this stage it is tempting to interpret the geometry as D3-D5 self-similar spikes, with $N^{\textrm{D3} \, (i)}_{\textrm{D5}}$ D3-branes ending on $N^{(i)}_{\textrm{D5}}$ D5-branes. 
The sign of $\delta_1^{(i)}$ controls whether the D3-branes end on or emerge from the D5-branes (to preserve supersymmetry all D3-branes should have the same orientation). This intuition is correct and supported by the change of coordinates and the probe computation. 

The D5-branes are at $v=0$, which, on the Riemann surface, corresponds to $y=i\frac{\pi}{2}$. Recall from \eqref{eq: Page charge D3 brane} that 
\begin{equation}
    \frac{ Q_{\text{Page} ,  \, \text{D5}}^{\text{D3}\,(i)}}{ Q_{\text{D5}}^{(i)}} = - 8 \pi \,  h_2^D \left( \frac{i \pi}{2} + \delta_1^{(i)} \right) \,,
\end{equation} 
so we can interpret the (smooth) function $h_2^D \left( \frac{i \pi}{2} + x \right)$ for $x \in \mathbb{R}$ as giving the effective ratio of charges at point $x$
\begin{equation}
        \frac{ Q_{\text{Page} ,  \, \text{D5}}^{\text{D3}\,(i)}}{ Q_{\text{D5}}^{(i)}} (x) = - 8 \pi \,  h_2^D \left( \frac{i \pi}{2} + x \right) = - 4 \pi \, z u  |_{y = \frac{\pi}{2}} \, . 
\end{equation}
This equation should be thought of as determining the position of the five-brane, given its D3 Page charge/linking number.

\begin{figure}
    \centering
    \includegraphics[width=0.5\linewidth]{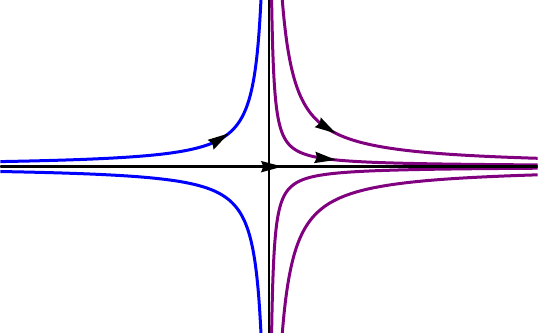}
    \caption{The D3-branes ending and emerging out of a D5-brane wall. In black some D3-branes which go through without ending on the D5. In blue the D3 branes ending on the D5 branes, $\frac{Q_{\text{D3}}}{Q_{\text{D5}}}=-3$, and in purple two self similar spikes corresponding to D3-branes emerging out of the D5 branes. These purple spikes have  $\frac{Q_{\text{D3}}}{Q_{\text{D5}}}=+1$ and $\frac{Q_{\text{D3}}}{Q_{\text{D5}}}=+5$. The arrows represent the orientation of the D3 branes.}
    \label{fig:D3 D5 spike two sides}
\end{figure}

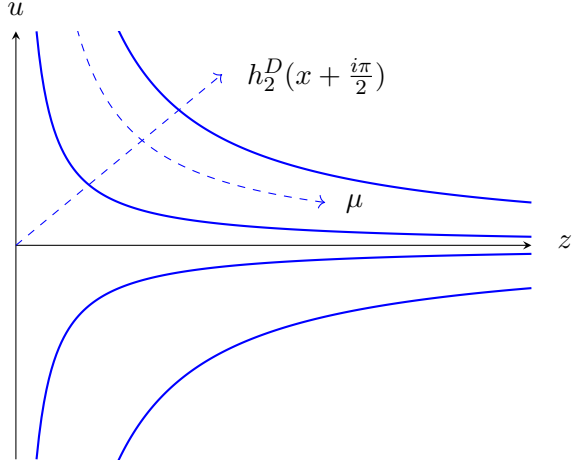
\begin{figure}
    \centering
\begin{tikzpicture}
  \begin{axis}[
    axis lines = middle,
    xmin = 0, xmax = 5,
    ymin = -5, ymax = 5,
    samples = 200,
    xtick = \empty,
    ytick = \empty
  ]
    \addplot[blue, thick, domain=0.2:5] {5/x};
    \addplot[blue, thick, domain=0.2:5] {-1/x};
    \addplot[blue, thick, domain=0.2:5] {-5/x};
    \addplot[blue, thick, domain=0.2:5] {1/x};
    \addplot[blue, dashed, ->, domain=0.2:3] {3/x};
    \addplot[blue, dashed, ->, domain=0:2] {2*x};
  \end{axis}
    \node at (4,5.1) {$h_2^D (x+ \frac{i\pi}{2})$};
    \node at (4.5,3.4) {$\mu$};
    \node at (7.3,2.9) {$z$};
    \node at (0,6) {$u$};
\end{tikzpicture}
    \caption{The D3-branes end on a D5-brane wall at $v=0$. 
In blue are two self-similar spikes corresponding to D3-branes emerging out of the D5-branes. There are $\frac{Q_{\text{D3}}}{Q_{\text{D5}}}=+1$ and $\frac{Q_{\text{D3}}}{Q_{\text{D5}}}=+5$ of them in the figure. In dashed blue lines the change of variables $(u,z) \rightarrow (\mu , x)$ at $v=0$ is represented. }
    \label{fig:D3 D5 spike one side}
\end{figure}

This shows that the Riemann-surface coordinates, together with the AdS scaling coordinate, $\mu$, are adapted to the geometry of a D3–D5 spike.
The sign of $\delta_1^{(i)}$ determines whether the configuration corresponds to a D3-brane ending on a D5-brane or to a D3-brane expanding out from the D5-brane. For a single stack of D5s at $\delta=0$, we have $\sinh(\delta)=0$ and therefore both the Page charge 
and the difference between the asymptotic numbers of D3-branes vanish. This situation corresponds to D3-branes passing through the D5-branes without ending on them. 
The situation is summarized in Fig. \ref{fig:D3 D5 spike two sides}. 
Here $\delta=0$ appears special because we chose $\beta_1=\beta_2=0$. For $\beta=\beta_1=\beta_2$, the special value where D3-branes change from ending to emerging from the D5-branes would be $\delta=\beta$.

The same physical picture applies when one of the AdS$_5$ regions closes off. These solutions are expected to describe semi-infinite D3-branes that terminate on one of the D5-branes. We take the appropriate limit of the configuration described in Section~\ref{sec:One AdS5 region}. 
Reintroducing the $\beta$ parameters and taking the limit:
\begin{equation}
    \alpha_i\rightarrow 0,\, \beta_i\rightarrow \infty \qquad \textrm{with} \qquad \alpha_i\, e^{\beta_i}=2\theta_i \, \, \textrm{fixed}
\end{equation}
we find that the asymptotic number of D3 branes is:
\begin{equation}
    N_{\text{D3},-\infty}=\frac{16}{\pi(\alpha')^2} \left(\vert\theta_2\vert\sum_i\vert \gamma_1^{(i)}\vert e^{- \delta_1^{(i)}}\right)\,,\quad N_{\text{D3},+\infty}=0 \,.
\end{equation}
As before, each term in the sum can be understood microscopically as a number of D3-branes ending on $\frac{4}{\alpha'^2}\gamma_1^{(i)}$ D5-branes. This solution describes a collection of self-similar D3-D5 spikes. 

The brane realization of these spikes  is well understood from both the D5-brane and D3-brane perspectives. From the perspective of the D5s, whose worldvolume theory is $SU\left(N_{\text{D5}}=\frac{4}{(\alpha')^2}\sum_i\gamma^{(i)} \right)$ maximally supersymmetric Yang-Mills in $5+1$ dimensions, 
it corresponds to a collection of Abelian monopoles at the same position, each with charge $\left(\frac{Q_{\text{D3}}}{Q_{\text{D5}}}\right)^{(i)}$, and the gauge group is broken to $\prod_{i} SU\left(\frac{4}{(\alpha')^2}\gamma^{(i)}\right)$. From the $SU( N_{\text{D3},-\infty})$ $\mathcal{N}=4$ super Yang-Mills D3-brane perspective, it corresponds to solutions of the Nahm equation with different $SU(2)$ representations embedded in $SU( N_{\text{D3},-\infty})$ corresponding to fuzzy D5-branes. This time, the gauge group is broken to  $\prod_i SU\left(\frac{16}{\pi(\alpha')^2} \left(\vert\theta_2\vert\vert \gamma_1^{(i)}\vert e^{- \delta_1^{(i)}}\right)\right)$.

The supergravity solution is completely democratic between the roles played by the D5-branes and NS5-branes. As such, everything said so far for D3-D5 intersections also applies to D3-NS5 intersections.

\subsection{D3-branes ending on both types of five-branes with  AdS$_5\times S^5$ asymptotics}

In this subsection, we consider solutions with both D5 and NS5 sources that have either one or two asymptotic AdS$_5$ regions. We assume again that $\beta_1=\beta_2=0$ so that there is no Janus-interface deformation. To summarize the picture that has emerged for solutions with D5 branes only: 
the $\sinh(\delta^{(i)})$ factors appearing in the general expressions for the charges at infinity can be understood microscopically as the net number of D3-branes ending on or emerging from the five-branes. More precisely, they capture the difference in D3-brane charge between the two asymptotic regions. This na\"ive interpretation does not remain valid in the presence of both D5-branes and NS5-branes.

There are two main differences from the simpler situation with only one kind of five-brane. First, as seen in the computation of a probe D5-brane in the NS5-brane background, 
these five-branes interact non-trivially. Moreover, the ordering between the D5-branes and NS5-branes becomes non-trivial data. 
Recall the three-brane Page charges and the difference between the asymptotic numbers of D3-branes given by \eqref{eq:3-chargesPage} and \eqref{eq:diff ND3} with $\beta_1=\beta_2=0$. 

We see that provided $\xi_1=\xi_2$, we have 
\begin{equation}
    N_{\textrm{D3}\, ,+\infty}-N_{\textrm{D3}\, ,-\infty}= \sum_i  N_{\text{D5}}^{\text{D3}\,(i)}+ \sum_j N_{\text{NS5}}^{\text{D3}\,(j)}
\end{equation} 
as expected. From \eqref{eq:diff ND3} it is tempting to identify each $\frac{16}{\pi (\alpha ')^2}\vert \alpha_2\, \gamma_1^{(i)}\vert \sinh(\delta_1^{(i)})$ as the number of D3-branes ending on the $i^{\text{th}}$ stack of D5-branes, 
and $\frac{16}{\pi (\alpha ')^2}\vert \alpha_1 \,\gamma_2\vert \sinh(\delta_2^{(j)})$ as the number of D3-branes ending on the $j^{\text{th}}$ stack of NS5-branes. Unfortunately this microscopic intuition only works when one kind of five-branes is involved. When both species of five-branes are present, their interaction makes it impossible to identify how many D3 branes end on a particular five-brane. Indeed, this number can be changed by Hanany-Witten moves, which are irrelevant for the field theory in the IR. 

The only quantity that is independent of these moves, and is well-defined both in the absence and presence of back-reaction, is the Page charge. This determines the bending of the five-branes in the scaling region, and it receives contributions both from the D3 branes ending on the five-branes and from the induced D3 charge coming from the D5-NS5 interactions.

To explore the D5 singularities of these solutions one needs to go to $v=0$, $y=i\frac{\pi}{2}$ on the upper boundary of the strip. From the change of variables in Section \ref{ss:mapping} and the probe calculation in Section \ref{sec:Probing}, one can see that a probe D5-D3 spike that has a minimum at $v=0$ and at a given value of $x$ must have the DBI world-volume charge (proportional to the supergravity Page charge) $\frac{1}{4\pi}\frac{Q^{D3\,,(i)}_{\text{Page}}}{Q_{D5}^{(i)}}\big(x\big)=2 h_2^{D}\big(\frac{i\pi}{2}+x\big)$. Hence, the value of $x$ where the probe sits is determined by the strength of the pulling exerted by the D3-branes ending on the D5-branes. The self-similar spikes are spanned by the AdS radial coordinate $\mu$, and they sit at fixed values of the $\mu$-independent quantity, $u ~ z$, evaluated at a given value of $x$. This gives the steepness of the spike:
\begin{equation}
    u~z \sim \frac{ Q^{\text{D3},\, (i)}_{\text{Page},\,\text{D5}}}{Q_{\text{D5}}^{(i)}}\,.
\end{equation}
Similarly, one can study the pulling of NS5-branes at $u=0$ by D3-branes and the interaction with the fluxes sourced by the D5-branes by changing the role of $(v,u,z)$ to $(u,v,w)$ and studying the spikes, parameterized by the $\mu$-independent combination of the spike  coordinates, $v\, w$.

We also note that  in the supergravity solutions the five-brane sources are ordered by their Page charges. This suggests that supergravity has a preferred Hanany-Witten frame, in which the linking numbers of {\em both} the D5 and NS5 branes are monotonically increasing. This frame is the one depicted at the bottom of Fig. \ref{fig:summaryBranes}.

\subsection{Solutions with two AdS$_4\times B_6$ caps}

In this subsection we focus on solutions in which the warp factor transforms the Riemann surface in a compact two-dimensional manifold. The fully back-reacted solutions have an AdS$_4\times B_6$ region at both ends of the Riemann strip, and are holographically dual to certain 3-dimensional $\mathcal{N}=4$ superconformal field theories. From Section \ref{sec:General Intersection} and Section \ref{sec:Probing}, we saw that the self-similar spikes that source these solutions have an AdS$_4$ world-volume, and, since in these solutions the $\mu$ coordinate always extends from $0$ to $\infty$, these spikes appear to be semi-infinite. 

Here, we explain how such semi-infinite sources are compatible with a solution arising from  D3 branes stretched between D5 and NS5 branes at {\em finite} distance, and describing holographically a 2+1 dimensional SCFT.  

The simplest solution in this class has a single D5-brane source and a single NS5-brane source and exhibits the essential physical features of the full geometry.  The solution is given by the two harmonic functions:
\begin{align}
    h_1&=- \gamma_{1}\log\left(\tanh\left(-\frac{\RSz-\delta_1-i\frac{\pi}{2}}{2}\right)\right)+\text{c.c} \,,\\
    h_2&= -\sum_i \gamma_{2} \log\left(\tanh\left(\frac{\RSz-\delta_2}{2}\right)\right)+\text{c.c}\,. 
\end{align}
Remember that $\frac{4}{(\alpha ')^2}\gamma_1$ and $\frac{4}{(\alpha ')^2}\gamma_2$ are the numbers of D5 and NS5 branes respectively. The D3 Page charges of the five-branes are:

\begin{align}
Q^{D3,\, }_{\text{Page},\,\text{NS5}}  &=  2^8 \pi^3\,  \left(
- 2\,  \gamma_1\gamma_2 \arctan (e^{\delta_2- \delta_1})-\xi_2\right)
 \ ,   \\
Q^{D3,\, }_{\text{Page},\,~\text{D5} }&=  2^8 \pi^3\,  \left(
  + 2\,
\gamma_2\gamma_1\left( \arctan (e^{\delta_2 - \delta_1})-\xi_1\right)
 \right) 
\end{align}
and were identified with linking numbers in \cite{Assel:2011xz}. We have shown in Section~\ref{sec:Probing} that the Page charge of a given five-brane is the same as the steepness number that governs the asymptotic bending of this brane (or the bulge) and {\em not} the number of D3 branes ending on it.

\begin{figure} 
    \centering
    \begin{tabular}{cc}
    \includegraphics[width=0.35\linewidth,trim=0cm 0cm 0cm 0.65cm, clip]{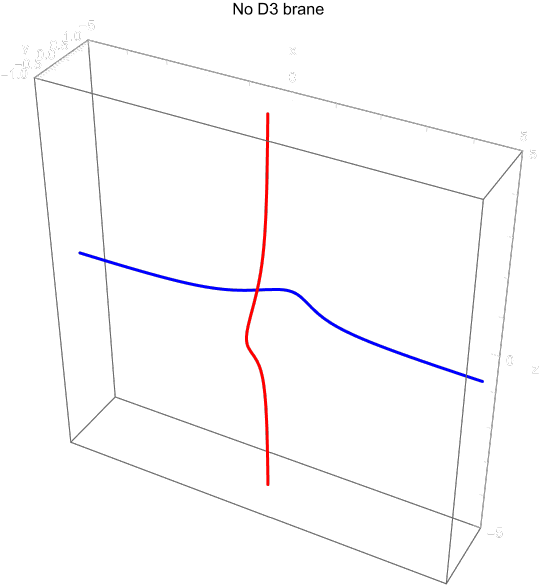} &     \includegraphics[width=0.35\linewidth,trim=0cm 0cm 0cm 0.65cm, clip]{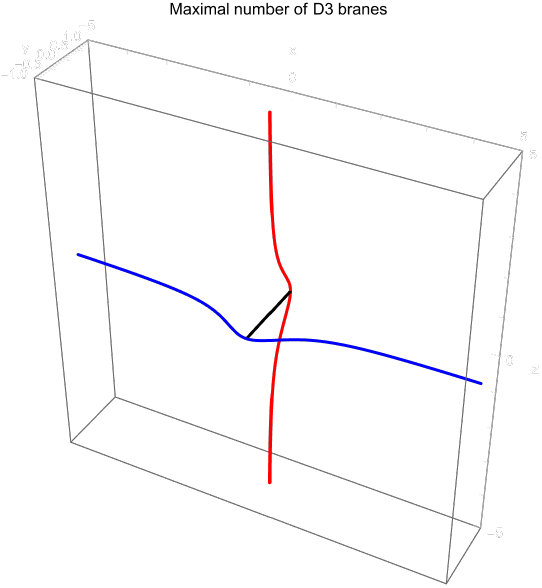}
    \end{tabular}
    \caption{In blue the D5 branes, and in red the NS5 branes, when there are no D3 branes stretched between them (left) or $N_{NS5} N_{D5}$ D3 stretched between them (right). The bulge is given by $\frac{N_{NS5} N_{D5}}{2}$ in both configurations, but with a different orientation. This means that in the scaling limit we have $z= \pm \frac{N_{NS5} N_{D5}}{2}\frac{1}{u}$. Of course from a gauge-theory point of view, the theory is trivial, but the bending of the branes illustrates the general physics.}
    \label{fig:No D3}
\end{figure}

First, recall the rules derived from the probe analysis in Section \ref{regime-2} summarized in Fig. \ref{fig:PelcVariousPsi}. In the scaling limit, we found that for an NS5-brane and a D5-brane, the steepness number  (bulge) is given by $\frac{N_{NS5} N_{D5}}{2}$ when no D3-branes stretch between them. The bulge vanishes when exactly $\frac{N_{NS5} N_{D5}}{2}$ D3-branes are stretched between the D5 and the NS5 branes. Finally, the bulge becomes $-\frac{N_{NS5} N_{D5}}{2}$ when $N_{NS5} N_{D5}$ D3-branes stretch between them. The two extreme configurations are depicted in Figure~\ref{fig:No D3}.

Since both the steepness numbers/bulges and the Page charges are gauge dependent, in order for the Page charges to reproduce the bulges of branes, we must shift $\xi_1$ and $\xi_2$ in \eqref{eq:harmonic-conjs}  to a symmetric gauge \eqref{eq:symmetricGauge}:
\begin{align}
Q^{\rm{D3,~symmetric}}_{\text{Page},\,\text{NS5}}  &=  -2^8 \pi^3\,  \left(
2\,  \gamma_1\gamma_2 \right)\left(\arctan (e^{\delta_2- \delta_1})-\frac{\pi}{4}\right)
 \ ,   \\
Q^{\rm{D3,~symmetric}}_{\text{Page},\,~\text{D5} }&=  ~~~2^8 \pi^3\,  \left(
  2\,
\gamma_2\gamma_1\right)\left( \arctan (e^{\delta_2 - \delta_1})-\frac{\pi}{4}\right)\ .
\end{align}

These shifts correspond to gauge transformations in the $B_2$ and $C_2$ fields at infinity, and not to brane Hanany–Witten moves \cite{Assel:2011xz}, which leave the linking numbers invariant.  As shown in Section \eqref{sec:Probe D5 in NS5 background}, supergravity naturally implements a symmetric gauge with no right-left asymmetry, see \eqref{eq:symmetricGaugeLinkingNumbers}. 

The probe D5-brane in an NS5 background exhibits a scaling region in which $z\sim \frac{Q^{{\rm D3}, (i)}_{\text{steepness, D5}}}{u}$, where the steepness number, $Q^{{\rm D3}, (i)}_{\text{steepness, D5}}$ that governs the asymptotic behavior at  $u\rightarrow\infty, z\rightarrow 0$ is the same as the ``symmetric'' Page charge, $Q^{{\rm D3}, (i)}_{\text{Page, D5}}$.  This again confirms that supergravity captures the scaling regions of the five-branes.

As illustrated in Fig. \ref{fig:Zoom in}, the degrees of freedom of the SCFT live on these scaling regions, which appear to be infinite in supergravity. Taking the infrared limit in supergravity collapses the non-scaling portions of the brane configuration to zero size. As a result, the supergravity dual of a 3-dimensional SCFT appears to contain semi-infinite D3-branes ending on D5 and NS5 branes. The SCFT is then reconstructed from the steepness numbers of the branes, which are the same as the Page charges and the linking numbers. 

\begin{figure} 
    \centering
    \includegraphics[width=0.35\linewidth,trim=0cm 0cm 0cm 0.65cm, clip]{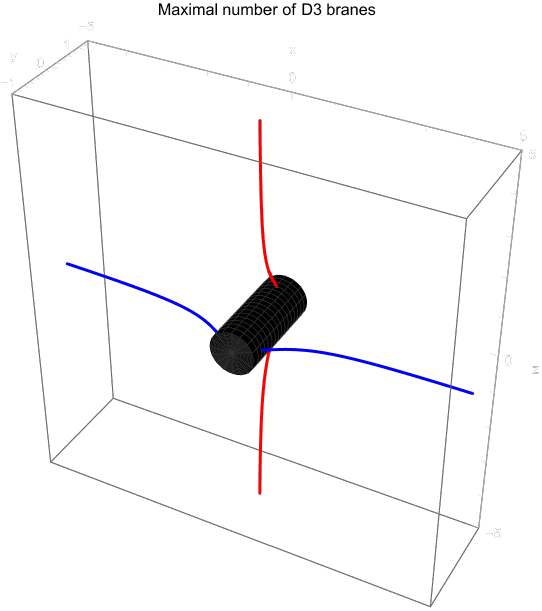}
    \caption{Fig. \ref{fig:No D3}  
     drawn from the viewpoint of the $AdS_4$ supergravity solution. The only portion of the configuration this solution captures  is the {\em scaling region} outside the black box. Our scaling limit shrinks the black region to zero size and, upon zooming into the scaling region, the scaling regions of the five-branes appear as semi-infinite D3-D5 and D3-NS5 spikes. This is the mechanism through which the holographic dual of a genuinely 3-dimensional theory comes to look like a collection of infinite self-similar spikes. The SCFT linking numbers are the same as the steepness numbers controlling the scaling regions of the branes.}
    \label{fig:Zoom  in}
\end{figure}

We now study the most general solution with two AdS$_4\times B_6$ asymptotic regions given in \eqref{eq:most general sol with two closing}. For an arbitrary number of stacks of branes, this large gauge transformation transforms the charges to:
\begin{equation}
\begin{aligned} 
Q^{D3,\, (i)}_{\text{Page},\,\text{D5}}  &= - 2^8 \pi^3\,  \left(
   2\,  \gamma_1^{(i)}\right) \sum_{j} \gamma_2^{(j)} \left(\arctan (e^{\delta_2^{(j)} - \delta_1^{(i)}})-\frac{\pi}{4}
  \right)  \ ,   \\
Q^{D3,\, (i)}_{\text{Page},\,\text{NS5} }&=  ~~~2^8 \pi^3\,  \left(
 2\,
\gamma_2^{(i)} \right)\sum_{j}\gamma_1^{(j)}\left( \arctan (e^{\delta_2^{(i)} - \delta_1^{(j)}})-\frac{\pi}{4}\right)
 \ .
\end{aligned}
\end{equation}

In particular, consider two stacks of NS5-branes and a single stack of D5-branes. For definiteness, take $\delta_1 = \delta_2^{(1)}$, which ensures that there are exactly $\frac{N_{NS5} N_{D5}}{2}$ D3-branes stretched between the first two five-branes, and therefore the bulge contribution from the D5-NS5 interaction cancels that from the D3 branes. Hence this D5 brane has no bulge, and the steepness number characterizing its asymptotic region vanishes,  $Q^{{\rm D3}, (1)}_{\text{steepness},\,\text{NS5}} = Q^{{\rm D3}, (1)}_{\text{Page},\,\text{NS5}} = 0$. The two remaining Page/steepness numbers are then given by:
\begin{equation}
\begin{aligned}
Q^{D3,\, (1)}_{\text{Page},\,\text{D5}}  &= -2^8 \pi^3\,\left(
 2\,
\gamma_2^{(2)} \right)\gamma_1^{(1)}\left( \arctan (e^{\delta_2^{(2)} - \delta_1^{(1)}})-\frac{\pi}{4}\right) \ ,   \\
Q^{D3,\, (2)}_{\text{Page},\,\text{NS5} }&=  ~~~2^8 \pi^3\,  \left(
 2\,
\gamma_2^{(2)} \right)\gamma_1^{(1)}\left( \arctan (e^{\delta_2^{(2)} - \delta_1^{(1)}})-\frac{\pi}{4}\right)
 \ .
\end{aligned}
\end{equation}

\begin{figure}[h!]
     \centering
     \includegraphics[width=0.35\linewidth,trim=0cm 0cm 0cm 0.65cm, clip]{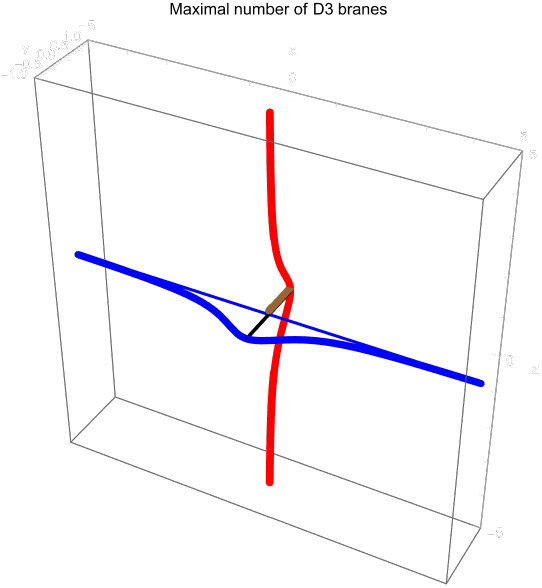}
     \caption{Two stacks of D5 branes in blue, one stack of NS5 branes in red. In brown and black we represent the D3 branes; the different thicknesses of these lines are proportional to the numbers of D3 branes ending on each D5 brane.}
     \label{fig:D3 2 D5 NS5}
 \end{figure}

We draw the brane picture dual to the aforementioned configuration for $\arctan(e^{\delta_2^{(2)} - \delta_1^{(1)}})=\frac{\pi}{8}$ in figure. \ref{fig:D3 2 D5 NS5}.

It is now evident that supergravity collapses the non-scaling regions to zero size, capturing only a zoomed-in view of the scaling regions. This gives the impression that all D5 and NS5 sources of the supergravity solutions are semi-infinite spikes. 
It is only the asymptotic behavior at the two boundaries of the Riemann-surface strip that can distinguish between solutions dual to 3-dimensional SCFTs (two asymptotic AdS$_4 \times B^6$ regions), solutions dual to superconformal boundary conditions of $\mathcal{N}=4$ super Yang–Mills (one asymptotic AdS$_5$ and one asymptotic AdS$_4 \times B^6$ region) and solutions dual to Janus domain walls of $\mathcal{N}=4$ super Yang–Mills (two asymptotic AdS$_5 \times S^5$ regions). In particular, for brane configurations giving rise to 3-dimensional SCFTs on the D3 worldvolume, the corresponding supergravity solution captures the scaling regions away from the D3 branes. So one can intuitively say that, when going to the infrared, the degrees of freedom migrate from the D3 region to the scaling region, where they are captured by the supergravity solution. It would be interesting to try to see if this phenomenon can be captured purely in the field theory.

\subsection{Janus interfaces}
\label{Janus-interfaces}

It remains to describe the branes that give rise to the Janus-interface solution, which has two asymptotic $AdS_5 \times S^5$ regions and the central charge on both sides of the Janus interface is the same.  Naively, this suggests that there are no 5-branes present. However, as explained above, the supersymmetries of this solution are precisely those of the D3-D5-NS5 system.  Moreover, there are non-trivial $C_2$ and $B_2$ flux profiles near the interface, and the dilaton jumps between two different constant asymptotic values. 

Given the  $SO(3) \times SO(3)$-isometry of these solutions one can try to deduce the distribution of five-branes that could be responsible for creating the Janus solution:

Consider for simplicity a very large number of NS5 branes\footnote{One can make a similar argument with D5 branes.} smeared over a finite-size $S^2$ at $u=u_0$, and also smeared uniformly over half the $z$ axis ($z>0$). 

At large $z$, this solution resembles the solution sourced by NS5 branes smeared on $S^2 \times \IR$. This solution is very simple. Outside the sphere ($u>u_0$) it is identical (by the Gauss-Birkhoff theorem) to the solution of NS5 branes smeared on the $z$ axis, with a harmonic function that decays as $1/u$. Inside the sphere ($u<u_0$) the NS5 harmonic function is constant. This means that the field strength of the $B_2$ field vanishes inside the sphere, and the dilaton, $\Phi_{\rm Inside}$, is constant, but larger that its value, $\Phi_{\infty}$, far from the NS5 branes. In this way we can engineer a dilaton jump along the $z$ axis and a $B_2$ field that only has a non-trivial profile around $z=0$.

We can now place at $u=0$ a very large number of D3 branes extending along the $z$ direction. These D3 branes are mutually supersymmetric with the NS5 branes smeared on the semi-infinite cylinder, $S^2 \times \IR^+$. The fluxes of these D3 branes will interact with the non-trivial $B_2$ profile around $z=0$ to create a $C_2$ flux, which will also be localized around $z=0$. 

The Janus solution emerges in the near D3-brane limit of this configuration and all that is left of the 5-branes are non-normalizable sources of $B_2$ and $C_2$ at the  boundary.  At $z \ll 0$, the D3 branes are far away from the semi-infinite cylinder of smeared NS5 branes, and their near-horizon geometry is   $AdS_5 \times S^5$ with a dilaton equal to $\Phi_{\infty}$. At  $z \gg 0$ the  D3 brane are deep inside the semi-infinite cylinder, where there are no $B_2$ or $C_2$ fields, but the dilaton is constant at $\Phi_{\rm Inside}$. Hence, the near-horizon  geometry in this region is $AdS_5 \times S^5$, with a dilaton equal to $\Phi_{\rm Inside}$. This description of the Janus interface solution will be fully analyzed in \cite{Dulac2026}.  

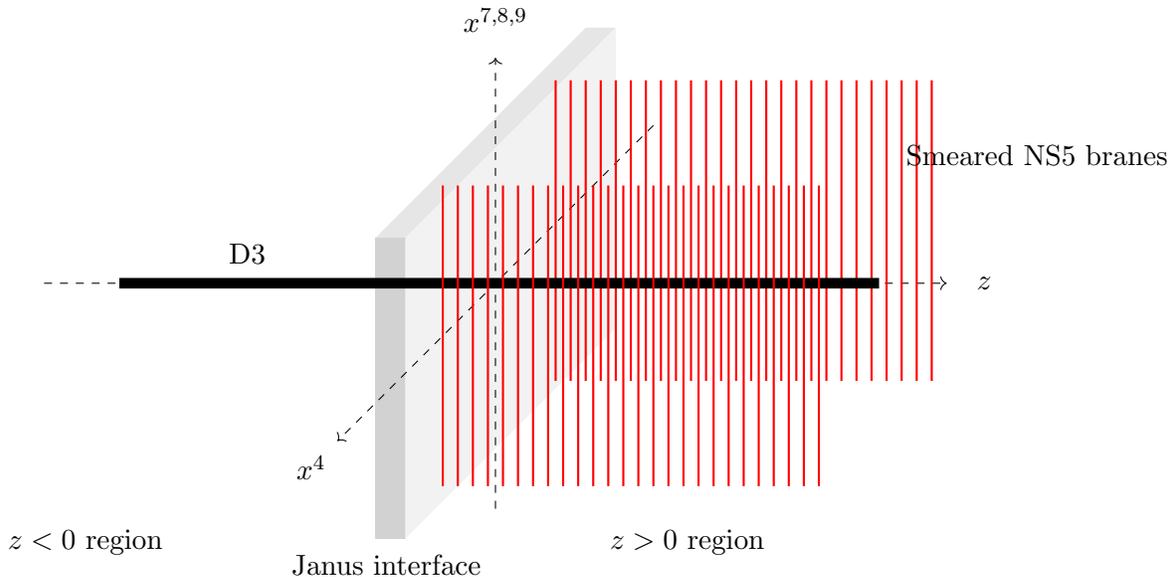
\begin{figure}[t]
    \centering
\begin{tikzpicture}[
    x={(-0.7cm,-0.7cm)},
    y={(1cm,0cm)},
    z={(0cm,1cm)}
] 
\def\xo{-2} \def\yo{-.2} \def\zo{-2}  
\def\Lx{4} \def\Ly{.4} \def\Lz{4}
\begin{scope}[shift={(\xo,\yo,\zo)}]
\fill[gray!20] (0,0,\Lz) -- (\Lx,0,\Lz) -- (\Lx,\Ly,\Lz) -- (0,\Ly,\Lz) -- cycle; 
\fill[gray!35] (\Lx,0,0) -- (\Lx,\Ly,0) -- (\Lx,\Ly,\Lz) -- (\Lx,0,\Lz) -- cycle; 
\fill[gray!10] (0,\Ly,0) -- (\Lx,\Ly,0) -- (\Lx,\Ly,\Lz) -- (0,\Ly,\Lz) -- cycle;         
\end{scope}
\draw[->,dashed] (-3,0,0)--(3,0,0);
\draw[->,dashed] (0,0,-3)--(0,0,3);
\draw[->,dashed] (0,-6,0)--(0,6,0);
\node at (0,6.5,0) {$z$};
\node at (3.5,0,0) {$x^4$};
\node at (0,0,3.5) {$x^{7,8,9}$};
\node at (3.5,5,-1) {$z>0$ region};
\node at (3.5,-3,-1) {$z<0$ region};
\node at (3.5,1,-1.3) {Janus interface};
\foreach \i in {0,...,25}{
    \draw[NS5] (-1,{0.1+0.2*\i},-2) -- (-1,{0.1+0.2*\i},2);
}
\draw[line width=4pt] (0,-5,0)--(0,5.1,0);
\foreach \i in {0,...,25}{
    \draw[NS5] (1,{0.2*\i},-2) -- (1,{0.2*\i},2);
}
\node at (-1,6.5,1) {Smeared NS5 branes};
\node at (-1,-4,-.3) {D3};
\end{tikzpicture}
    \caption{Janus Interface, classical view. The red lines are the NS5 branes, we represent only an $S^0 \subset S^2$ given by the intersection with the $x^4$ axis. The black line is the stack of D3 branes. The Janus interface is the gray zone around $z=0$. }
    \label{fig:JanusInterface}
\end{figure}

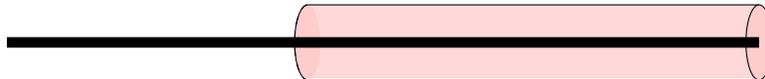
\begin{figure}
    \centering
\begin{tikzpicture}
\node at (0,0) {\cylNS{.5}};
\draw[line width=4pt] (-7,0)--(3,0);
\end{tikzpicture}
    \caption{Schematic depiction of Fig. \ref{fig:JanusInterface}. The red cylinder depicts the $S^2 \times \mathbb{R}_{z>0}$ cylinder over which the NS5 branes are smeared ($S^1 \subset S^2$ is represented).  }
    \label{fig:JanusInterfaceSchematic}
\end{figure}
 
\subsection{Multi-Janus solutions}
\label{Multi-Janus}

Given our understanding of the supersymmetries, it is also tempting to speculate on the possible D3-D5-NS5 brane configurations that give rise to multi-Janus solutions \cite{DHoker:2007hhe}.

A solution with multiple $AdS_5$ throats preserving the same Killing spinors as flat orthogonal D3, D5 and NS5 branes must necessarily come from multiple D3 brane stacks. The five-form flux of each $AdS_5$ throat should give the number of D3 branes in the stack.  However, it is not hard to see that the D3 branes cannot preserve the 
$SO(3) \times SO(3)$ isometry of the solutions if one uses more than one stack of coincident branes, and this only gives a solution with two $AdS_5$ throats - the Janus interface solution.

Hence, to obtain the $SO(3) \times SO(3)$-invariant multi-Janus solutions one needs to start from multiple stacks of D3 branes, which necessarily do not preserve this isometry, and to recover this isometry after taking the near-brane limit. 

Such a possible starting configuration involves multiple stacks of D3 branes, at distances of order $\epsilon$, each semi-infinite portion of which is surrounded by a semi-infinite cylinder of smeared NS5 branes, as in the Janus-interface solution. Note that in this solution we did not specify the radius $u_0$ of the semi-infinite cylinder, but only asked that it create a region of constant dilaton, $\Phi_{\rm Inside}$, inside at $z \gg 0$. This can be done either by many NS5 branes on a semi-infinite cylinder with large $u_0$, or by fewer NS5 branes on a cylinder of small $u_0$.

For the multi-Janus configuration we have to choose all the semi-infinite cylinders surrounding the D3 brane clusters to have $U_0\leq\epsilon$, and then to take the limit $\epsilon \rightarrow 0$ keeping the values of the dilaton inside every semi-infinite cylinder constant. It is not hard to see that the near-horizon limit of the D3 branes is compatible with this limit, and that each stack will give rise to its own semi-infinite $AdS_5$ asymptotic regions. 
It would be interesting to see if there are any properties or constraints of the \AdSSS multi-Janus solution that can be used to validate this  brane-interpretation proposal.

\begin{figure}
    \centering
\begin{tikzpicture}
\node at (-8,0) {\cylNS{.5}};
\node at (-8,2.3) {\cylNS{1.1}};
\node at (-8,4) {\cylNS{.3}};
\node at (0,0) {\cylNS{.3}};
\node at (0,1.5) {\cylNS{.7}};
\node at (0,3.5) {\cylNS{1}};
\draw[line width=4pt] (-11,0)--(3,0);
\draw[line width=4pt] (-11,1.5)--(3,1.5);
\draw[line width=4pt] (-11,3)--(3,3);
\draw[line width=4pt] (-11,4)--(3,4);
\draw[<->] (-4,0.2)--(-4,1.3);
\draw[<->] (-4,1.7)--(-4,2.8);
\draw[<->] (-4,3.2)--(-4,3.8);
\draw[<->] (3,1.6)--(3,2.1);
\node at (-4.3,0.75) {$\epsilon_1$};
\node at (-4.3,2.25) {$\epsilon_2$};
\node at (-4.3,3.5) {$\epsilon_3$};
\node at (3.6,1.9) {$U_0$};
\end{tikzpicture}
    \caption{Using the schematic depiction of Fig. \ref{fig:JanusInterfaceSchematic}, we depict the brane configuration whose $\epsilon, U_0 \rightarrow 0$ limit is conjectured to give rise to a multi-Janus solution with six asymptotic $AdS_5$ regions.}
    \label{fig:JanusMulti}
\end{figure}
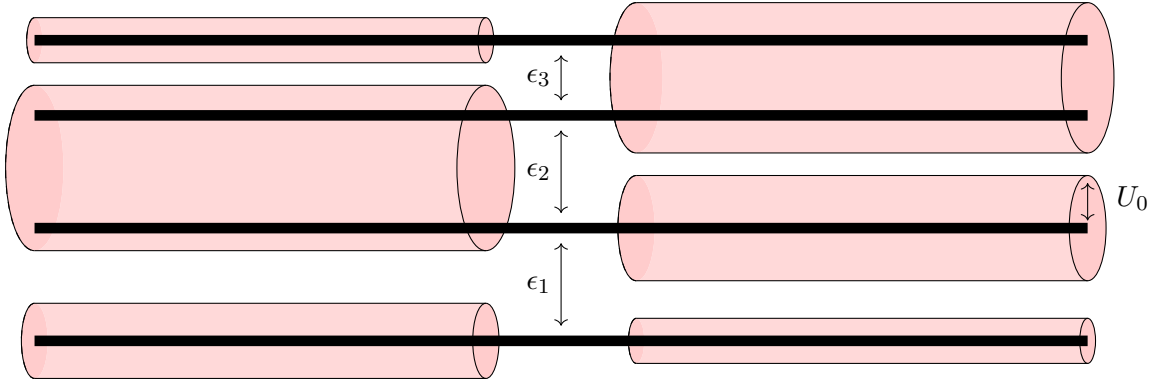

\section{Conclusions and Future Directions}
\label{sec:Discussion}

In this paper we have shown that \AdSSS solutions \cite{DHoker:2007zhm,DHoker:2007hhe} come from zooming in on certain scaling regions of solutions preserving the same 8 Killing spinors as orthogonal intersecting D3, D5 and NS5 branes \cite{Lunin:2007mj}.
We have provided an explicit map between the \AdSSS coordinates and the orthogonal-brane coordinates, and have shown that the D5 and NS5 sources of {\em all} the \AdSSS solutions correspond in orthogonal-brane coordinates to infinite D3-D5 and D3-NS5 spikes. 

We have argued that, as one turns on $g_s$ and moves from the non-back-reacted regime {\bf 1} where the branes are flat and orthogonal, to the regime {\bf 2} where some of the branes back-react and some do not, and then to the fully-back-reacted regime {\bf 3}, the D5 and NS5 branes develop a {\em scaling region}, and that the \AdSSS solutions come from zooming in on this region.

We have shown that the steepness numbers describing this scaling region are the same as the regime {\bf 1} linking numbers and the regime {\bf 3} Page charges, and this has allowed us to explain the origin of the good-bad-ugly classification of quivers. Good quivers are those where branes are ordered by linking numbers, and these back-react into ordered scaling spikes in regime {\bf 2} and into \AdSSS solutions (where the sources are ordered by Page charge). In contrast, in the bad and ugly quivers, the branes need to cross each other, and this indicates that one must perform one or more Seiberg-like dualities in order to reach an ordering of the branes that back-reacts into self-similar spikes with an \AdSSS in the infrared.

We have argued that exactly the same phenomenon happens for solutions dual to boundary conformal field theories and to Janus domain walls. These solutions come from semi-infinite D3 branes ending on D5 and NS5 branes, possibly superposed (for Janus domain walls) with infinite D3 branes. When the five-branes are ordered by linking number, we expect there to be a one-to-one correspondence between these brane configurations and \AdSSS solutions. However, when the five-branes are not ordered, we expect them to cross each other to a configuration of ordered five-branes, which then give rise to an \AdSSS solution. 

It would be very interesting to prove explicitly this correspondence for boundary and Janus interfaces. Also, the crossing of the branes needed to transform ``bad'' boundary conditions and ``bad'' Janus walls into good ones is expected to correspond to a type of {\em boundary Seiberg duality} and {\em defect Seiberg duality}, which would be very interesting to establish purely in field theory.

Our link between \AdSSS solutions and branes also allowed us to argue that Janus interface solutions come from the near-horizon of D3 branes traversing a dilaton kink created by D5 or NS5 branes smeared over a semi-infinite cylinder, and to illustrate that probe D3 branes in these solutions have a six-dimensional moduli space, consistent with the masslessness of the six scalars of maximally-supersymmetric conformal interfaces \cite{DHoker:2006qeo}. This will be discussed in further detail in \cite{Dulac2026}. 

Our realization that conformal interfaces come from D3 branes that traverse a dilaton kink may help in elucidating the recent proposal that conformal interfaces are related to marginal operators \cite{Komatsu:2025cai}.  It would be interesting to see if one can realize other holographic domain interfaces in this way, and whether one can find a relation between the interface displacement operator and the branes that realize this conformal interface. It would be also fascinating to see if one can put on firmer ground the multi-Janus brane-realization proposal in Section \ref{Multi-Janus}.

The most obvious question our work raises is whether other $AdS \times S \times S $ solutions warped over a Riemann surface also come from scaling regions of solutions corresponding to orthogonal D-branes and NS5 branes. In an upcoming paper \cite{Bena2026toappear}  we will show that the answer is a resounding {\em yes}. Besides the $AdS_3 \times S_3 \times S_3 \times \Sigma$ solutions considered in \cite{Bena:2024dre, Bena:2025hxt,Bena:2023rzm} and the \AdSSS solutions considered in this paper, this has also been shown for $AdS_7 \times S_2 \times {\rm Line}$ solutions \cite{Apruzzi:2013yva,Cremonesi:2015bld,Apruzzi:2015wna}, which were related in \cite{Apruzzi:2025znw} to solutions with Killing spinors corresponding to orthogonal D8-D6-NS5 branes \cite{Imamura:2001cr}. We will show in \cite{Bena2026toappear} that this happens for known $AdS \times S \times S $ solutions, and also conjecture the existence of some types of $AdS \times S \times S $ solutions that have not been yet constructed.

The second important question is whether one can find the most general asymptotically flat D3-D5-NS5 solutions whose scaling limits give the \AdSSS geometries. The asymptotically flat solutions are governed by a nonlinear Monge-Ampère equation, which is hard to solve. It would be interesting to ascertain whether there exist other coordinate changes and limits, besides the one considered in this paper, which can render this equation linear.

We certainly expect to have Monge-Ampère solutions that do not have $AdS_4$ factors. The field theories on D3 branes suspended between NS5 and D5 branes can flow to the infrared either to conformal fixed points, which are dual to \AdSSS solutions, or to confining vacua, which must be capped in the infrared, and hence {\em cannot} be dual to such solutions. Finding such capped geometries (possibly in the spirit of I-branes \cite{Nunez:2023nnl,Barbosa:2025lvd}) would be very interesting, not only for providing new supergravity duals to confining theories but also for building horizonless super-maze geometries \cite{Bena:2022wpl,Bena:2023fjx,Bena:2024qed}, which are expected to capture a finite fraction of the entropy of supersymmetric black-holes.

\vspace{-6mm}

\section*{Acknowledgments}
\vspace{-3mm}
We would like to thank Costas Bachas, Nikolay Bobev, Julius Grimminger, Pierre Heidmann, Matthew Heydeman, Eric D'Hoker, Anthony Houppe, Daniel Jafferis, Hora\c tiu N\u astase, Radu Roiban, Alessandro Tomasiello, and Christoph Uhlemann for helpful comments and interesting discussions. The work of IB and NW was supported in part by the ERC Grant 787320 - QBH Structure. The work of DT is supported by the Israel Science Foundation (grant No. 1417/21), by the German Research Foundation through a German-Israeli Project Cooperation (DIP) grant “Holography and the Swampland”, by Carole and Marcus Weinstein through the BGU Presidential Faculty Recruitment Fund, by the ISF Center of Excellence for theoretical high energy physics, by the VATAT Research Hub in the Field of quantum computing, by a scholarship by the Ministry of Foreign Affairs of Israel and by the ERC starting Grant dSHologQI (project number 101117338). The work of NW was also supported in part by the DOE grant DE-SC0011687.

\bigskip
\bigskip
\appendix
\leftline{\LARGE \bf Appendices}

\section{Dualities from M theory to Type IIB supergravity}
\label{app:IIBtoM}

In this appendix we explain the dualities that take us from the most general M-theory solution describing intersecting M2, M5 and M5' branes to type IIB supergravity solutions describing intersecting D3, D5 and NS5 branes.

The most general M-theory solution describing intersections of M2-M5-M5' branes is:
\begin{align}
    \d s_{11}^2 ~=~  e^{2  A_0}\, \Big( - \d t^2 &~+~  \d y^2 ~+~ e^{-3  A_0} \, (-\partial_z w )^{-\frac{1}{2}}\, \d \vec u \cdot \d \vec u  ~+~ e^{-3  A_0} \, (-\partial_z w )^{\frac{1}{2}}\, \d \vec v \cdot \d \vec v \,    \\
  & ~+~  (-\partial_z w ) \, \big(  \d z ~+~(\partial_z w )^{-1}\,   (\vec \nabla_{\vec u} \, w )  \cdot  \d \vec u \big)^2  \Big)\,.
\end{align}
The gauge fields are:
\begin{equation}
C^{(3)} ~=~   - e^0 \wedge e^1 \wedge e^2 ~+~ \frac{1}{3!}\, \epsilon_{ijk\ell} \,  \big((\partial_z w )^{-1}\, (\partial_{u_\ell} w) \,  \d u^i \wedge \d u^j \wedge \d u^k ~-~ (\partial_{v_\ell} w)  \, \d v^i \wedge \d v^j \wedge \d v^k  \big)  \,.
 \label{C3gen}
\end{equation}
We first smear along one of the directions of $\mathbb{R}^4_u$, so we make the solution completely independent of $u_4$, and then dimensionally reduce along this direction to obtain:
\begin{align}
    \d s_{10}^2 = e^{\frac{3}{2}A_0}\big( -\partial_z w \big)^{-\frac{1}{4}} \Bigg( \,  - & \d t^2 +  \d y^2   + e^{-3A_0} \Big( \big( -\partial_z w \big)^{-\frac{1}{2}} \, \d \vec{ u} \cdot \d \vec{u} ~+~ \big( -\partial_z w \big)^{\frac{1}{2}} \, \d \vec{v} \cdot \d \vec{v} \, \Big)  \nonumber \\
    &  + \big( -\partial_z w \big) \,\left(  \d z + \big( \partial_z w \big)^{-1} \, \big( \nabla_{\vec{u}} w \cdot \d \vec{ u})\right) ^2\Bigg)\,,
\end{align}
with gauge fields and dilaton given by:
\begin{align}
    &e^{2 \Phi}=e^{-\frac{3}{2}A_0}(-\partial_z w)^{-3/4} \,,\quad C^{(3)}=-\frac{\varepsilon_{ijkl}}{6}(\partial_{v_i} w) \, \d v^j \wedge \d v^k\wedge \d v^l\,,\quad B^{(2)}= \frac{\varepsilon_{ijk}}{2} \frac{\partial_{u_i} w}{\partial_z w} \,\d u^j \wedge \d u^k\\
    & C^{(3)}=-e^{3 A_0} (-\partial_z w)^{1/2} dt \wedge   \d y \wedge \left( \d z + (\partial_z w)^{-1}  \big( \vec{\nabla} w \big)\cdot \d \vec{u}\right)\,. 
\end{align}
Now we smear the solution along one of the directions of $\mathbb{R}^4_v$, say $v_4$ and we T-dualize along this direction and obtain:
\begin{align}
    \d s_{10}^2 = e^{\frac{3}{2}A_0}\big( -\partial_z w \big)^{-\frac{1}{4}} \Bigg( \,  & -  \d t^2  +  \d y^2  +   \d v_4^2  +  e^{-3A_0} \Big( \big( -\partial_z w \big)^{-\frac{1}{2}} \, \d \vec{ u} \cdot \d \vec{u} ~+~ \big( -\partial_z w \big)^{\frac{1}{2}} \, \d \vec{v} \cdot \d \vec{v} \, \Big) \nonumber \\
    &+ \big( -\partial_z w \big) \,\left(  \d z + \big( \partial_z w \big)^{-1} \, \big( \nabla_{\vec{u}} w \cdot \d \vec{ u})\right) ^2\Bigg)\,,
\end{align}
where now $\vec{u},\,\vec{v}$ have only indices $1,2,3$ . The gauge fields and dilaton are given by:
\begin{align}
    &e^{2 \Phi}=(-\partial_z w)^{-1}\,,\quad C^{(2)}=-\frac{\varepsilon_{ijk}}{2}(\partial_{v_i} w) \, \d v^j \wedge \d v^k\,,\quad B^{(2)}= \frac{\varepsilon_{ijk}}{2} \frac{\partial_{u_i} w}{\partial_z w} \,\d u^j \wedge \d u^k \,,\\
    & C^{(4)}=-e^{3 A_0} (-\partial_z w)^{1/2} dt \wedge   \d x_1 \wedge  \d x_2\wedge  \left( \d z + (\partial_z w)^{-1}  \big( \vec{\nabla} w \big)\cdot \d \vec{u}\right)\,.     
\end{align}

\section{Supersymmetries in M theory}
\label{app:susies}

We can use the results of Section \ref{ss:mapping} to verify directly that the solutions do indeed satisfy the BPS equations.  This exercise is much simpler in M-theory.

Since the coordinates $(u,v,z)$ are the brane directions, this leads to a natural system of frames:
\begin{equation}
\begin{aligned}
e^0 ~=~  &\big(\mu \, f_4 \big)^{4/3} \, dt \,, \qquad \quad e^1~=~ \big(\mu \, f_4 \big)^{4/3} \, \d y \,, \\
e^2 ~=~  & \big(\mu \, f_4 \big)^{4/3} \, e^{-2\phi} \,\Big(  \d z + \big( \partial_z w \big)^{-1} \, \big( \partial_u w\big)\,   \d u\, \Big) \,,  \\
e^3~=~  &  \big(\mu \, f_4 \big)^{-2/3}   \,\bigg(\frac{N_2}{N_1} \bigg)^{1/4} \,        \d u\,,  \qquad \quad e^4~=~ \big(\mu \, f_4 \big)^{-2/3}   \,\bigg(\frac{N_1}{N_2} \bigg)^{1/4} \,   \d v\,, \\
 e^5~=~ & \big(\mu \, f_4 \big)^{-2/3} \, \bigg(\frac{N_2}{N_1} \bigg)^{1/4} \, u \, \d \theta_1 \,, \qquad \quad e^6~=~  \big(\mu \, f_4 \big)^{-2/3} \, \bigg(\frac{N_2}{N_1} \bigg)^{1/4} \, u \, \sin \theta_1 \,  \d \phi_1 \,,  \\
e^{7}~=~  &  \big(\mu \, f_4 \big)^{-2/3} \, \bigg(\frac{N_2}{N_1} \bigg)^{1/4} \,  \d \lambda_1   \,, \qquad \quad e^8~=~  \big(\mu \, f_4 \big)^{-2/3} \, \bigg(\frac{N_1}{N_2} \bigg)^{1/4} \, v \, \d \theta_2 \,, \\
e^{9}~=~  &  \big(\mu \, f_4 \big)^{-2/3} \, \bigg(\frac{N_1}{N_2} \bigg)^{1/4} \,v \, \sin \theta_2  \, \d \phi_2  \,, \qquad \quad e^{10} ~=~  \big(\mu \, f_4 \big)^{-2/3} \, \bigg(\frac{N_1}{N_2} \bigg)^{1/4} \, \d \lambda_2 \,,
\end{aligned}
 \label{11frames2}
\end{equation}
where one uses (\ref{uvzmap1}) to replace $(u,v,z)$ along with $w$ and its derivatives.  

Note that $e^3, e^5, e^6, e^7$ and $e^4, e^8, e^9, e^{10}$ are conformal multiples of frames on two copies of a flat $\IR^4$. These two $\IR^4$'s are spanned by the coordinates, $(u,\theta_1,\phi_1, \lambda_1)$ and $(v,\theta_2,\phi_2, \lambda_2)$.  One of the flat directions defined by $\lambda_j$ is used to compactify to IIA supergravity, while the other direction is used for the T-duality to get to IIB. This second direction is then promoted to a Poincar\'e AdS coordinate, extending $(t, y)$ to yield  Poincar\'e AdS$_4$.

The analysis of the supersymmetries proceeds exactly as in \cite{Bena:2023rzm}.  The crucial difference here is that the eleven-dimensional metric has explicit powers of $\mu$ which implies that this metric is only conformally AdS invariant.  This is an artifact of the compactification and dualization. The IIB metric is, of course, AdS invariant.

The eight Killing spinors of the system are  defined in terms of the frame components along the M2 and M5 directions: 
\begin{equation}
 \Gamma^{012} \, \varepsilon  ~=~ \eta_1  \, \varepsilon \,,   \qquad  \Gamma^{013567} \, \varepsilon  ~=~ \eta_2 \, \varepsilon \,.
 \label{projs1}
\end{equation}
The parameters $\eta_j$ are real with $|\eta_j|=1$ and set the brane orientations and so can be chosen at will.

Also note that in eleven dimensions one has 
\begin{equation}
 \Gamma^{0123456789\,10}    ~=~ \oneone \,,  
 \label{prodgammas}
\end{equation}
so equation  (\ref{projs1}) implies that
\begin{equation}
 \Gamma^{01489\,10} \, \varepsilon  ~=~ -\eta_1\, \eta_2\,\varepsilon \,,
 \label{projs2}
\end{equation}
and hence adding a second  set of M5 branes along $01789\,10$, which we denoted as  M5', does not break supersymmetry any further.

One then solves the gravitino equation 
\begin{equation}
\delta \psi_\mu ~\equiv~ \nabla_\mu \, \epsilon ~+~ \frac{1}{288}\,
\Big({\Gamma_\mu}^{\nu \rho \lambda \sigma} ~-~ 8\, \delta_\mu^\nu  \, 
\Gamma^{\rho \lambda \sigma} \Big)\, F_{\nu \rho \lambda \sigma} ~=~ 0 
\label{11dgravvar}
\end{equation}
to determine the metric and three-form vector potential of this system. 

One then finds that there are indeed eight supersymmetries for the metric defined by (\ref{11frames2}) and for the flux:
\begin{equation}
C^{(3)} ~=~   \eta_1  \, e^0 \wedge e^1 \wedge e^2 ~+~ \eta_1  \,\eta_2  \,b_1 \, \sin \theta_1 \, \d \theta_1 \wedge \d \phi_1 \wedge \d \lambda_1 ~-~ \eta_2  \,b_2 \, \sin \theta_2 \, \d \theta_2 \wedge \d \phi_2  \wedge \d \lambda_2 \,.
 \label{C3gen2}
\end{equation}
The Killing spinor is then given by:
\begin{equation}
\varepsilon~=~   \big(\mu \, f_4 \big)^{-2/3} \, \varepsilon_0 \,,
\end{equation}
where $\varepsilon_0$ is a constant spinor satisfying the projection conditions (\ref{projs1}).  Note that this is consistent with the well-established result:
\begin{equation}
\bar \varepsilon \,\Gamma^\mu \, \varepsilon \, \frac{\partial}{\partial x^\mu}~=~  c\, \frac{\partial}{\partial t} \,,
\end{equation}
for some constant, $c$.

\section{Probe Bulges : D2-D6 and D1-D7 systems}
\label{appendix:Probe bulges}
The results we present apply to all ``Hanany-Witten'' brane systems, and can be illustrated in different duality frames. When the intersecting branes are D-branes with 8 Neumann-Dirichlet boundary conditions for open strings, the Hanany-Witten effect produces a fundamental string when the branes cross. Consider, for instance, a system of D2-branes extended along $(x_7,x_8)$ in the presence of D6-branes spanning $(x_1,x_2,x_3,x_4,x_5,x_6)$. Although this configuration preserves $\frac{1}{4}$ of the supersymmetry, a D2-brane probing the background of a D6-brane experiences a repulsive force and requires half the tension of a fundamental string to remain at a finite distance from the D6-brane.

This brane system is particularly illuminating because it has a natural uplift to M-theory, where it corresponds to an M2-brane probing the background of a Kaluza-Klein monopole. In this description, the M2-brane follows a holomorphic curve in the Taub-NUT space, which can be explicitly integrated \cite{Marolf:2000ci,Kubota:2000qn,Bachas:1999um}. We review this in subsection \ref{D2D6}.

Another interesting brane system, which we review in subsection \ref{D1D7}, consists of D1 branes in the D7 geometry, and admits a clear pictorial representation providing valuable insight into the underlying physics.

\subsection{D2 probe in a D6 background}
\label{D2D6}
We begin by introducing complex coordinates on the Kaluza–Klein monopole background, which will be crucial for the M2-brane probe analysis that follows.
\subsubsection*{D2 branes in D6 background}
The D6 brane can be described in M-theory as a Kaluza-Klein monopole, described by the Taub-NUT metric, times a seven-dimensional Minkowski space. The M-theory metric is:
\begin{equation}
ds_{11}^2 =\;  -(dt)^2\,  +\,\sum_{i=1}^6 ( \d x_i)^2\;  +\,  V\,
 dx_pdx^p + V^{-1}
(  \d x_{10} + {A_p}d{x^p})^2\ , 
\end{equation}
with $p=7,8,9$ spanning the $\textbf{R}^3$ base space, and $V$ a harmonic function on this base, with poles at the location of the KK-monopole:

\begin{equation}
     V = 1 + \frac{1}{2\sqrt{x_px^p}} \, .
\end{equation}
As is well known for Taub-NUT geometries, $x_{10}$ is $2\pi$ periodic (this is a choice). The one form $A$
satisfies ${\vec \nabla} \times
{\vec A} = {\vec\nabla} V$.

The holomorphic structure in Taub-NUT depends on the choice of one specific direction in the $\textbf{R}^3$ base that we choose to be $x_9$, and write the orthogonal plane as $\mathbb{C}$. With these choices, the holomorphic coordinates on Taub-NUT are:
\begin{equation}
    v = x_7+ix_8\, ,\quad
    w  = e^{-(x_9+ix_{10})} \left( - x_9 + \sqrt{ x_9^2 + \vert v^2 \vert} \right)^{1/2}.
    \label{singleD6}
\end{equation}
In the presence of multiple D6-branes, the metric takes the same Gibbons-Hawking form with:
\begin{equation}
    V+1+\sum_{i}^{Nf}\frac{1}{2 |x-x_i|}\,.
\end{equation}
The holomorphic coordinates on such GH space are given by:
\begin{equation}
    v = x_7+ix_8\, ,\quad\,,w  = e^{-(x_9+ix_{10})} \Pi_{i}^{Nf}\left( \mid \vec{x}-\vec{x}_i\mid - (x_9-x_{9\,i})
  \right)^{1/2} \,,
\end{equation}
with $\vert \vec{x}-\vec{x}_i\vert=\sqrt{\vert v-v_i\vert^2+(x_9-x_{9\,i})^2}$.

\subsubsection*{One D2 in a single D6 background}

We first probe the background sourced by a single D6 brane working in eleven dimensions where the physics is most clear. The M2 brane probe is required by supersymmetry to follow a holomorphic curve in the Kaluza-Klein monopole background.

If there is only one D6 brane and one D2 brane, there are two possible choices of holomorphic curve. The first choice is $w=e^{-b}$, with $b$ a real constant. For this solution $x_{10}$ is constant. The second holomorphic curve is $w=e^{-b}v$, and now $x_{10}=\text{arg}\left(x_7+i x_8\right)$.  When going around the position of the D6 brane, the M2 brane winds once around $x_{10}$. Note that there is a coordinate singularity on the positive $x_{9}$ axis at the position of the D6 brane on the $x_9$ axis.

It is easy to find $|v|$ for the solution that does not wind around $x_{10}$:
\begin{equation}
    | v|= \sqrt{e^{4(x_9-b)}+2 x_9 e^{2(x_9-b)}}\,.
\end{equation}
From this expression, we see that the D2 brane extends along the $x_9$ axis all the way from infinity, with asymptotics $|v|\sim e^{2x_9}$. For some finite negative value of $x_9$, the argument of the square-root becomes negative and is ill-defined. This is the minimal value of $x_9$ reached by the probe M2 brane. 

In the winding solution:
\begin{equation}
    |v|=\sqrt{e^{-4(x_9-b)}-2x_9e^{-2(x_9-b)}}\,,
\end{equation}
where $x_9$ now ranges from $-\infty$, where $|v|\sim e^{-2x_9}$, up to a finite positive value. Because there is one unit of winding along $x_{10}$, the Type IIA corresponding solution has one fundamental string stretched between the D6 brane and D2 brane. 

To summarize, in one solution the D2 is pulled to the right, while in the other it is pulled to the left. In both solutions it is not possible to have a straight asymptotic D2 brane. This would require a D2 brane with zero linking number, and this can only be realized if half a fundamental string is stretched between the D6 and the D2 brane.

In the next subsection we will engineer a solution with zero linking number and hence straight D2 asymptotics by considering a D2 brane probe in the solution sourced by two D6 branes at $x_9=\mid v\mid=0$. 

\subsubsection*{Zero linking number and flat D2 asymptotics}
To engineer a situation with two D6 branes and one D2 brane, with only one fundamental string stretched between them, we need to choose a holomorphic curve given by a polynomial that is first order in both $v$ and $w$. The holomorphic curve is
\begin{equation}
    w  = e^{-(x_9+ix_{10})}\left( \sqrt{\mid v\mid^2+(x_9)^2} - (x_9)
  \right)^{1}=e^{-b} v.
\end{equation}
where we display the power of the bracket, $(\quad )^1$, explicitly, to highlight the difference with the solution in which there is a single D6 brane \eqref{singleD6}. The holomorphic curve winding once around $x_{10}$ is
\begin{equation}
    \mid v\mid= \frac{2 x_9 e^{x_9-b}}{\mid 1-e^{2(x_9-b)}\mid}
\end{equation}
and we clearly see that $\mid v\mid \rightarrow \infty$ occurs at a finite value of $x_9$, which indicates an asymptotically flat D2 brane, with no asymptotic bending. This makes sense, as there is a single F1 string running between the D6 branes and the D2 brane, so the D2 brane has zero linking number. Note that as one changes the distance between the branes, the shape of the D2 changes, but the asymptotic bending remains always the same.

\subsection{D1-D7: a pictorial story}
\label{D1D7}

We can also see the physics of probe branes with Hanany-Witten strings by considering a probe D1 brane in a background sourced by D7 branes with  transverse coordinates $(x,y)$. The effect of a D7 brane is to create a deficit angle, and we can choose the orientation of the monodromy cut. This cut causes D1 strings to be bent by the presence of the D7 brane, and to acquire an F1 charge. The picture is: 
\begin{equation}
    \raisebox{-.5\height}{\begin{tikzpicture}
\draw[dotted] (0,0)--(2,0);
\node[D5] at (0,0) {};
\draw [red] (1,-1)--(1,0)--(0,1);
\node at (3,0) {\footnotesize $\left( \begin{array}{cc}
1 & -1 \\ 0 & 1
\end{array} \right)$};
\node at (1,-1.5) {\color{red} \footnotesize $\left( \begin{array}{c}
0 \\ 1
\end{array} \right)$};
\node at (0,1.5) {\color{red} \footnotesize $\left( \begin{array}{c}
-1 \\ 1
\end{array} \right)$};
\end{tikzpicture}}
\end{equation}

Let $N \geq 1$ be an integer, $\theta_{\max} \in [0 , \frac{\pi}{2}]$ and $g_s >0$. 
We place $2N$ D7 branes at the origin $(x,y) = (0,0)$, with monodromy cuts spread uniformly within the angular sector $[-\theta_{\max} , \theta_{\max}]$. 
These monodromy cuts play the same role as the NS5 blast in the NS5-D5 probe calculation of section \ref{sec:Probe D5 in NS5 background}.
The shape of a D1 brane in this background is piecewise linear, with kinks at the intersection with the monodromy cuts. For instance, if we take $N=4$ and $\theta_{\max} = \frac{\pi}{3}$, we get for $g_s = 1$
\begin{center}
    \includegraphics[scale=.4]{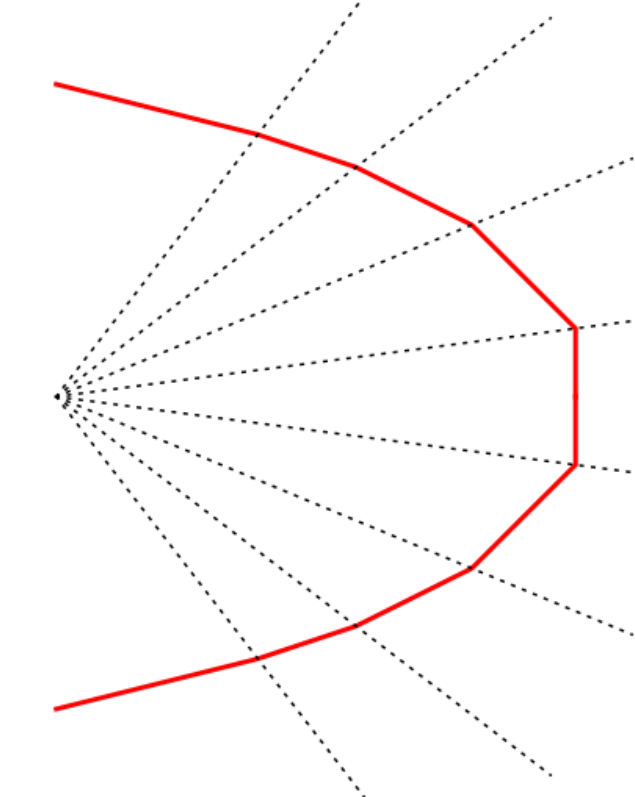}
\end{center}

A simple computation shows that the broken line joins points of coordinates $(x_n , y_n)$ for $n = -N , \dots , +N$ which can be obtained by solving recursively the system of equations 
\begin{equation}
    \frac{x_{n+1}-x_n}{y_{n+1}-y_n} = - g_s n \, , \qquad \frac{y_n}{x_n} = \tan \left[ \left( n - \frac{1}{2}\right) \frac{\theta_{\max}}{N} \right] \,.
\end{equation}

Letting $N$ be large, the piecewise linear function becomes smooth in the angular domain. 
The equation of the curve can be found by taking the continuum limit of the above system. We consider the equation of the curve to be $x = f(y)$. The second equation gives in the large $N$ limit 
\begin{equation}
    n \sim n + \frac{1}{2} = \frac{N}{\theta_{\max}} \arctan \frac{y}{f(y)} \,,
\end{equation}
which we insert in the first equation to find 
\begin{equation}
 f'(y) = - g_s \frac{N}{\theta_{\max}} \arctan \frac{y}{f(y)} 
\end{equation}
with the initial condition $f(0) = 1$. 
We are not aware of a closed form solution to this equation. 

We can still compare with the exact result by numerically solving the equation. For instance, for $N=20$, $g_s = \frac{1}{5}$ and $\theta_{\max} = \frac{\pi}{3}$ we find 
\begin{center}
    \includegraphics[scale=.6]{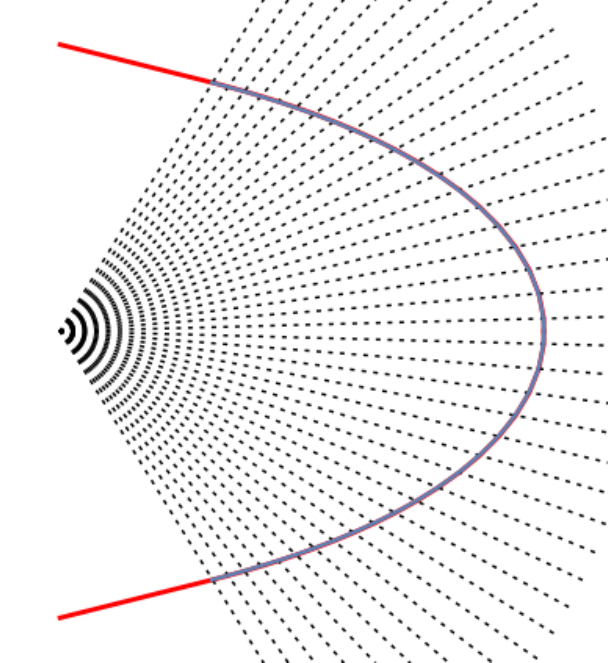}
\end{center}
where the red curve is the exact solution and the blue curve is the numerical solution to the differential equation above in the angular domain.

\bibliographystyle{utphys}      

\bibliography{references}      

\end{document}